\documentclass[times]{aastex701}
\usepackage{amssymb}
\usepackage[normalem]{ulem}
\usepackage{xcolor}
\usepackage{lipsum}
\usepackage{supertabular}
\usepackage{tikz}
\usetikzlibrary{svg.path}
\definecolor{orcidlogocol}{HTML}{A6CE39}
\tikzset{orcidlogo/.pic={
  \fill[orcidlogocol] svg{M256,128c0,70.7-57.3,128-128,128C57.3,256,0,198.7,0,128C0,57.3,57.3,0,128,0C198.7,0,256,57.3,256,128z};
  \fill[white]
    svg{M86.3,186.2H70.9V79.1h15.4v48.4V186.2z}
    svg{M108.9,79.1h41.6c39.6,0,57,28.3,57,53.6c0,27.5-21.5,53.6-56.8,53.6h-41.8V79.1z M124.3,172.4h24.5c34.9,0,42.9-26.5,42.9-39.7c0-21.5-13.7-39.7-43.7-39.7h-23.7V172.4z}
    svg{M88.7,56.8c0,5.5-4.5,10.1-10.1,10.1c-5.6,0-10.1-4.6-10.1-10.1c0-5.6,4.5-10.1,10.1-10.1C84.2,46.7,88.7,51.3,88.7,56.8z};
}}
\newlength{\orcidXht}
\newcommand{\orcidicon}{%
  \setlength{\orcidXht}{\fontcharht\font`X}%
  \mbox{\begin{tikzpicture}[
    yscale=-0.00390625\orcidXht,
    xscale= 0.00390625\orcidXht,
    transform shape]
    \pic{orcidlogo};
  \end{tikzpicture}}%
}

\usepackage{adjustbox}
\usetikzlibrary{shapes.geometric, arrows, positioning}
\usepackage{booktabs}
\usepackage{tabularx}
\usepackage{makecell}
\usepackage{amsmath}
\usepackage{hyperref}
\usepackage{float}

\usepackage{titlesec}
\titleformat{\paragraph}
{\normalfont\normalsize}   
{\theparagraph}{1em}{}
\titlespacing*{\paragraph}
{0pt}{3.25ex plus 1ex minus .2ex}{1ex}

\begin{document}

\title{Physics-Informed Neural Networks and Data-Driven Models for GRB X-ray Light-Curve Gap Reconstruction}

\author[orcid=0009-0006-0569-9051,gname=Ayush,sname=Garg]{A. Garg}
\altaffiliation{Corresponding author 2}
\affiliation{Department of Astronomy, Astrophysics \& Space Engineering, Indian Institute of Technology, Khandwa Road, Simrol, Indore, Madhya Pradesh, 453552, India}
\email[show]{ayushg827@gmail.com} 

\author[orcid=0009-0009-0399-9140]{Ritik Kumar}
\affiliation{Department of Physics, Indian Institute of Science, Malleshwaram, Bengaluru, Karnataka, 560012, India}
\email{hrithikk048@gmail.com} 

\author[orcid=0000-0003-4442-8546,gname=Maria,sname=Dainotti]{M. G. Dainotti}
\altaffiliation{Corresponding author 1}
\affiliation{Division of Science, National Astronomical Observatory of Japan, 2-21-1 Osawa, Mitaka, Tokyo, 181-8588, Japan}

\affiliation{The Graduate University for Advanced Studies (SOKENDAI), Shonankokusaimura, Hayama, Miura District, Kanagawa, 240-0115, Japan}

\affiliation{Space Science Institute, 4765 Walnut St Ste B, Boulder, 80301, CO, USA}

\affiliation{Nevada Center for Astrophysics, University of Nevada, 4505 Maryland Parkway, Las Vegas, 89154, NV, USA}
\email[show]{mariagiovannadainotti@yahoo.it} 

\author[orcid=0009-0009-0964-3524]{V. Sharma}
\affiliation{Indian Institute of Science Education and Research Bhopal, Bhopal Bypass Road, Bhauri, Bhopal, 462066, Madhya Pradesh, India}
\email{vipul22@iiserb.ac.in} 

\author[orcid=0000-0002-5656-2657]{A. Shukla}
\affiliation{Department of Astronomy, Astrophysics \& Space Engineering, Indian Institute of Technology, Khandwa Road, Simrol, Indore, Madhya Pradesh, 453552, India}
\email{amit.shula@iiti.ac.in} 
                
\author[orcid=0000-0002-8028-0991]{D. H. Hartmann}
\affiliation{Department of Physics and Astronomy, Clemson University, Clemson, SC 29634, USA}
\email{hdieter@g.clemson.edu} 

%
%
\begin{abstract}
Temporal gaps in Swift-XRT X-ray afterglows of gamma-ray bursts (GRBs) limit how precisely the standard empirical plateau function constrains three parameters: the plateau end time $T_a$, the plateau flux $F_a$, and the post-plateau decay index $\alpha$. 
These parameters underpin the luminosity-time and fundamental-plane correlations of GRB plateaus so that smaller measurement uncertainties can improve the statistical power of GRBs as cosmological probes. 
As the fifth study in a series on light-curve reconstruction, this work benchmarks four models on a parent sample of 545 Swift-XRT GRBs: a Physics-Informed Neural Network (PINN) with seven afterglow-motivated priors, ReFANN, a Siamese dual-branch network with Monte Carlo dropout, and Polynomial Quantile Regression (PQR). 
All four reduce the fractional uncertainties in $\log T_a$, $\log F_a$, and $\alpha$ relative to the original observations. 
The PINN with a broken power-law prior gives the largest reductions ($\sim$41–49\%), at a higher outlier rate ($\sim$15–20\%), while ReFANN, Siamese network and PQR give more modest reductions ($\sim$19–27\%) with outlier fractions $\lesssim$4\%, across a wider range of light-curve morphologies. 
A reduced-$\chi^2$ prior-selection scheme recovers a morphological classification consistent with independent labelling, though it does not match the best single-prior reduction. 
Together, these results provide a systematic benchmark of physics-informed and data-driven reconstruction strategies for irregularly sampled GRB afterglows.
\end{abstract}
%
%
\maketitle
%
\section{Introduction} \label{sec:intro}
The increasing availability of time-domain observations has transformed modern astrophysics, enabling detailed studies of transient phenomena across a wide range of temporal and spatial scales \cite{Graham2019, Ivezi2019}.
Gamma-ray bursts (GRBs) represent one of the most challenging classes of transient events for time-series analysis. 
Their X-ray afterglows exhibit diverse temporal behaviours, including steep decay phases, plateau emission, flares, and late-time transitions spanning several orders of magnitude in both flux and time \citep{OBrien2006, Nousek2006, Zhang2006}.

These behaviours are broadly explained within the relativistic fireball framework \citep{PIRAN1999}, in which the afterglow emission arises from synchrotron radiation produced by electrons accelerated in the forward shock expanding into the circumburst medium.
Accurate characterisation of these features is essential for understanding the physics that drives afterglow evolution and for exploiting empirical plateau correlations. 
The relation between plateau luminosity and rest-frame plateau duration \citep{Dainotti2008, Dainotti2010}, whose slope survives selection-bias correction \citep{Dainotti2013}, and its extension to a fundamental plane that includes the prompt peak luminosity \citep{Dainotti2016, Dainotti2020} have been used to standardise GRBs; \citet{Wang2015} and \citet{Bargiacchi2025} gave the reviews of GRB cosmology, and \citet{Khadka2021} constrained the independent test, and an application of the fundamental plane as a cosmological tool by \citet{Dainotti2023a}.

Observational gaps in GRB X-ray light-curves may prevent the identification of plateau phases, flares, and temporal breaks, thereby affecting the measurement of necessary light-curve parameters and the interpretation of the underlying physical processes \citep{Zhang2006, Nousek2006}; \cite{Willingale2007}(hereafter, W07); \citep{Racusin2009, Dainotti2021}.
Standard data-driven interpolation methods do not enforce the physical structure of synchrotron emission in an external shock; they are therefore free to produce reconstructions that are locally smooth but physically implausible, for example, recovering a positive flux gradient in a sparsely sampled region where synchrotron theory demands a decline.

PINNs address this by embedding governing equations directly into the training objective, penalising solutions that violate the chosen physical prior while still remaining free to fit the data \citep{Raissi2019, Karniadakis2021, Cuomo2022a}.
This approach has been demonstrated across computational astrophysics: coronal magnetic-field extrapolation \citep{Jarolim2023, Baty2024}, astrophysical shock modelling \citep{Moschou2023}, and stellar structure equations \citep{Baty2023}, but, to our knowledge, it has not previously been applied to GRB afterglow reconstruction.

Neural network (NN)-based function reconstruction has also been applied successfully to cosmological datasets \citep{Dialektopoulos2023, Dialektopoulos2024}, demonstrating that embedding physical constraints into a network's training objective improves parameter recovery over purely data-driven interpolation, the same principle motivating our PINN approach in the GRB context.
A closely related model-independent neural-network calibration of the GRB Hubble diagram itself has recently been demonstrated by \cite{Mukherjee2026}, illustrating a direct cosmological application for the uncertainty-reduced $T_a$, $F_a$, and $\alpha$ values produced by reconstruction pipelines such as ours.


Consequently, the development of reliable light-curve reconstruction (LCR) techniques has become increasingly important for analysing irregularly sampled astronomical time series \citep{VanderPlas2018}. 
A robust reconstruction method should accurately recover the temporal evolution of the source, preserve key light-curve features, and provide reliable uncertainty estimates in sparsely sampled regions.

Recent advances in machine learning and probabilistic modelling have provided powerful tools for reconstructing incomplete astrophysical time series. 
Methods based on Gaussian Processes (GPs), recurrent NNs, and other deep learning architectures have demonstrated promising performance in recovering missing observations while quantifying predictive uncertainty \citep{Hochreiter1997, Roberts2013, Aigrain2023}.

Prior work in this series has progressively expanded the reconstruction toolkit. 
\cite{Dainotti2023} introduced the stochastic reconstruction framework on 207 good GRBs. 
\cite{Manchanda2025} extended this to 521 GRBs, testing deep learning architectures including multilayer perceptron (MLP), Bi-Mamba, , bidirectional long short-term memory (LSTM) Fourier transform, Gaussian process-random forest hybrid (GP--RF), conditional generative adversarial networks (CGAN), SARIMAX-based Kalman filter, Kolmogorov–Arnold networks (KAN), and Attention U-Net.
\cite{Kaushal2026} further systematised this comparison across the same expanded sample, adding Deep Gaussian Process (DGP), Temporal Convolutional Network (TCN), Hybrid CNN with Bidirectional Long Short-Term Memory (CNN-BiLSTM), Bayesian Neural Network (BNN), Polynomial Curve Fitting, Isotonic Regression, and Quartic Smoothing Spline (QSS).
Most recently, \cite{Gupta2026} introduced five additional models: Kernel Ridge Regression (KRR), Cubic Smoothing Spline (CSS), Gaussian Process via ANN (GP via ANN), Artificial Neural Networks (ANN), and Symbolic Regression (SR), finding that the CSS model achieved the highest uncertainty reduction across all three parameters among all models tested to that point, including a 0\% outlier rate, with 521 GRB samples, for the ``Good GRBs" subclass.
A complementary study by \citet{Zhou2025} applied stochastic reconstruction to 35 plateau GRBs, recalibrated three luminosity correlations and constrained flat and non-flat $\Lambda$CDM; the added light-curve points left the intrinsic scatter and the cosmological constraints unchanged within uncertainties, suggesting that sample size matters more than light-curve sampling. 

Building upon these four preceding studies, the present work extends the LCR framework by introducing four additional models: Physics-Informed Neural Networks (PINNs), Reconstruct Functions with Artificial Neural Networks (ReFANN), a modified Siamese NN, and PQR model, on a 545-burst parent sample. 
While \cite{Gupta2026} demonstrated that the data-driven CSS model can surpass the 47.5\% uncertainty-reduction threshold identified by \cite{Dainotti2022a}
as sufficient to match the cosmological precision of Type Ia supernovae, it does so without incorporating any physical knowledge of GRB afterglow dynamics into the reconstruction process.
The present work investigates whether physics-informed architectures, which embed the governing equations of synchrotron afterglow evolution directly into the training objective, can achieve comparable or superior uncertainty reductions while also providing physically interpretable reconstructions in sparsely observed regions.

ReFANN is a neural-network-based reconstruction framework that combines predictive uncertainty estimation with physics-inspired gradient regularisation to recover incomplete light-curves without assuming a predefined analytical model. 
The modified Siamese NN employs a shared-weight dual-branch architecture together with physics-informed monotonicity regularisation to learn robust latent representations of GRB light-curves.
These methods represent complementary approaches to non-linear function approximation, uncertainty-aware learning, and representation learning.
By evaluating them within a common reconstruction framework, we aim to further explore the capabilities of modern techniques for recovering missing segments of GRB X-ray light-curves and preserving their underlying temporal behaviour.

This paper extends our earlier work in this series,
and is organised as follows: Section \S\ref{sec:methods} provides an in-depth description of the dataset and the models used to reconstruct GRB LCs.
\S\ref{section:results} reports the uncertainty, performance, and outlier statistics. 
\S\ref{section:Discussion} discusses these results in the context of prior GRB reconstruction work, and \S\ref{section:conclusion} provides the synopsis and conclusions.

\section{Methodology} \label{sec:methods}

This section presents the architectural designs, physical foundations, mathematical formulations, and multi-stage optimisation framework developed for GRB X-ray light-curve reconstruction.
This study systematically evaluates the performance of the PINN architecture alongside ReFANN, Siamese NNs, and the PQR model under an identical experimental framework. Particular emphasis is placed on incorporating seven empirically grounded physics-based priors governing GRB afterglow evolution into the PINN model, thereby enabling a consistent comparison across all reconstruction approaches.

\subsection{Motivation for Physics-Informed, Parametric and Non-Parameteric Reconstruction} \label{Motivation}

X-ray afterglow observations acquired by space-based observatories such as the Swift X-ray Telescope (XRT) \citep{Gehrels2004, Burrows2005}exhibit temporal variability across six orders of magnitude in time ($10^1 \text{--} 10^7\text{ s}$) and flux ($10^{-14} \text{--} 10^{-8}\text{ erg cm}^{-2}\text{ s}^{-1}$) \citep{Nousek2006}; W07. 
These light-curves feature complex morphological structures, including early steep decays, extended plateau phases, energetic flares, relativistic jet breaks, and late-time power-law (PL) decays \citep{Zhang2006}.

Standard parametric fitting routines require a priori selection of a rigid functional template. 
When orbital gaps, Earth-occultation periods, or low signal-to-noise passages obscure key transitions (such as the plateau break time $T_a$), unconstrained parametric fits frequently converge to unphysical parameter spaces. 
Conversely, purely data-driven, non-parametric interpolations (e.g., standard cubic splines or unconstrained NNs) risk generating arbitrary oscillations within observational gaps.

To address these limitations, a hybrid framework is introduced. 
By embedding analytical afterglow physics regularisers directly into the objective function of PINN \citep{Raissi2019}, missing flux sequences are constrained to satisfy hydrodynamical decay laws while preserving the flexible interpolation capacity of deep NNs. 
To rigorously benchmark the performance of the PINN, three non-parametric models are evaluated on the exact same dataset.

\subsubsection{PINN} \label{PINN motivation}

The core motivation for including a PINN in this framework is threefold. 
Firstly, GRB afterglows obey well-understood PL decay laws derived from synchrotron emission theory; the PINN can exploit this structure to constrain reconstructions in data-sparse gaps rather than reverting to an unconstrained smooth interpolant.

In addition, unlike purely data-driven models (e.g., QSS, isotonic regression, or polynomial fitting), the PINN explicitly models the physical slope of the light-curve at every point, which is particularly valuable near the plateau endpoint $T_a$, where small errors in slope translate directly into large errors in the recovered Willingale parameters: $\log T_{a}$, $\log F_{a}$, and $\alpha$.
Furthermore, the architecture is fully differentiable end-to-end, so the physics constraint can be imposed as a soft regularizer without requiring prior knowledge of which physical regime (plateau, normal decay, jet break, etc.) applies to a given burst.

We adopt a value-space physics regularisation rather than the classical ODE-derivative-matching form (\cite{Raissi2019}) for the reasons detailed in Section~\ref{PINN pre-processing}. 
The resulting model interpolates temporal gaps with physically plausible trajectories while preserving observed flux values and their uncertainties.

\subsubsection{Siamese} \label{Siamese}

Siamese NNs use two parallel branches with shared parameters to learn patterns from the input data \citep{Bromley1993}.
We expect this architecture to perform well for GRB X-ray light-curve reconstruction because it can capture important temporal features while maintaining a consistent representation of the data. 
The shared structure may help the network better handle irregular sampling and observational gaps, which are common in GRB light-curves. 
In addition, Monte Carlo (MC) dropout enables the model to estimate the uncertainty of its predictions, making it a promising approach for reconstructing missing portions of GRB light-curves \citep{Gal2016}.

\subsubsection{ReFANN} \label{ReFANN}

ReFANN (Reconstruct Functions with Artificial Neural Network) is a data-driven method that reconstructs light-curves without assuming any predefined mathematical form \citep{Wang2020}. 
We expect it to perform well for GRB X-ray light-curves because these light-curves often show complex behaviours, such as plateau phases, flares, and changes in decay rate, that can be difficult to describe with simple models.
Instead of forcing the data to follow a specific function, ReFANN allows the observations themselves to determine the shape of the reconstructed light-curve \citep{Wang2020}. Its ability to work with irregularly sampled data makes it a promising approach for recovering missing portions of GRB light-curves \citep{Wang2020}.

\subsubsection{Quantile} \label{Quantile}

Gamma-Ray Burst afterglows frequently exhibit asymmetric dispersion and heteroscedastic noise that evolve over time. Traditional regression models, which predict a single conditional mean under an assumption of constant variance, are ill-suited to this behaviour and are sensitive to outliers in irregularly sampled astronomical data \citep{VanderPlas2018}. PQR model was selected because it directly estimates several points of the conditional distribution of the flux rather than only its mean, yielding a robust, distribution-free characterisation of the light-curve together with physically motivated upper and lower bounds \citep{Koenker2005}.

\subsection{Dataset Selection, Processing \& Gap Identification} \label{Dataset Selection}

\subsubsection{Sample Selection} \label{Sample Selection}
We selected the GRBs used in this study from observations available in the Swift BAT-XRT repository \citep{Evans2007, Evans2009}. 
The parent sample comprises 545 GRBs in total: 455 \textit{Swift} GRBs recorded between 2005 and 2019 \citep{Dainotti2020, Srinivasaragavan2020}; four GRBs observed by \textit{Fermi}-LAT, with redshifts inferred via machine-learning ensemble methods \citep{Dainotti2024, Dainotti2024a}; 62 \textit{Swift} GRBs observed during the 2019--2023 period \citep{Narendra2025}; and 24 additional \textit{Swift} XRT GRBs with available W07 plateau parameters retrieved directly from the Swift BAT-XRT repository \citep{Evans2007, Evans2009} that fall outside the coverage of the catalogues above.

This study utilises the same parent dataset introduced in our earlier work in this series.
Unlike earlier papers in this series, we do not impose a fixed, common number of GRBs across all models. 
Instead, every model processes the complete parent sample of 545 GRBs, and the number of GRBs that survive to the final error-reduction statistics is allowed to vary from model to model, depending on whether a convergent W07 fit could be obtained for that GRB's reconstructed light-curve. 
This ranges from $N_{min} = 434~ (\textrm{Energy-injection prior})$ to $N_{max} = 524~ (\textrm{Siamese})$ out of 545 across models (Table~\ref{tab:reconstruction_comparison}).

We adopt this approach because forcing a uniform sample size across architecturally different models would either discard GRBs that some models handle perfectly well or artificially keep GRBs that a given model cannot reconstruct reliably, in both cases, biasing the reported improvement statistics. 
Each model's effective sample is therefore determined solely by its own fit convergence, using the same convergence and covariance-validity criteria for all models.
Within this parent set, 230 GRBs possess confirmed spectroscopic or photometric redshifts $z$; machine-learning classifiers have been developed to extend redshift coverage to GRBs currently lacking spectroscopic follow-up \citep{Dainotti2025a, Narendra2025}.

We use the GRB X-ray afterglow light-curve sample from our earlier studies.
Following the same pre-processing and selection criteria, we analyse the subset of GRBs for which the proposed reconstruction framework produces successful reconstructions and parameter estimation. 
The final sample used in this work, therefore, varies between models, depending on the reconstruction framework employed, as discussed above. 
As in our previous studies, the GRBs are grouped into four morphological classes based on their X-ray afterglow light-curves: (i) Good GRBs, (ii) Flares/Bumps, (iii) Break, and (iv) Flares/Bumps + Double Break. 
These classes are adopted from our previous studies and are used only to present representative reconstruction examples in this work.

\begin{table*}[!htp]
\centering
\renewcommand{\arraystretch}{1.2}

\begin{adjustbox}{width=0.9\textwidth, center}
\begin{tabular}{lcccccc}

\hline 
\textbf{Reconstruction Model} & \multicolumn{3}{c}{\textbf{Uncertainty Decrease}} & \multicolumn{3}{c}{\textbf{\% Outliers}} \\
\hline
 & \% log$_{10}$(T$_a$) & \% log$_{10}$(F$_a$) & \% $\alpha$ & \% logT$_a$ & \% logF$_a$ & \% $\alpha$ \\
\hline
\multicolumn{7}{c}{\textbf{Our Models}} \\
\hline
PINN &  &  &  &  &  &   \\
PL (436 GRBs) & -37.79 & -40.51 & -47.01 & 19.15 & 19.71 & 5.52  \\
BPL (450 GRBs) & \textbf{-40.65} & \textbf{-44.68} & \textbf{-48.83} & 14.79 & 16.82 & 3.33  \\
SBPL (441 GRBs) & -39.31 & -43.83 & -48.76 & 16.39 & 18.78 & 5.16  \\
MSM (445 GRBs) & -39.41 & -42.70 & -46.63 & 16.02 & 18.05 & 4.05  \\
JBM (443 GRBs) & -35.83 & -37.40 & -41.70 & 16.94 & 18.42 & 5.71  \\
EI (434 GRBs) & -38.37 & -40.70 & -46.98 & 19.34 & 20.07 & 5.52  \\
Synchrotron (439 GRBs) & -36.98 & -41.27 & -47.10 & 17.13 & 19.15 & 5.52  \\
Siamese (524 GRBs) & -22.88 & -22.83 & -26.93 & 3.12 & \textbf{3.68} & \textbf{1.10} \\
ReFANN (522 GRBs) & -19.83 & -21.43 & -26.51 & 3.31 & 4.04 & 1.65 \\
Quantile (522 GRBs) & -18.92 & -20.04 & -24.42 & \textbf{2.94} & 4.04 & 1.28 \\

\hline
\multicolumn{7}{c}{\textbf{521 Good GRBs \citep{Gupta2026}}} \\
\hline
Symbolic Regression & -32.8 & -33.4 & -46.8 & 3.071 & 4.414 & 1.535  \\
KRR & -28.5 & -29.4 & -33.9 & 2.687 & 2.879 & 1.151  \\
Cubic Smoothing Spline &\textbf{ -44.2} & \textbf{-44.5} & \textbf{-50.4} & 2.69 & 3.07 & \textbf{0.384}  \\
GP via ANN & -24.3 & -25.2 & -29.0 & 2.687 & 3.455 & 1.343  \\
ANN & -28.8 & -29.6 & -33.6 & 2.303 & 2.495 & 1.151  \\
QSS & -43.5 & -43.2 & -48.3 & 2.69 & 3.84 & 0.960  \\
Polynomial Curve Fitting & -20.8 & -21.6 & -27.4 & 2.88 & 2.30 & \textbf{1.15}  \\
Attention U-Net & -37.9 & -38.5 & -41.4 & 1.73 & 2.50 & \textbf{1.34}  \\
MLP & -37.2 & -38.0 & -41.2 & 1.73 & 2.30 & \textbf{1.34}  \\
GP (W07) & -16.9 & -18.6 & -24.3 & 3.07 & 3.45 & 1.54  \\
W07 model (10\%) & -18.0 & -19.1 & -25.2 & 2.30 & \textbf{2.30} & 2.11  \\
W07 model (20\%) & -15.8 & -17.8 & -23.8 & 2.30 & 2.69 & 2.30  \\

\hline

\multicolumn{7}{c}{\textbf{521 GRBs \citep{Kaushal2026}}} \\
\hline
QSS & \textbf{-43.5} & \textbf{-43.2} & \textbf{-48.3} & 2.69 & 3.84 &\textbf{0.960} \\
Polynomial Curve Fitting & -20.8 & -21.6 & -27.4 & 2.88 & 2.30 & 1.15 \\
CNN-BiLSTM          & -20.3 & -20.9 & -25.1 & 3.07 & 2.69 & \textbf{0.768} \\
Isotonic Regression & -18.0 & -18.5 & -24.0 & 3.07 & 2.88 & \textbf{0.960} \\
BNN              & -10.9 & -17.6 & -15.9 & 5.95 & 5.76 & 2.69 \\
DGP             & -11.6 & -12.3 & -15.9 & 6.91 & 5.95 & 1.92 \\
TCN             & -5.31 & -12.7 & -16.2 & 18.5 & 14.6 & 4.62 \\
Attention U-Net  & -37.9 & -38.5 & -41.4 & \textbf{1.73} & 2.50 & 1.34 \\
MLP              & -37.2 & -38.0 & -41.2 & \textbf{1.73} & 2.30 & 1.34 \\
GP (W07)         & -16.9 & -18.6 & -24.3 & 3.07 & 3.45 & 1.54 \\
W07 model (10\%) & -18.0 & -19.1 & -25.2 & 2.30 & 2.30 & 2.11 \\
W07 model (20\%) & -15.8 & -17.8 & -23.8 & 2.30 & 2.69 & 2.30 \\
\hline

\end{tabular}
\end{adjustbox}

\end{table*}

\begin{table*}[!htp] 
\centering
\renewcommand{\arraystretch}{1.2}
\begin{adjustbox}{width=0.9\linewidth, center}
\begin{tabular}{lcccccc}

\hline 
\textbf{Reconstruction Model} & \multicolumn{3}{c}{\textbf{Uncertainty Decrease}} & \multicolumn{3}{c}{\textbf{\% Outliers}} \\
\hline
& \% log$_{10}$(T$_a$) & \% log$_{10}$(F$_a$) & \% $\alpha$ & \% logT$_a$ & \% logF$_a$ & \% $\alpha$ \\
\hline
\multicolumn{7}{c}{\textbf{521 GRBs \citep{Manchanda2025}}} \\
\hline
MLP             & -37.2 & -38.0 & -41.2 & \textbf{1.73} & 2.30 & \textbf{1.34} \\
Fourier Transform & -14.9 & -15.2 & -20.6 & 4.61 & 4.22 & 2.50 \\
CGAN             & -14.5 & -15.4 & -20.0 & 3.26 & 3.45 & \textbf{1.34} \\
Bi-LSTM          & -21.2 & -21.5 & -26.1 & 2.50 & 2.88 & \textbf{1.34} \\
SARIMAX-based Kalman & -9.70 & -13.3 & -17.3 & 3.26 & 3.84 & \textbf{1.34} \\
Bi-Mamba         & -20.8 & -21.2 & -27.6 & 2.88 & 2.30 & \textbf{1.50} \\
GP-RF            & -16.3 & -19.5 & -23.2 & 4.03 & 4.03 & 2.30 \\
Attention U-Net  & \textbf{-37.9} & \textbf{-38.5} & \textbf{-41.4} & \textbf{1.73} & 2.50 & \textbf{1.34} \\
KAN              & -4.74 & -14.2 & -11.6 & 4.99 & \textbf{1.92} & 1.73 \\
GP (W07)         & -16.9 & -18.6 & -24.3 & 3.07 & 3.45 & 1.54 \\
W07 model (10\%) & -18.0 & -19.1 & -25.2 & 2.30 & 2.30 & 2.11 \\
W07 model (20\%) & -15.8 & -17.8 & -23.8 & 2.30 & 2.69 & 2.30 \\
\hline

\multicolumn{7}{c}{\textbf{207 Good GRBs \citep{Dainotti2023}}} \\
\hline
Attention U-Net  & \textbf{-38.8} & \textbf{-40.3} & \textbf{-44.0} & 0.483 & 0.483 & \textbf{0} \\
MLP              & -38.1 & -39.3 & -43.9 & 0.483 & 0.483 & \textbf{0} \\
GP (W07)         & -17.2 & -20.0 & -28.8 & 1.93 & 1.45 & \textbf{0} \\
W07 model (10\%) & -23.0 & -24.9 & -30.8 & 0.966 & 0.966 & \textbf{0} \\ 
W07 model (20\%) & -21.2 & -23.2 & -28.9 & \textbf{0} & \textbf{0} & \textbf{0} \\
\hline

\end{tabular}
\end{adjustbox}
\caption{Summary of the average reduction in parameter uncertainties and the corresponding outlier fractions for the different reconstruction methods, evaluated over the GRB sample and the subset of good GRBs. Bold marks the best value in each column, within a block
(largest uncertainty decrease, smallest outlier fraction). Blocks correspond to different GRB samples and priors and are not directly comparable.
\label{tab:reconstruction_comparison}}
\end{table*}

\subsubsection{The Willingale model} \label{The Willingale model}

We model the GRB light-curves using the W07 function (W07), given by\\
\begin{equation}
f(t) = \left \{
\begin{array}{ll}
\displaystyle{F_i \exp{\left ( \alpha_i \left( 1 - \frac{t}{T_i} \right) \right )} \exp{\left (
- \frac{t_i}{t} \right )}} {\hspace{1cm}\rm for} \ \ t < T_i, \\
\\
\displaystyle{F_i \left ( \frac{t}{T_i} \right )^{-\alpha_i}
\exp{\left ( - \frac{t_i}{t} \right )}} {\hspace{1cm}\rm for} \ \ t \ge T_i. \\
\end{array}
\right\}
\label{eqn1}
\end{equation}
Here, $T_i$ and $F_i$ denote the time and flux at the end of the plateau phase, and $\alpha_i$ is the corresponding temporal decay index. The parameter $t_i$ marks the beginning of the rising phase. For the afterglow component, $i=a$, giving the parameters $T_a$, $F_a$, and $t_a$.

\subsubsection{Data Processing} \label{Data Processing}

We use the pre-processing and training steps described in previous work in this series
for all models.
This allows us to use the same procedure and make a fair comparison between the different methods. 
After obtaining the required hyperparameters, models that use automatic hyperparameter tuning are trained separately for each GRB light-curve (LC). We use the $\log_{10}$-transformed time and flux values for training. The data are then scaled to the range [0,1] using min--max normalisation, given by:

\begin{equation}
\label{eq: minmax-scaling}
    X_{i} = \frac{X_{i} - X_{min}}{X_{max} - X_{min}};  i \in D,
\end{equation}
where $X_{min}$ and $ X_{max}$ represent the minimum and the maximum value in the training dataset $D$.

Given the large dynamic ranges in observational time $t$ (seconds) and unabsorbed flux $F$ ($\text{erg cm}^{-2}\text{ s}^{-1}$), all light-curves are transformed into logarithmic coordinates:
  $$\tau = \log_{10}\left(\frac{t}{\text{s}}\right), \quad y = \log_{10}\left(\frac{F}{\text{erg cm}^{-2}\text{ s}^{-1}}\right)$$
The models then apply specific normalisation strategies to these logarithmic coordinates.
For PINN scaling, temporal and flux coordinates are normalised using min-max scaling into $[0, 1]$ using eq.~\ref{eq: minmax-scaling}.
$X_{\min}$ and $X_{\max}$ denote the empirical limits of the individual light-curve.
Furthermore, time values are standardised using sample statistics $\bar{\tau} = \langle \log_{10} t \rangle$ and $\sigma_\tau$ via Z-score normalization: ($\hat{\tau} = (\tau - \bar{\tau})/\sigma_\tau$). 
Centring and scaling inputs prior to network training is a well-established practice for accelerating convergence and preventing saturation of non-linear activations \citep{LeCun1998, Klambauer2017}.

For the Siamese and ReFANN models, Z-score normalisation was applied in log-space, performed independently for each GRB light-curve:
\begin{equation}
\label{eq:standard-scaling-grb}
    X_{i} = \frac{X_{i} - X_{\mathrm{mean}}}{X_{\mathrm{std}}}; \quad i \in \{t, F\},
\end{equation}
where $X_i$ denotes $\log_{10}(t)$ or $\log_{10}(F)$, and $X_{\mathrm{mean}}$ and $X_{\mathrm{std}}$ are the mean and standard deviation that are computed from the data points of that individual GRB light-curve. 
Unlike QSS and DGP, where these statistics are derived from the full training sample, here each GRB is normalised using only its own light-curve.

For the PQR model, Z-score normalisation is applied globally in log-space. Consistent with the treatment of the QSS and DGP models, the statistical parameters are derived from the full training sample rather than individual light-curves:
\begin{equation}
    X_i = \frac{X_i - \mu_{\text{global}}}{\sigma_{\text{global}}}; \quad X_i \in \{\tau, y\}
\end{equation}
where $\mu_{\text{global}}$ and $\sigma_{\text{global}}$ represent the mean and standard deviation calculated from all data points across the GRB light-curves in the dataset. This global scaling ensures methodological consistency with baseline statistical models and preserves the relative magnitude differences across the entire GRB population.

\subsubsection{Observational Gap Identification} \label{Observational Gap Identification}

An observational gap is formally identified whenever the logarithmic temporal spacing between consecutive data points exceeds a set threshold $d_\tau$:
  $$\Delta \tau_i = \tau_{i+1} - \tau_i > d_\tau, \quad \text{where } d_\tau = 0.05\text{ dex}$$
This threshold corresponds to a fractional temporal gap $\Delta t / t \approx 12.2\%$, effectively identifying orbital data gaps while skipping densely sampled observational intervals.
\\

\subsection{The Machine-Learning Approach} \label{ML Approach}

\subsubsection{PINN} \label{PINN pre-processing}

\paragraph{Network Architecture} \label{Network Architecture}

The PINN is implemented as a feed-forward neural network with fully connected layers that takes the logarithmic time $\tau = \log_{10}(t/\mathrm{s})$ as input and predicts the logarithmic flux $\hat{y} = \log_{10}(F)$.
Both the input and output reside in log–log space, which linearises the underlying PL dynamics and compresses the dynamic range of GRB light-curves by several orders of magnitude. 
The architecture consists of an input layer, $N_h$ hidden layers each of width $h$ (neurones per layer), and a single scalar output. 
The network applies GELU activation and dropout to its hidden layers, while using a linear activation function in the final layer.

Input normalisation is applied internally. 
The network buffers the sample mean $\bar{\tau} = \langle \log_{10}(t) \rangle$ and standard deviation $\sigma_\tau$ of the observed timestamps, computed once from the training set and maps every input as $\hat{\tau} = (\tau - \bar{\tau})/\sigma_\tau$ before the first linear layer. 
This guarantees that the GELU units receive inputs in approximately the $[-2, 2]$ range rather than in the raw $\log_{10}(t) \sim [2, 7]$ range, which would push Tanh activations (used in comparable architectures such as the Siamese and ReFANN models) into saturation and suppress gradients \citep{LeCun1998, Klambauer2017}. 
The output-layer bias is initialised to $\bar{y} = \langle \log_{10}(F) \rangle$, so the network begins near the observed mean flux rather than near zero; this prevents the large initial residuals from triggering the gradient-clipping spike guard on every batch.

The architecture hyperparameters: hidden dimension $h \in [32, 512]$, number of hidden layers $N_h \in [2, 8]$, learning rate $\eta \in [10^{-5}, 10^{-2}]$, physics regularisation weight $\lambda \in [0.05, 1.0]$, dropout probability $p_{\text{drop}} \in [0.05, 0.30]$, and weight decay $w_d \in [10^{-7}, 10^{-3}]$, are optimised per GRB using the Optuna framework (\cite{Akiba2019}) with 80 trials. 
The search objective is a 5-fold cross-validated sigma-weighted mean-squared error (MSE) evaluated on held-out data points, making the architecture selection directly proportional to goodness of fit in units of reduced $\chi^2$, consistent with the metrics reported in Table~\ref{tab:reconstruction_comparison}.

The minimum hidden dimension of 32 and minimum depth of 2 were chosen after verifying that single-hidden-layer networks with $\lesssim 17$ neurones (the typical outcome of unconstrained searches) produce systematic $\sim 0.2~\rm{dex}$ level biases attributable to insufficient model capacity. 
Networks with $h \geq 32$ and $N_h \geq 2$ have enough representational power to interpolate the curvature of plateau-to-decay transitions within the observational error bars.

\paragraph{Physics Prior: Seven Analytical Models} \label{Physics Prior: Seven Analytical Models}

Rather than imposing a single analytical form, we run the PINN seven times per GRB, each time with a different physics model as the regularising prior.
All models operate in log–log space, with independent variable $\tau \equiv \log_{10}(t/\mathrm{s})$ and dependent variable $\hat{y} \equiv \log_{10}(F / \mathrm{erg\,cm^{-2}\,s^{-1}})$.
The reference intercept $F_0 \equiv \log_{10} F$ is evaluated at $\tau = 0$ (i.e.\ $t = 1\,\mathrm{s}$), not at the break time, avoiding the $\gtrsim 4\,\mathrm{dex}$ offset between the polynomial-fit intercept and a break-centred $F_0$ that otherwise causes \texttt{curve\_fit} divergence.
The parameter search bounds are motivated by the canonical four-segment Swift-XRT afterglow template of \citet{Zhang2006} and \citet{Nousek2006}: steep decay ($\alpha \gtrsim 3$), plateau ($\alpha \approx 0.3$–$0.8$), normal decay ($\alpha \approx 1$–$1.5$), and post-jet ($\alpha \gtrsim 2$).

\medskip
\noindent\textbf{1. Simple Power Law.} \citep{Sari1998}
\begin{equation}
    \hat{y}(\tau) = F_0 - \alpha\,\tau,
    \qquad \alpha \in [0.3,\,3.0].
    \label{eq:pl}
\end{equation}
$F_0$: log-flux intercept at $t = 1\,\mathrm{s}$ (free).
$\alpha$: temporal decay index; the lower bound $\alpha_{\min}=0.3$ encompasses the plateau and shallow-decay regime ($\alpha \approx 0.3$–$0.8$); $0.8$–$1.5$ is the normal-decay segment; and $1.5$–$3.0$ covers the steep-decay and post-jet phases.

\medskip
\noindent\textbf{2. Broken Power Law.} \citep{Zhang2006, Nousek2006} 
\begin{equation}
    \hat{y}(\tau) =
    \begin{cases}
        F_0 - \alpha_1\,\tau,
            & \tau < \tau_b, \\[3pt]
        F_0 + (\alpha_2-\alpha_1)\,\tau_b - \alpha_2\,\tau,
            & \tau \geq \tau_b,
    \end{cases}
    \label{eq:bpl}
\end{equation}
where flux continuity is enforced algebraically at $\tau_b$.
Free parameters: $F_0$; $\alpha_1 \in [0.05,\,5.0]$ (pre-break); $\alpha_2 \in [0.05,\,2.5]$ (post-break); $\tau_b \in [\tau_{\min},\,\tau_{\max}]$ (break time searched within the observed $\log_{10}t$ span).

\medskip
\noindent\textbf{3. Smoothly Broken Power Law.} \citep{Beuermann1999, Evans2009}
\begin{equation}
    \hat{y}(\tau)
    = F_0 - \alpha_1\,\tau
      - \frac{\alpha_2-\alpha_1}{s}\,
        \log_{10}\!\bigl[1 + 10^{\,s(\tau-\tau_b)}\bigr],
    \label{eq:sbpl}
\end{equation}
evaluated with the numerically stable decomposition $\log_{10}(1+10^u) = \max(u,0) + \log_{10}(1+10^{-|u|})$, clamped to $|u| \leq 50$ for float32 stability.
The asymptotes are $\hat{y}\to F_0-\alpha_1\tau$ for $\tau\ll\tau_b$ and $\hat{y}\to F_0-\alpha_2\tau+(\alpha_2-\alpha_1)\tau_b$ for $\tau\gg\tau_b$.
The break argument $u = s(\tau-\tau_b)$ does not depend on the sign of $(\alpha_2-\alpha_1)$; using $u = s(\alpha_2-\alpha_1)(\tau-\tau_b)$ instead would cause the break contribution to vanish for flattening breaks ($\alpha_2 < \alpha_1$), collapsing the model to single-slope behaviour for exactly the plateau-end and energy-injection cases this prior is meant to capture.
Free parameters: $F_0$; $\alpha_1,\alpha_2 \in [0.05,\,5.0]$; $\tau_b \in [\tau_{\min},\,\tau_{\max}]$; $s \in [1,\,10]$ (break sharpness; $s=1$ is broad and smooth, $s=10$ is nearly sharp).
 
\medskip
\noindent\textbf{4. Multi-Segment (Three-Segment) Power Law.}
\citep{Zhang2006, Nousek2006}
\begin{equation}
    \hat{y}(\tau) =
    \begin{cases}
        F_0 - \alpha_1\,\tau,
            & \tau < \tau_{b1}, \\[3pt]
        F_0 + (\alpha_2-\alpha_1)\,\tau_{b1} - \alpha_2\,\tau,
            & \tau_{b1} \leq \tau < \tau_{b2}, \\[3pt]
        F_0 + (\alpha_2-\alpha_1)\,\tau_{b1}
            + (\alpha_3-\alpha_2)\,\tau_{b2} - \alpha_3\,\tau,
            & \tau \geq \tau_{b2},
    \end{cases}
    \label{eq:mspl}
\end{equation}
with flux continuity enforced algebraically at both $\tau_{b1}$ and $\tau_{b2}$ ($\tau_{b1} < \tau_{b2}$ required).
Free parameters: $F_0$; $\alpha_1 \in [0.05,\,5.0]$; $\alpha_2,\alpha_3 \in [0.05,\,3.0]$; $\tau_{b1},\tau_{b2} \in [\tau_{\min},\,\tau_{\max}]$.

\medskip
\noindent\textbf{5. Jet-Break Model.} \citep{Rhoads1999,Sari1999}
\begin{equation}
    \hat{y}(\tau) =
    \begin{cases}
        F_0 - \alpha_1\,\tau,
            & \tau < \tau_j, \\[3pt]
        F_0 + (\alpha_2-\alpha_1)\,\tau_j - \alpha_2\,\tau,
            & \tau \geq \tau_j,
    \end{cases}
    \label{eq:jbm}
\end{equation}
identical in form to Eq.~\eqref{eq:bpl} but with the post-break index constrained to $\alpha_2 \in [2.0,\,3.0]$, enforced as hard Optuna bounds. 
After a beamed jet break, $\alpha \simeq p$ (the electron spectral index) in the fast-cooling regime \citep{Rhoads1999, Sari1999}; typical values $p \in [2,3]$ motivate these bounds \citep{Racusin2009}.
Free parameters: $F_0$; $\alpha_1 \in [0.05,\,5.0]$; $\alpha_2 \in [2.0,\,3.0]$; $\tau_j \in [\tau_{\min},\,\tau_{\max}]$.

\medskip
\noindent\textbf{6. Energy Injection Model.}
\citep{Dai1998, Zhang2001}
\begin{equation}
    \hat{y}(\tau) = F_0 - (\alpha - q)\,\tau,
    \qquad q \in [0,\,0.99),\quad q < \alpha.
    \label{eq:ei}
\end{equation}
$q$: injection index; a long-lived central engine injecting energy into the forward shock flattens the observed decay from $\alpha$ to the effective index $(\alpha - q)$. 
The constraint $q < \alpha$ (enforced in Optuna bounds) ensures the effective decay remains positive.
$q = 0$ recovers Eq.~\eqref{eq:pl}.
Free parameters: $F_0$; $\alpha \in [0.3,\,3.0]$; $q \in [0.0,\,0.99)$.

\medskip
\noindent\textbf{7. Synchrotron Cooling-Break Model.}
\citep{Sari1998, Beuermann1999}
\begin{equation}
    \hat{y}(\tau)
    = F_0 - \alpha_1\,\tau
      - \frac{\Delta\alpha}{s}\,
        \log_{10}\!\bigl[1 + 10^{\,s\,\Delta\alpha\,(\tau-\tau_c)}\bigr],
    \qquad \Delta\alpha = 0.25\ \text{(fixed)}.
    \label{eq:sync}
\end{equation}
Unlike Eq.~\eqref{eq:sbpl}, the break argument here is $u = s\,\Delta\alpha\,(\tau-\tau_c)$, i.e.\ it carries an extra factor of $\Delta\alpha$. 
Because $\Delta\alpha=1/4$ is fixed and always positive, the effective log-time width of the synchrotron break is $1/(s\,\Delta\alpha)$ -- four times wider than a Model~3 break at the same sharpness parameter $s$. 
$\Delta\alpha$ is physically fixed by the shift in synchrotron spectral index as the cooling-break frequency crosses the observing band \citep{Sari1998}; the post-break slope $\alpha_2 = \alpha_1+0.25$ is therefore not a free parameter.
Free parameters: $F_0$; $\alpha_1 \in [0.5,\,3.0]$; $\tau_c \in [\tau_{\min},\,\tau_{\max}]$ (cooling-break time); $s \in [1,\,10]$.

\begin{figure}[t]
\centering
\includegraphics[width=\linewidth]{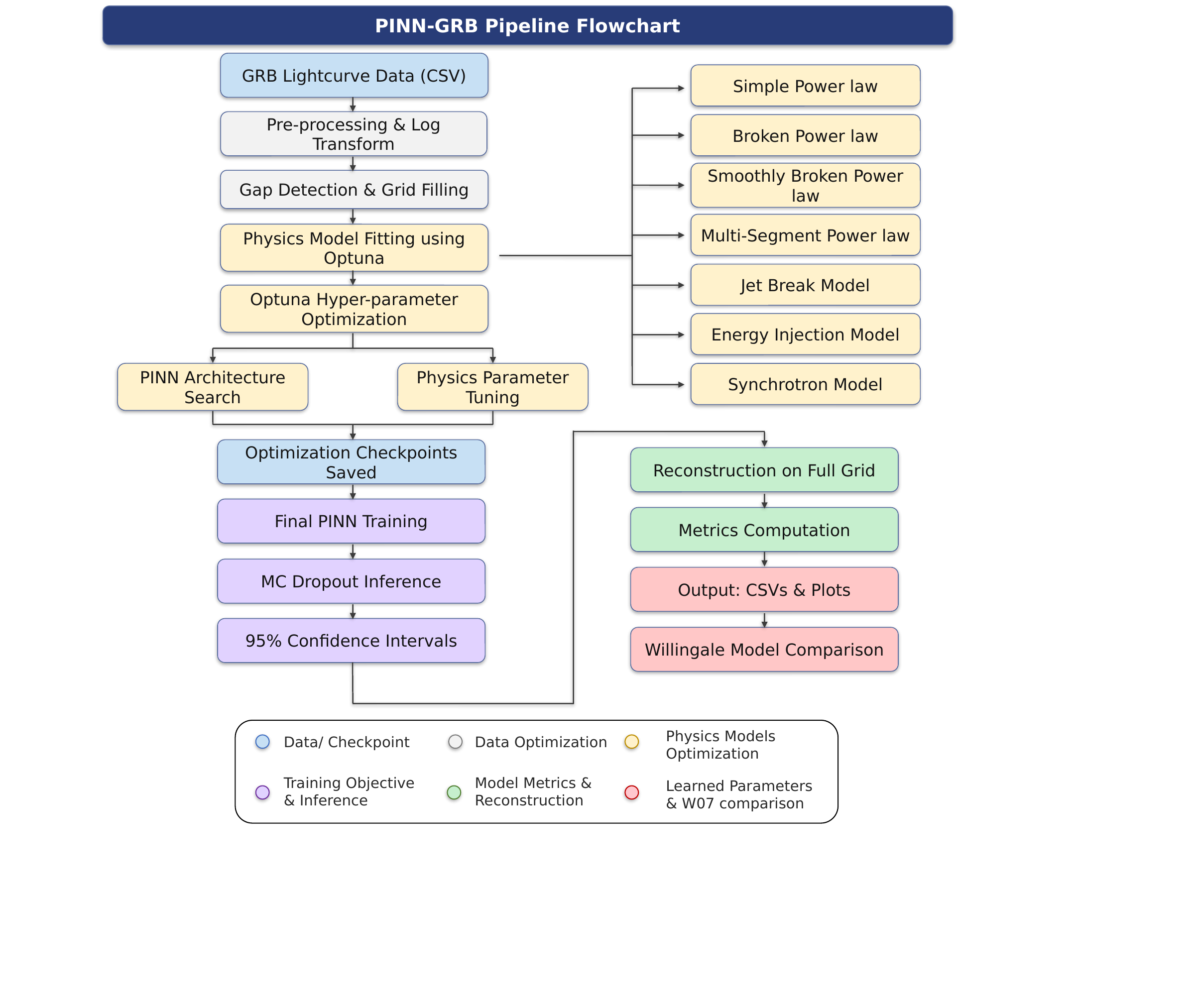}
\caption{Schematic overview of the PINN pipeline for GRB X-ray afterglow light-curve reconstruction. Using raw input Swift-XRT light-curve data (blue), the pipeline proceeds through log-space pre-processing and gap detection, followed by simultaneous Optuna-based optimisation of both the PINN architecture and the physics model parameters (yellow) across seven afterglow priors. Optimisation checkpoints are saved at each stage to enable resumable runs. The best-configuration network then undergoes final training (purple), after which MC dropout inference is used to construct 95\% confidence intervals on the reconstructed flux. The full-grid reconstruction (green) is passed to the metrics computation module, which evaluates reduced $\chi^2$, AIC, BIC, MSE, and correlation coefficient before the augmented light-curve is fitted with the W07 model to extract the plateau parameters $\log_{10} T_a,~ \log_{10} F_a,~ \textrm{and} \alpha$ and quantify the percentage decrease in their associated uncertainties relative to the original observed-data fit (red).
\label{fig:pinn_architecture}}
\end{figure}

\paragraph{Physics Model Parameter Optimisation} \label{Physics Model Parameter Optimisation}

For each of the seven physical models, initial parameter estimates are obtained via a non-linear least-squares fit \citep{Virtanen2020} with $\sigma-$weighted residuals applied to the observed $(\log_{10} t,~\log_{10} F)$ data. 
This gives the best-fit values of the model parameters, together with the $1-\sigma$ uncertainties estimated from the diagonal elements of the covariance matrix.
The Optuna search \citep{Akiba2019} for each model is then restricted to the window $\mathrm{best} \pm 4\sigma$ per parameter, with the search clamped to the physically motivated hard limits described above ($N_{trials} = 80$). 
This replaces the arbitrary fixed windows used in earlier implementations and allows Optuna to refine the curve-fit solution within a physically meaningful range rather than 
wandering through 
a vast, uninformed parameter space \citep{Bergstra2012}.

A two-level degeneracy check is applied before final training: 
(i) if the intercept parameter $\log F_0$ lies outside $\langle \log_{10} F \rangle \pm 15$~dex, it is reset to $\langle \log_{10} F \rangle$;
(ii) if the physics model evaluated at the observed timestamps deviates by more than 5~dex from the observed $\log_{10} F$ at any point, the parameter set is considered stale and discarded in favour of a fresh non-linear fit. 
Neither check affects training once a valid physics model is in hand.

\paragraph{Training Procedure} \label{Training Procedure}

Training is performed for 1500 epochs using the AdamW optimiser \citep{Loshchilov2017}. 
The learning rate is gradually reduced from $\eta$ to $\eta_{\min}=10^{-6}$ using cosine annealing, with $T_{\max}=1500$ epochs.
Gradient descent is applied over the full observed dataset at each step (batch size = number of data points, $N_{obs}$), eliminating mini-batch noise that in our experiments caused epoch-to-epoch loss oscillations and occasionally triggered gradient spikes even with clipping.
Gradients are clipped to a $\ell_2-$norm of 1.0 at every step \citep{Pascanu2012}.

Lambda warmup: The physics weight $\lambda$ is linearly ramped from 0 to its Optuna-selected value over the first 20\% of epochs ($\approx 300$~epochs). 
Without this warm-up, the physics term acts at full strength from the first epoch, before the data loss has converged, pulling the network toward the physics prior rather than the observed data; the warm-up ensures the network first learns the data shape and is then gently guided by the physics prior.
The gradient flow imbalance between data and physics loss terms, in which the physics gradient can overwhelm the data gradient early in training, has been studied systematically by \citet{Wang2021}, who showed that adaptive loss weighting and warmup scheduling are necessary to prevent physics-induced gradient pathologies. 
Our $\lambda$ warm-up and EMA(exponential moving average) normalisation directly address the pathologies identified by \citet{Wang2021}.

\begin{itemize}
    \item Data loss: When per-point flux uncertainties $\sigma_i = \log_{10}(F + \Delta F) - \log_{10}(F)$ are available (all GRBs in this sample), the data loss is a $\sigma-$weighted Huber loss:
    $\mathcal{L}_\mathrm{data} = \frac{1}{N}\sum_{i=1}^N w_i \cdot H_\delta(\hat{y}_i - y_i), \quad w_i = \frac{1/\sigma_i^2}{\langle 1/\sigma^2 \rangle}$,
    where $H_\delta$ the Huber loss is with $\delta = 0.5$, and the weights normalised so that $\langle w \rangle = 1$. 
    A floor $\sigma_\mathrm{min} = 0.05$~dex is imposed to cap the maximum weight at $1/0.05^2 = 400$ before normalisation; without this floor, points with $\sigma \approx 0.001$~dex would receive weights $\sim 10^6$ and trigger the gradient spike guard on every batch. T
    The combination of $\ sigma$-weighting and normalisation drives the optimiser to reduce each residual toward its own error bar, thereby pushing the reduction $\chi^2$ toward unity.

    \item Physics loss: The physics regularisation is value-space rather than derivative-matching. For each observed timestamp $\tau_i$, the physics model $P(\tau_i)$ is evaluated in no-gradient mode, and the physics loss is
    $$\mathcal{L}_\mathrm{phys} = \frac{1}{N}\sum_{i=1}^N [\hat{y}_i - P(\tau_i)]^2.$$
    This penalises the PINN for deviating from the analytical model in absolute flux units. 
    The classical ODE derivative form $d\hat{y}/d\tau = dP/d\tau$ was found to produce degenerate reconstructions in which the network, constrained to match the physics slope everywhere, converges to a constant output (trivially satisfying $d(\mathrm{const})/d\tau = 0$) rather than fitting the data. 
    Value-space regularisation avoids this by allowing the PINN to deviate from the physics model wherever the data demand it, while still being guided toward the analytical curve in poorly observed regions.
    
    \item EMA-normalised physics contribution: The physics loss is normalised by an exponential moving average of its own magnitude ($\alpha_\mathrm{EMA} = 0.95$), keeping the normalised physics term $\mathcal{L}_\mathrm{phys} / \mathrm{EMA} \approx 1$ throughout training \citep{Loshchilov2017}. 
    The effective total loss is then:
    \begin{equation}
        \lambda_{\mathrm{eff}} = \lambda \cdot \mathrm{sg}(L_{\mathrm{data}}),
        \qquad
        L = L_{\mathrm{data}} + \lambda_{\mathrm{eff}} \cdot \frac{L_{\mathrm{phys}}}{\mathrm{EMA}(L_{\mathrm{phys}})},
        \label{eq:totalloss}
    \end{equation}
    where $\mathrm{sg}(\cdot)$ denotes the stop-gradient (detach) operator applied to $L_{\mathrm{data}}$, so that $\lambda_{\mathrm{eff}}$ is treated as a constant scale factor during back-propagation rather than as a differentiable term. 
    This additive form keeps the physics contribution bounded by a fraction $\lambda < 1$ of the current data loss magnitude without conflating the two terms multiplicatively.

\end{itemize}

A collapse detection step runs after training: if the PINN prediction range across the reconstruction grid is less than 0.25~dex (indicating a degenerate flat solution), training is restarted with $\lambda$ reduced by a factor of five to allow the data loss to dominate. 
This single retry is sufficient in practice because collapse is caused by an overly strong physics prior, not by data quality.\\

\subsubsection{Siamese} \label{Siamese pre-processing}

\begin{figure*}[t]
\centering
\includegraphics[width=\textwidth]{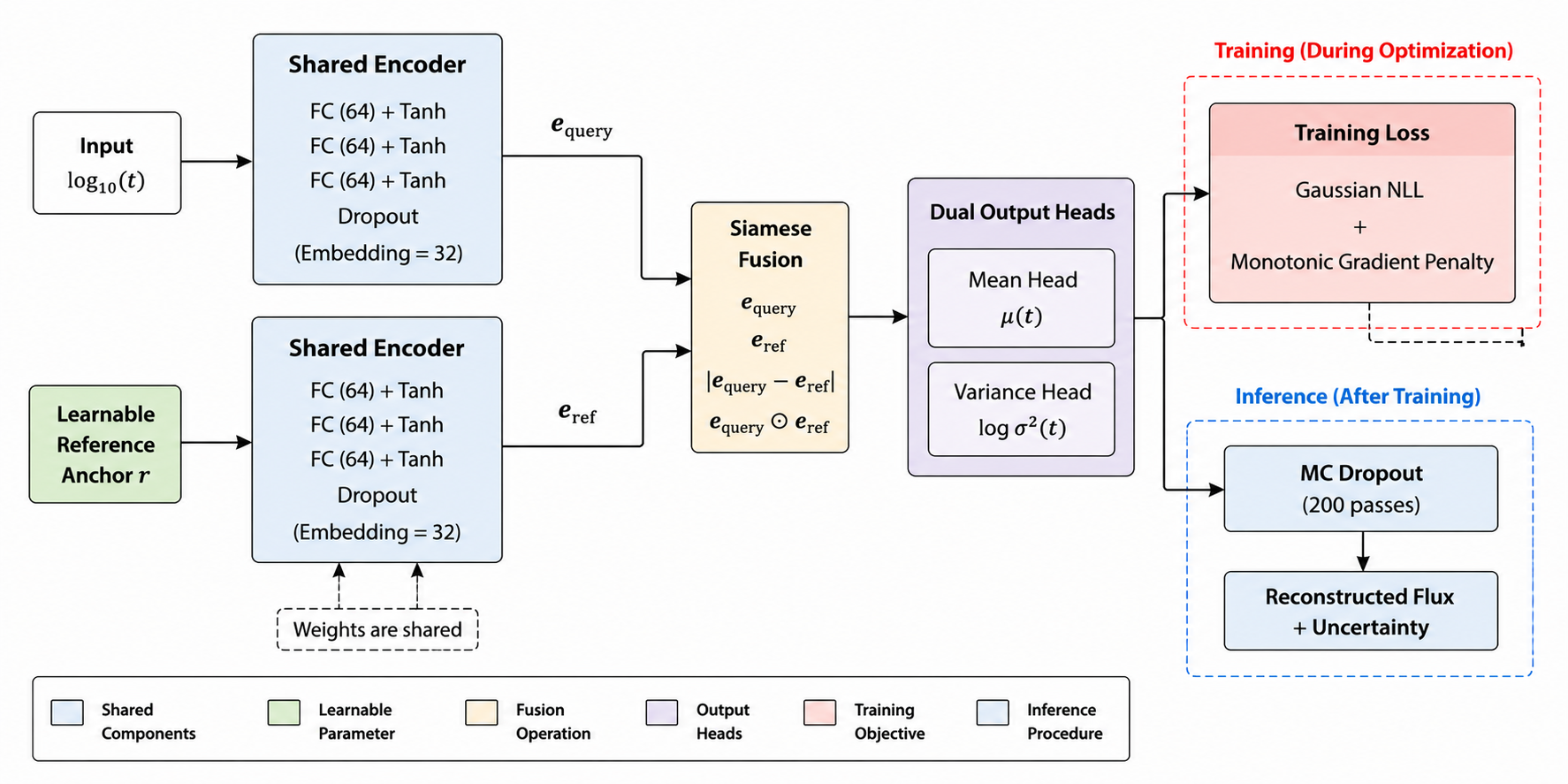}
\caption{
Architecture of the proposed Siamese network for GRB
X-ray light-curve reconstruction. The first branch
processes the logarithmic time coordinate, while the second branch receives a learnable reference anchor.
Both branches share a common encoder network and
produce latent embeddings that are combined through
a Siamese fusion operation. The fused representation
is passed to dual output heads that predict the
reconstructed logarithmic flux and its associated
variance. During training, the network is optimised
using a Gaussian negative log-likelihood loss together
with a physics-inspired regularisation term. During
inference, predictive uncertainties are estimated using
MC dropout with 200 stochastic forward passes. \label{fig:siamese_architecture}}
\end{figure*}

The Siamese network is designed to reconstruct GRB X-ray light-curves by learning a continuous mapping between logarithmic time and logarithmic flux while simultaneously estimating predictive uncertainties. Prior to training, both the logarithmic time and logarithmic flux values are standardised to zero mean and unit variance. This normalisation improves numerical stability and ensures that all inputs contribute comparably during optimisation.\\
Figure~\ref{fig:siamese_architecture} shows the architecture of the Siamese NN used to reconstruct GRB X-ray light-curves. The model is designed to learn a continuous mapping between the logarithm of time and the logarithm of the observed flux while simultaneously estimating predictive uncertainties.

Prior to training, the observed time values are transformed to a logarithmic scale and standardised to have zero mean and unit variance. Since GRB afterglows vary over several orders of magnitude in time, the logarithmic transformation compresses the input's dynamic range, allowing the network to learn temporal features more effectively. Standardisation further improves numerical stability and accelerates convergence during optimisation.\\
The encoder uses three fully connected hidden layers, with 64 neurons in each layer. The Tanh activation function is used, and a dropout rate of 0.1 is applied. Unlike conventional Siamese architectures, which compare two independent inputs, the second branch is supplied with a learnable reference anchor, $\mathbf{r}$, initialised at zero and optimised jointly with the network parameters during training. This reference anchor serves as an adaptive baseline in the latent space, enabling the model to learn temporal relationships relative to a self-learned reference point rather than a fixed, manually selected value.\\
The same encoder is applied to both branches, ensuring that both inputs are represented within a common feature space. The encoder consists of three fully connected hidden layers containing 64 neurons per layer, followed by Tanh activation functions and a dropout rate of 0.1 \citep{Srivastava2014}. The shared weights enforce a consistent feature representation across both branches while reducing the number of trainable parameters. The final encoder layer produces a 32-dimensional embedding vector,

\begin{equation}
\mathbf{e}_{\rm query} = f_{\theta}(x)
\end{equation}

for the input time coordinate, and

\begin{equation}
\mathbf{e}_{\rm ref} = f_{\theta}(\mathbf{r})
\end{equation}

for the learnable reference anchor, where $f_{\theta}$ denotes the shared encoder network.

To capture both absolute and relative temporal information, the two embeddings are combined using four complementary representations: the query embedding, the reference embedding, their element-wise absolute difference, and their element-wise product. These quantities are concatenated to form a 128-dimensional fusion vector,
\begin{equation}
\mathbf{z}
=
\mathrm{Concat}
\left(
\mathbf{e}_{\rm query},
\mathbf{e}_{\rm ref},
\left|\mathbf{e}_{\rm query}-\mathbf{e}_{\rm ref}\right|,
\mathbf{e}_{\rm query}\odot\mathbf{e}_{\rm ref}
\right)
\end{equation}
where $\odot$ denotes element-wise multiplication. This fusion strategy enables the network to simultaneously learn similarities, differences, and interactions between the query representation and the learned reference representation.
The fused representation is subsequently passed to two independent output heads. 
The first head predicts the mean reconstructed logarithmic flux, $\mu(t)$, while the second predicts the logarithm of the predictive variance, $\log\sigma^2(t)$. 
The variance head is initialised with zero weights and a bias of $-3.0$, corresponding to a moderate initial uncertainty level. 
To avoid unstable uncertainty estimates during training, the predicted log-variance is constrained to the interval $[-6,-3]$. 
This dual-head architecture enables simultaneous reconstruction of the light-curve and estimation of predictive uncertainty \citep{Kendall2017}.
The use of the absolute difference $|e_{query} - e_{ref}|$ as a fusion feature follows the contrastive embedding paradigm of \citet{Hadsell2006}, who showed that learning an invariant mapping between pairs improves robustness to intra-class variability, here the variability in GRB afterglow temporal morphology.
The four-component concatenation: query embedding, reference embedding, absolute difference, and element-wise product follows the bilateral multi-perspective matching strategy of \citet{Wang2017}, which demonstrated that combining similarity and difference representations in a shared latent space captures richer relational information than any single representation alone.
The model uses a Gaussian negative log-likelihood (NLL) loss to include the observational uncertainties during training as given
\begin{equation}
\mathcal{L}_{\rm NLL} = \frac{1}{N} \sum_{i=1}^{N} \frac{1}{2} \left[\log \sigma_i^{2} + \frac{(y_i-\mu_i)^2} {\sigma_i^{2}} \right]
\end{equation}
where $y_i$ denotes the observed logarithmic flux, $\mu_i$ is predicted mean flux, and $\sigma_i^2$ is the predicted variance.

In addition to the data-fitting term, a physics-inspired monotonicity regularisation term is introduced to discourage positive gradients in the reconstructed light-curve,
\begin{equation}
\mathcal{L}_{\rm mono} = \frac{1}{N} \sum_{i=1}^{N}
\left[ \max\left(0,\frac{\partial \mu_i}{\partial x_i}\right) \right]^2.
\label{eq:regularisation}
\end{equation}
where $x_i$ denotes the normalised logarithmic time coordinate. 
Since $\max(0,z)$ corresponds to the ReLU activation function, only positive gradients contribute to the regularisation term, while negative gradients incur no penalty. 
This regularisation reflects the widely observed behaviour that GRB X-ray afterglows exhibit an overall decay in flux over time. 
By penalising positive slopes while leaving negative slopes unconstrained, the model is encouraged to follow physically plausible temporal evolution without imposing a rigid functional form. 
The total loss function is therefore given by

\begin{equation}
\mathcal{L} = \mathcal{L}_{\rm NLL} + \lambda\,\mathcal{L}_{\rm mono}
\end{equation}

where $\lambda = 10^{-3}$ controls the contribution of the physics monotonic regularisation term. 
This value was chosen to balance the influence of the monotonicity regularisation and the data-fitting objective, allowing physically plausible reconstructions while preserving agreement with the observations. 
The regularisation weight was further investigated by varying $\lambda$ over the range $0 \le \lambda \le 1$. 
Although some values yielded slightly improved performance for individual metrics, $\lambda = 10^{-3}$ provided the best overall (while dropout fixed to 0.1), as can be seen in (Table~\ref{tab:lambda_study}).

\begin{table*}[t]
\centering
\footnotesize
\caption{Effect of the physics-loss regularisation parameter $\lambda$ on the reconstruction performance while dropout was fixed to 0.1 in the Siamese model.
\label{tab:lambda_study}}

\begin{tabular*}{\textwidth}{@{\extracolsep{\fill}}ccccccccc}
\toprule
No. &
$\lambda$ &
$\%\log F_a$ &
$\%\log T_a$ &
$\%\alpha$ &
\makecell{Outliers\\($F_a$)} &
\makecell{Outliers\\($T_a$)} &
\makecell{Outliers\\($\alpha$)} &
GRBs \\
\midrule

1 & $0$        & -23.29 & -23.60 & -26.81 & 4.59 & 4.04 & 1.65 & 519 \\
2 & $10^{-6}$ & -22.66 & -22.63 & -27.34 & 3.86 & 3.67 & 1.29 & 523 \\
3 & $10^{-5}$ & -23.37 & -23.33 & -27.08 & 4.41 & 3.68 & 1.47 & 520 \\
4 & $10^{-4}$ & -23.08 & -22.94 & -27.38 & 4.04 & 2.94 & 1.65 & 522 \\
5 & $\mathbf{10^{-3}}$ & $\mathbf{-22.83}$ & $\mathbf{-22.88}$ &
$\mathbf{-26.93}$ & $\mathbf{3.68}$ & $\mathbf{3.12}$ &
$\mathbf{1.10}$ & $\mathbf{524}$ \\
6 & $10^{-2}$ & -23.02 & -22.80 & -27.26 & 4.04 & 3.68 & 1.10 & 522 \\

\bottomrule
\end{tabular*}
\end{table*}

Model parameters are optimised using the Adam optimiser with an initial learning rate of 0.01. A StepLR scheduler is employed during training, reducing the learning rate by a factor of 0.5 every 200 iterations to improve convergence stability. The network is trained for 500 iterations.

Predictive uncertainties are estimated using MC Dropout \citep{Gal2016}, where dropout layers remain active during inference and 200 stochastic forward passes are performed. The variance among these predictions provides an estimate of the epistemic uncertainty,
Epistemic uncertainty applied in all the models refers to uncertainty arising from the model's limited knowledge of the true underlying function, in principle reducible with more data or a better architecture, while aleatoric uncertainty refers to the irreducible instrumental scatter intrinsic to the observations themselves (photon-counting noise, calibration, etc.).

\begin{equation}
\sigma_{\rm epi}^{2} = {\rm Var} \left(\mu^{(k)} \right)
\end{equation}
where $k$ denotes the MC realisation.

The network simultaneously predicts the aleatoric uncertainty,

\begin{equation}
\sigma_{\rm ale}^{2} = \left\langle \exp\!\left(\log\sigma^{2}\right) \right\rangle
\end{equation}

which accounts for the uncertainty associated with the observational data. The total predictive uncertainty is obtained by combining both contributions in quadrature,

\begin{equation}
\sigma_{\rm tot} = \sqrt{\sigma_{\rm epi}^{2} + \sigma_{\rm ale}^{2}}
\end{equation}
The corresponding 95\% confidence interval is computed as
\begin{equation}
\mu \pm 1.96,\sigma_{\rm tot}.
\end{equation}

To improve reconstruction in sparsely sampled regions, a gap-aware reconstruction strategy is adopted. Additional reconstruction points are inserted only within temporal gaps exceeding a minimum logarithmic width of 0.05. The number of inserted points is adjusted based on the density of the original light-curve, allowing the model to concentrate predictions in poorly sampled regions while avoiding unnecessary interpolation in well-observed segments.

Finally, model performance is evaluated using five-fold cross-validation. Training and validation mean-squared errors are computed for each fold and then averaged to assess the model's generalisation capability across different GRB light-curves.

\subsubsection{ReFANN} \label{ReFANN pre-processing}

The Reconstruct Functions with Artificial NNs (ReFANN) model is designed to learn a continuous mapping between the logarithm of time and the logarithm of the observed flux without assuming any predefined functional form. The input time values are first transformed to a logarithmic scale and standardised to improve training stability. We used 3 hidden layers with 64 neurons, which provides sufficient model capacity to capture non-linear temporal evolution while avoiding excessive model complexity and over-fitting. Tanh activation functions were selected because they provide smooth, non-linear mappings well-suited to reconstructing continuous astrophysical light-curves. A dropout rate of 0.1 was employed to reduce over-fitting \citep{Srivastava2014} and improve generalisation. During training, approximately 10\% of neurons are randomly deactivated in each forward pass, encouraging the network to learn more robust representations. 
Dropout is applied after each hidden layer, before the network branches into two linear output heads that predict the mean and log-variance of the flux distribution at each time point. The log-variance head is initialised with zero weights and a bias of $-3.0$, and its output is clamped to the range $[-6.0, -3.0]$ during every forward pass. This stabilises training and keeps the aleatoric noise floor within physically reasonable limits based on the typical Swift/BAT flux uncertainties. Rather than generating a single prediction, the model uses two output heads to simultaneously estimate the reconstructed flux and its predictive uncertainty by learning the mean and variance of the output distribution. \citep{Kendall2017}.

To account for observational uncertainties, the network is trained using a Gaussian negative log-likelihood (NLL) loss,
\begin{equation}
\mathcal{L}_{\rm NLL} = \frac{1}{N} \sum_{i=1}^{N} \frac{1}{2} \left[
\log\sigma_i^2 + \frac{(y_i-\mu_i)^2}{\sigma_i^2}\right],
\end{equation}

In addition, a physics-inspired regularisation term based on the squared gradient of the reconstructed light-curve is included to encourage smooth solutions and suppress unphysical fluctuations. 
\begin{equation}
\mathcal{L}_{\rm physics}
=
\frac{1}{N}
\sum_{i=1}^{N}
\left(
\frac{\partial \mu_i}{\partial x_i}
\right)^2.
\end{equation}

The total loss function is therefore a combination of the NLL loss and the physics regularisation term,
\begin{equation}
\mathcal{L} = \mathcal{L}_{\rm NLL} + \lambda \mathcal{L}_{\rm physics},
\end{equation}
where $\lambda = 10^{-3}$ in the present work. 
This value was chosen to ensure that the regularisation term guides the reconstruction toward smooth solutions without dominating the data-fitting objective. 
Similar regularisation strategies are commonly employed in physics-informed and scientific machine-learning applications to balance observational fidelity and physically motivated constraints \citep{Raissi2019, Karniadakis2021}. 

\begin{figure*}[htbp]
    \centering
    \includegraphics[width=\textwidth]{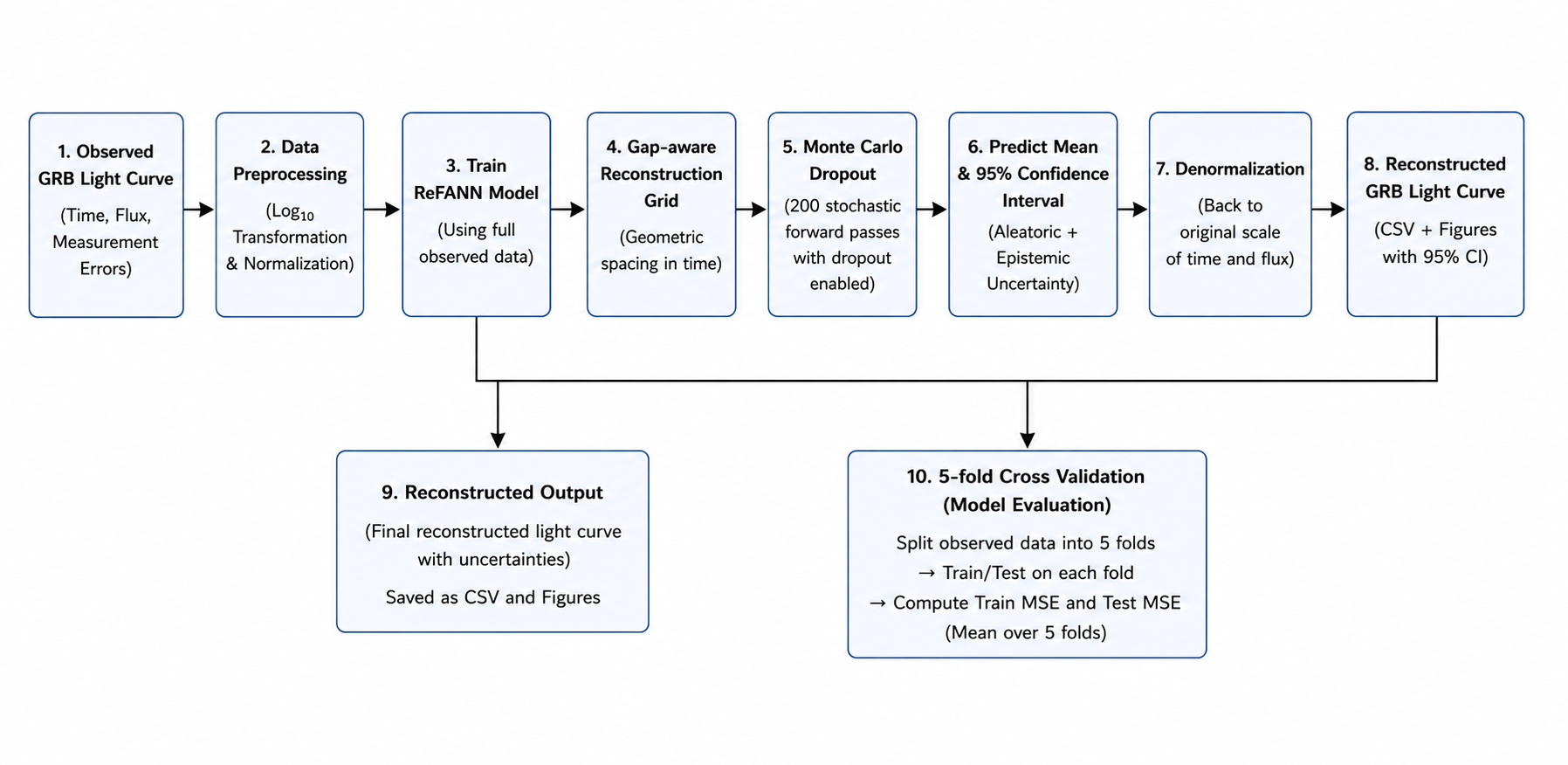}
    \caption{
Workflow of the proposed ReFANN framework for reconstructing GRB light-curves. The observed light-curves are first log-transformed and normalised before being used to train the ReFANN model. The trained model then produces gap-aware reconstructions, after which Monte Carlo dropout is used to estimate predictive uncertainty and construct 95\% confidence intervals. The reconstructed light-curves are subsequently denormalised and saved for further analysis. Independently, five-fold cross-validation is carried out to assess the model’s predictive performance on the observed data, with the mean training and testing mean squared errors (MSEs) reported.
    \label{fig:refann_workflow}}
\end{figure*}

The model parameters are optimised using the Adam optimiser with an initial learning rate of 0.01. A StepLR scheduler is used during training to support better convergence, and the network is trained for 500 iterations.

Predictive uncertainties are estimated using MC Dropout, where dropout layers remain active during inference and 200 stochastic forward passes are performed \citep{Gal2016}. The variability among these predictions provides an estimate of the epistemic uncertainty, which reflects uncertainty in the model parameters. In addition, the network directly predicts the output variance, providing an estimate of the aleatoric uncertainty associated with the observational data \citep{Kendall2017}. The total predictive uncertainty is obtained by combining the epistemic and aleatoric components in quadrature. The resulting 95\% confidence interval is constructed as
\[
\hat{\mu} \pm 1.96\,\sigma_{\mathrm{tot}},
\]
Here, $\sigma\_{\mathrm{tot}}$ represents the total predictive standard deviation. To improve reconstruction in sparsely sampled regions, a gap-aware strategy is used. Additional points are inserted within gaps in the observed log-time grid larger than 0.05 dex, with the number of points added to each gap scaled according to its size. The total number of reconstructed points is set to
\[\max(20,\, f \times N),\]
Here, \(N\) denotes the number of observed data points, while \(f\) is a fraction that depends on the sample size: \(f = 0.05\) when \(N > 500\), \(0.1\) when \(N > 250\), \(0.3\) when \(N > 100\), and \(0.4\) otherwise. This approach allocates denser sampling to light-curves that are more sparsely sampled. Reconstructed flux values at each new time point are obtained by drawing a single sample from the empirical distribution of the 200 MC dropout predictions, recentred on the predictive mean, to inject realistic point-to-point scatter consistent with the model's estimated uncertainty, rather than reporting only the smooth mean curve.
Finally, the model's performance is assessed using five-fold cross-validation and reconstruction metrics computed for each GRB light-curve.\\
A sensitivity analysis was performed by varying the physics-regularisation weight over the range $0 \le \lambda \le 1$. Although some values yielded slightly improved performance for individual metrics, $\lambda = 10^{-3}$ provided the best overall balance between parameter recovery, outlier suppression, and sample completeness. Consequently, this value was adopted throughout the remainder of the analysis. A complete architecture of the model is present in Figure~\ref{fig:refann_workflow}.

\subsubsection{Quantile} \label{sec:quantile}

The PQR model estimates the conditional distribution of $\log_{10}(\mathrm{flux})$ as a function of $\log_{10}(t)$ by independently fitting three quantiles: $q \in \{0.1, 0.5, 0.9\}$. These quantiles represent the lower bound, median, and upper bound of the reconstructed light-curve. For each GRB, the polynomial degree is adjusted according to the sample size rather than kept fixed, following
\begin{equation}
    d = \min(4,\, N_{\mathrm{obs}} - 2),
    \label{eq:quantile_degree}
\end{equation}
Here, \(N_{\mathrm{obs}}\) denotes the number of observed data points in the light-curve. GRBs with fewer than three observations are excluded because gap interpolation and variance estimation are not reliable below this threshold.
This criterion allows the model to use up to a quartic polynomial when sufficient data are available to resolve plateau phases and flares, while
automatically reducing the degree for sparser light-curves, ensuring that at least one degree of freedom remains for residual estimation.

Polynomial feature expansion is performed with
\texttt{sklearn.preprocessing.PolynomialFeatures}, and each of the three quantile models is optimised independently using the \texttt{QuantReg}
implementation in \texttt{statsmodels}, which minimises the asymmetric
Pinball Loss \citep{Koenker1978}:
\begin{equation}
    L_{q}(y, \hat{y}) = \frac{1}{N} \sum_{i=1}^{N}
    \max\bigl(q(y_i - \hat{y}_i),\, (q - 1)(y_i - \hat{y}_i)\bigr).
    \label{eq:pinball_loss}
\end{equation}

Robust regression methods for irregularly sampled astronomical light-curves have been explored by (\citet{Thieler2016}) in the context of periodogram estimation; the PQR model adopted here applies the same robustness principle, minimisation of the pinball loss rather than squared residuals, to GRB temporal reconstruction.

Gaps in each light-curve are defined as intervals where \(\Delta \log_{10} t > 0.05\). New reconstruction points are added only within these intervals, with their number proportional to the width of each gap.
The total number of inserted points is determined as a fraction of the original sample size: 5\% for light-curves with more than 500 points, 10\% for those with more than 250 and up to 500 points, 30\% for those with more than 100 and up to 250 points, and 40\% otherwise. A minimum of 20 points is always enforced, consistent with the gap-aware reconstruction strategy used throughout this work.
While this Gaussian-approximated interval is used for
visualisation, a fully distribution-free coverage guarantee can be obtained via conformal prediction applied to the quantile model output \citep{Romano2019}; we leave this extension to future work.

\begin{figure}[H]
    \centering
    \includegraphics[width=0.4\linewidth]{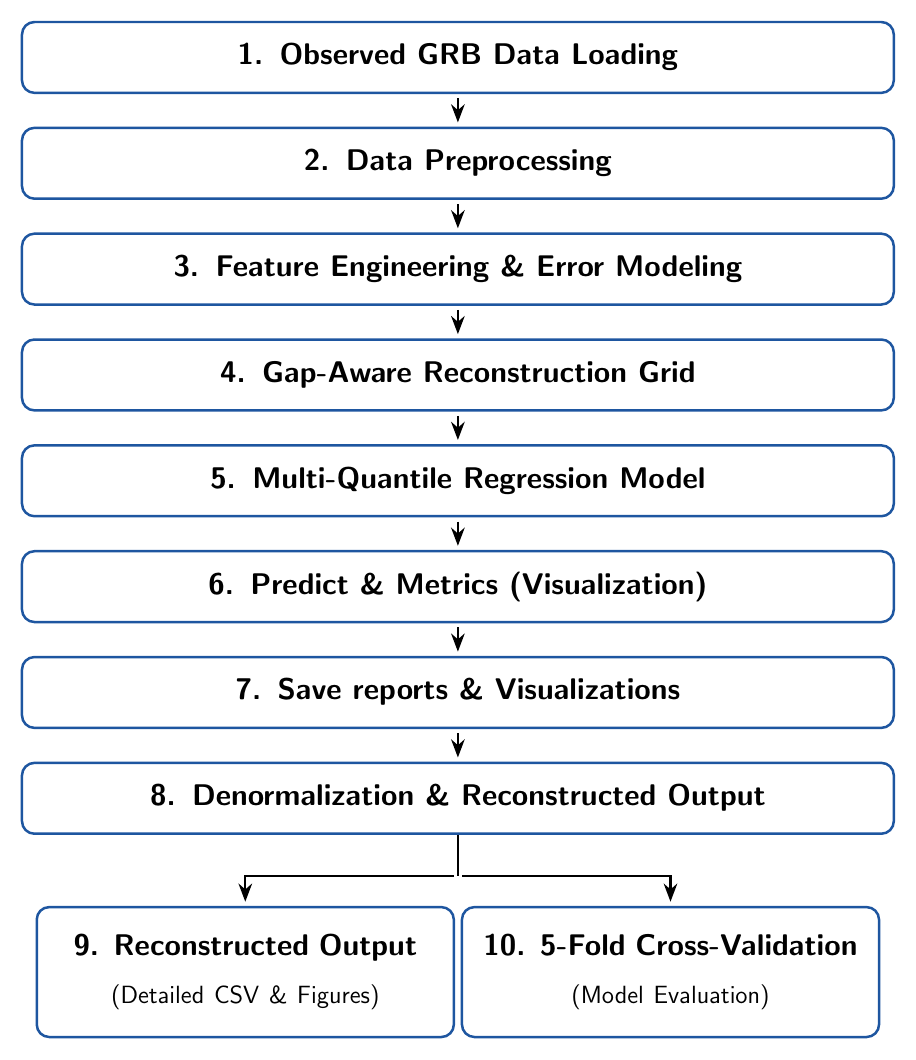}
    \caption{Schematic flowchart illustrating the multi-stage PQR model workflow for GRB light-curve reconstruction. The pipeline details the sequential process from raw data ingestion and logarithmic transformation to feature engineering, adaptive gap-aware grid generation, multi-quantile pinball loss minimisation ($q \in \{0.1, 0.5, 0.9\}$), cross-validation, and final distribution-free uncertainty estimation. 
    \label{fig:placeholder}}
\end{figure}

To account for observational uncertainties, the observed errors in
log-time and log-flux are characterised individually for each GRB by
fitting normal distributions to $\Delta t_{\mathrm{err}}/(t\ln 10)$ and
$\Delta F_{\mathrm{err}}/(F\ln 10)$, respectively. New time and flux
uncertainties at the reconstructed points are then drawn via MC
sampling from these fitted distributions (with negative draws clipped to
zero), and the resulting noise is added to the median-quantile ($q=0.5$)
prediction to obtain the final reconstructed flux values. This preserves
the statistical character of the original measurement errors rather than
introducing artificial smoothing. For visualisation, predictive
uncertainty is instead characterised by a static value, $\sigma$, taken
as the 68th percentile of the absolute observed log-flux errors for that
GRB; the corresponding 95\% confidence interval is computed as
$\hat{y} \pm 1.96\,\sigma$, where $\hat{y}$ is the quantile-regression
prediction evaluated on a dense grid in $\log_{10}(t)$.
Model performance is assessed using $K$-fold cross-validation, with up to five folds depending on the number of available points for each GRB. Evaluation is based on the median-quantile ($q=0.5$) fit and includes the Mean Squared Error (MSE), Mean Absolute Error (MAE), $R^2$, Mean Absolute Percentage Error (MAPE), Reduced $\chi^2$, and a quantile-adapted Pseudo-AIC \citep{Akaike1974}, computed as
\begin{equation}
    \mathrm{AIC} = N\ln\!\left(L_q + \epsilon\right) + 2(d+1),
    \label{eq:quantile_aic}
\end{equation}

where $L_q$ is the pinball loss for that quantile, $N$ is the number of
observed points, $d$ is the fitted polynomial degree, and
$\epsilon = 10^{-10}$ is a small constant added for numerical stability.
This provides a statistically grounded framework for parameter recovery
in the presence of complex, heteroscedastic light-curve morphology
\citep{Koenker1999}. This degree-capping criterion reflects the
classical bias-variance trade-off \citep{Hastie2009}: a polynomial of
degree $d$ has $d+1$ free parameters, so requiring
$N_{\mathrm{obs}} \geq d+2$ ensures at least one degree of freedom
remains for residual estimation.

\begin{table*}[ht] 
    \centering
    \renewcommand{\arraystretch}{1.5} 
    \begin{adjustbox}{width=0.95\textwidth, center}
    \begin{tabular}{|c|c|c|c|c|}
    \hline
    \textbf{Model / Sub-System} & \textbf{Parameter / Variable} & \textbf{Value /}  & \textbf{Physical / Algorithmic Role} & \textbf{Reference}  \\ 
     &  &  \textbf{Operational Range} &  &   \\ \hline
    \textbf{PINN Architecture} & Hidden Depth ($N_h$) & $[2, 8]$ layers & Fully connected layer count & \citep{Raissi2019}  \\ \hline
     & Layer Width ($h$) & $[32, 512]$ neurons & Hidden state capacity & \citep{Raissi2019}  \\ \hline
     & Activation Function & GELU & Non-linear activation & \citep{Hendrycks2016}  \\ \hline
     & Learning Rate ($\eta$) & $[10^{-5}, 10^{-2}]$ & Cosine annealing initial rate & \citep{Kingma2014}  \\ \hline
     & Optimiser / Epochs & AdamW /  & Loss optimization routine & \citep{Loshchilov2017}  \\ 
     &  &  1500 epochs &  &   \\ \hline
     & Huber Loss Delta ($\delta$) & $0.5$ & Outlier-resistant threshold & \citep{Huber1964}  \\ \hline
     & Error Floor ($\sigma_{\min}$) & $0.05$~dex & Weight capping threshold & \citep{Huber1964}, \textbf{this work}  \\ \hline
     & Physics Weight ($\lambda_{\text{phys}}$) & $[0.05, 1.0]$ & Dynamic regularisation weight & \citep{Raissi2019}  \\ \hline
     & EMA Momentum & $0.95$ & Loss Scale Normalisation & \citep{Loshchilov2017}  \\ \hline
     & Collapse Threshold & $< 0.25$~dex & Variance safety trigger & \citep{Raissi2019}  \\ \hline
    \textbf{1. Simple PL} & Decay Index ($\alpha$) & $[0.3, 3.0]$ & Temporal decay index & \citep{Zhang2006}  \\ \hline
    \textbf{2. Broken PL} & Pre-break Index ($\alpha_1$) & $[0.05, 5.0]$ & Early decay rate & \citep{Nousek2006}  \\ \hline
     & Post-break Index ($\alpha_2$) & $[0.05, 2.5]$ & Late decay rate & \citep{Nousek2006}  \\ \hline
     & Break Time ($\tau_b$) & $[\tau_{\min}, \tau_{\max}]$ & Transition time-stamp & \citep{Nousek2006}  \\ \hline
    \textbf{3. Smoothly Broken PL} & Sharpness Parameter ($s$) & $[1, 10]$ & Transition curvature factor & \citep{Beuermann1999}  \\ \hline
    \textbf{4. Multi-Segment PL} & Decay Indices ($\alpha_1, \alpha_2, \alpha_3$) & $\alpha_1 \in [0.05, 5.0]$,  & Multi-phase decay rates & \citep{Zhang2006}  \\ 
     &  & $\alpha_2, \alpha_3 \in [0.05, 3.0]$ &  &   \\ \hline
    \textbf{5. Jet Break Model} & Post-break Index ($\alpha_2$) & $[2.0, 3.0]$ & Jet expansion decay index & \citep{Rhoads1999, Sari1999}  \\ \hline
    \textbf{6. Energy Injection} & Injection Parameter ($q$) & $[0, 0.99]$ ($q < \alpha$) & Central engine power index & \citep{Dai1998};   \\ 
     &  &  &  & \citep{Zhang2001}  \\ \hline
    \textbf{7. Synchrotron} & Fixed Index Shift ($\Delta \alpha$) & $0.25$ (Fixed) & Cooling frequency shift & \citep{Sari1998}  \\ \hline
    \textbf{ReFANN Model} & Architecture / Activation & $3 \times 64$ / Tanh & Parametric baseline & Fully connected DNN  \\ \hline
     & Physics Weight ($\lambda$) & $10^{-3}$ & Derivative smoothness penalty & Derivative regularizer  \\ \hline
    \textbf{Siamese Model} & Shared Encoder / Fusion & $3 \times 64$ / $128$-dim & Dual-branch encoder & Shared weight network  \\ \hline
     & Output Heads & Dual ($\mu, \log \sigma^2$) & Predict mean \& variance & Heteroscedastic head  \\ \hline
    \textbf{Polynomial Quantile Regression} & Polynomial Degree ($d$) & $\min(4, N_{\text{obs}} - 2)$ & Non-parametric quantile fit & Pinball loss regression  \\ \hline

    \end{tabular}
    \end{adjustbox}
    \caption{Table summarises input values, parameter bounds, optimisation ranges, and literature sources for the PINN model and its seven empirical physics sub-models, along with comparative baseline architectures. \label{tab:sample}}
\end{table*}

\subsection{Reconstruction Methodology and Evaluation Metrics}

The reconstruction methodology adopted in this work follows that developed in \citet{Dainotti2023} and subsequently refined in \citet{Kaushal2026}. The observed GRB light-curves are augmented while preserving their intrinsic temporal morphology, thereby increasing the effective sampling without altering the underlying physical behaviour of the afterglow. Since the reconstruction framework is model-independent and does not assume any analytical description of the light-curve during training, it does not directly provide the physical parameters of the GRB afterglow. Therefore, both the original and reconstructed light-curves are independently fitted using the W07 model, providing a common and consistent basis for parameter estimation.

The performance of each reconstruction method is assessed by comparing the uncertainties in the W07 model parameters derived from the original and reconstructed light-curves. Following \citet{Srinivasaragavan2020}, the uncertainty associated with each fitted parameter, $p=(\log T\_{a}, \log F\_{a}, \alpha)$, is defined as

\begin{equation}
\Delta p=\frac{\max(p)-\min(p)}{2},
\end{equation}
As the number of available data points increases, the statistical uncertainty in the fitted parameters is expected to decrease approximately as $1/\sqrt{N}$, where $N$ represents the total number of observations. 
Data augmentation can, therefore, naturally improve parameter constraints. 
Because the reconstructed points are generated from a fit to the same light-curve, they are not independent of the observed ones. The formal uncertainty therefore decreases more slowly than $1/\sqrt{N}$ (with $N_{\rm eff}<N_{\rm obs}+N_{\rm rec}$), and part of the reduction reported below reflects refitting more points with the same model rather than new information.

However, increasing the sampling density indiscriminately may introduce systematic biases if the intrinsic temporal behaviour of the original light-curves is not preserved. 
To prevent this, the reconstruction strategy used in this work increases the data density while retaining the morphological characteristics of the observed afterglows. 
The reconstructed light-curves thus remain consistent with the analytical description provided by the W07 model, allowing for a meaningful comparison between the original and reconstructed datasets. 

The reconstructed light-curves may be regarded as representing an ideal observational scenario in which interruptions caused by satellite orbital gaps are effectively absent. Such a configuration may also approximate future observational capabilities that will be achieved through continuous monitoring or coordinated observations from multiple space-based missions. Throughout this work, the W07 model is used solely as a benchmark to assess reconstruction performance. Nevertheless, the same reconstruction framework can be readily applied to other analytical models, as demonstrated previously for the broken PL model in \citet{Dainotti2023}. Exploring additional fitting models or conducting simulations for future missions is beyond the scope of this work.

Incomplete temporal coverage is one of the main sources of uncertainty when determining the physical parameters of GRB afterglows. To assess the effectiveness of the reconstructed light-curves, we quantify the uncertainties in the fitted W07 parameters using the error fraction (EF), following the methodology of \citet{Dainotti2023}. This metric provides a consistent measure of the relative uncertainty before and after reconstruction, enabling a direct assessment of the improvement produced by the reconstruction process.

The effectiveness of the reconstruction is assessed by comparing the error fractions (EFs) of the W07 model parameters derived from the original and reconstructed light-curves.
The corresponding definitions are given below:
 \begin{equation}
EF_{\log_{10}(T_a)} = \left|\frac{\Delta \log_{10}(T_a)}
{\log_{10}(T_a)} \right|,
\end{equation}

\begin{equation}
EF_{\log_{10}(F_a)} = \left|\frac{\Delta \log_{10}(F_a)}
{\log_{10}(F_a)} \right|,
\end{equation}

\begin{equation}
EF_{\alpha} = \left| \frac{\Delta \alpha} {\alpha} \right|.
\end{equation}

$$\% \text{ Uncertainty Decrease} = \left( \frac{EF_{\text{original}} - EF_{\text{reconstructed}}}{EF_{\text{original}}} \right) \times 100$$

Consequently, the error fractions corresponding to the original light-curves ($EF_{\log_{10}(T_i)}$, $EF_{\log_{10}(F_i)}$, and $EF_{\alpha_i}$) remain unchanged throughout this study, as they are derived from the same W07 fits to the original observations. 
The complete list of GRBs together with their corresponding error fractions is presented in Table~\ref{tab:EF/RC}. 
Only the reconstructed light-curves differ between models, and therefore the corresponding reconstructed error fractions ($EF_{\mathrm{RC}}$) and the percentage change in parameter uncertainties are recomputed independently for each reconstruction method. 
This approach ensures that any differences in the reported performance arise solely from the reconstruction algorithm and not from variations in the underlying GRB sample.

\vspace{15mm}
\begin{table*}[b!]
\centering
\renewcommand{\arraystretch}{1.1}
\begin{tabular}{|l|c|c|c|c|c|c|c|c|c|}
\hline
\textbf{GRB ID} & $EF_{\log_{10}(T_i)}$ & $EF_{\log_{10}(F_i)}$ & $EF_{\alpha_i}$ & $EF_{\log_{10}(T_i)}$ RC & $EF_{\log_{10}(F_i)}$ RC & $EF_{\alpha_i}$ RC & \%$_{\log_{10}(T_i)}$ & \%$_{\log_{10}(F_i)}$ & \%$_{\alpha_i}$ \\
\hline
\multicolumn{10}{|c|}{\textbf{PINN - PL}} \\
\hline
070103 & 0.018877 & 0.006314 & 0.041296 & 0.006453 & 0.001447 & 0.014371 & -65.81 & -77.09 & -65.2  \\
070517 & 0.038593 & 0.01129 & 0.139547 & 0.026807 & 0.007595 & 0.094396 & -30.54 & -32.73 & -32.36  \\
080426 & 0.03746 & 0.010364 & 0.049361 & 0.011459 & 0.002916 & 0.006615 & -69.41 & -71.87 & -86.6  \\
090727 & 0.025308 & 0.008036 & 0.136314 & 0.007087 & 0.002338 & 0.038139 & -72 & -70.9 & -72.02  \\
120211A & 0.041991 & 0.012822 & 0.183032 & 0.03891 & 0.016623 & 0.227691 & -7.34 & 29.64 & 24.4  \\
180620B & 0.014386 & 0.003979 & 0.054122 & 0.015383 & 0.005067 & 0.04153 & 6.92 & 27.33 & -23.27  \\

\hline
\multicolumn{10}{|c|}{\textbf{PINN - BPL}} \\
\hline
070103 & 0.018877 & 0.006314 & 0.041296 & 0.005663 & 0.001268 & 0.012296 & -70 & -79.93 & -70.23  \\
070517 & 0.038593 & 0.01129 & 0.139547 & 0.026526 & 0.008169 & 0.097958 & -31.27 & -27.64 & -29.8  \\
080426 & 0.03746 & 0.010364 & 0.049361 & 0.010293 & 0.002288 & 0.013892 & -72.52 & -77.93 & -71.86  \\
090727 & 0.025308 & 0.008036 & 0.136314 & 0.006788 & 0.00222 & 0.035808 & -73.18 & -72.37 & -73.73  \\
120211A & 0.041991 & 0.012822 & 0.183032 & 0.023329 & 0.010331 & 0.146663 & -44.44 & -19.43 & -19.87  \\
180620B & 0.014386 & 0.003979 & 0.054122 & 0.01394 & 0.004621 & 0.040884 & -3.1 & 16.13 & -24.46  \\

\hline
\multicolumn{10}{|c|}{\textbf{PINN - SBPL}} \\
\hline
070103 & 0.018877 & 0.006314 & 0.041296 & 0.005645 & 0.001277 & 0.013026 & -70.09 & -79.78 & -68.46  \\
070517 & 0.038593 & 0.01129 & 0.139547 & 0.02609 & 0.007631 & 0.093721 & -32.4 & -32.41 & -32.84  \\
080426 & 0.03746 & 0.010364 & 0.049361 & 0.013354 & 0.003294 & 0.01046 & -64.35 & -68.21 & -78.81  \\
090727 & 0.025308 & 0.008036 & 0.136314 & 0.006874 & 0.002275 & 0.037057 & -72.84 & -71.69 & -72.82  \\
120211A & 0.041991 & 0.012822 & 0.183032 & 0.020343 & 0.008631 & 0.119351 & -51.55 & -32.69 & -34.79  \\
180620B & 0.014386 & 0.003979 & 0.054122 & 0.017216 & 0.005597 & 0.043945 & 19.67 & 40.67 & -18.8  \\

\hline
\multicolumn{10}{|c|}{\textbf{PINN - Multi-segment PL}} \\
\hline
070103 & 0.018877 & 0.006314 & 0.041296 & 0.005271 & 0.001196 & 0.011675 & -72.07 & -81.06 & -71.73  \\
070517 & 0.038593 & 0.01129 & 0.139547 & 0.02634 & 0.007395 & 0.092672 & -31.75 & -34.5 & -33.59  \\
080426 & 0.03746 & 0.010364 & 0.049361 & 0.014277 & 0.003733 & 0.006996 & -61.89 & -63.98 & -85.83  \\
090727 & 0.025308 & 0.008036 & 0.136314 & 0.006854 & 0.002262 & 0.036883 & -72.92 & -71.85 & -72.94  \\
120211A & 0.041991 & 0.012822 & 0.183032 & 0.043065 & 0.018015 & 0.25031 & 2.56 & 40.5 & 36.76  \\
180620B & 0.014386 & 0.003979 & 0.054122 & 0.015905 & 0.005279 & 0.045881 & 10.56 & 32.66 & -15.23  \\

\hline
\multicolumn{10}{|c|}{\textbf{PINN - Jet Break Model}} \\
\hline
070103 & 0.018877 & 0.006314 & 0.041296 & 0.005115 & 0.001178 & 0.011713 & -72.91 & -81.35 & -71.64  \\
070517 & 0.038593 & 0.01129 & 0.139547 & 0.026019 & 0.007572 & 0.092803 & -32.58 & -32.94 & -33.5  \\
080426 & 0.03746 & 0.010364 & 0.049361 & 0.01788 & 0.004336 & 0.034962 & -52.27 & -58.16 & -29.17  \\
090727 & 0.025308 & 0.008036 & 0.136314 & 0.006839 & 0.002442 & 0.039476 & -72.98 & -69.61 & -71.04  \\
120211A & 0.041991 & 0.012822 & 0.183032 & 0.023349 & 0.011147 & 0.164702 & -44.39 & -13.06 & -10.01  \\
180620B & 0.014386 & 0.003979 & 0.054122 & 0.016664 & 0.005566 & 0.049447 & 15.83 & 39.87 & -8.64  \\

\hline
\multicolumn{10}{|c|}{\textbf{PINN - Energy Injection Model}} \\
\hline 
070103 & 0.018877 & 0.006314 & 0.041296 & 0.006038 & 0.001356 & 0.013181 & -68.01 & -78.53 & -68.08  \\
070517 & 0.038593 & 0.01129 & 0.139547 & 0.027522 & 0.007812 & 0.096918 & -28.69 & -30.81 & -30.55  \\
080426 & 0.03746 & 0.010364 & 0.049361 & 0.012982 & 0.003351 & 0.0092 & -65.34 & -67.67 & -81.36  \\
090727 & 0.025308 & 0.008036 & 0.136314 & 0.006879 & 0.002317 & 0.037673 & -72.82 & -71.16 & -72.36  \\
120211A & 0.041991 & 0.012822 & 0.183032 & 0.044243 & 0.018679 & 0.271249 & 5.36 & 45.67 & 48.2  \\
180620B & 0.014386 & 0.003979 & 0.054122 & 0.016107 & 0.005362 & 0.042 & 11.96 & 34.75 & -22.4  \\

\hline 

\end{tabular}
\end{table*}
\begin{table*}[h!]
\centering
\renewcommand{\arraystretch}{1.1}
\begin{tabular}{|l|c|c|c|c|c|c|c|c|c|}
\hline
\textbf{GRB ID} & $EF_{\log_{10}(T_i)}$ & $EF_{\log_{10}(F_i)}$ & $EF_{\alpha_i}$ & $EF_{\log_{10}(T_i)}$ RC & $EF_{\log_{10}(F_i)}$ RC & $EF_{\alpha_i}$ RC & \%$_{\log_{10}(T_i)}$ & \%$_{\log_{10}(F_i)}$ & \%$_{\alpha_i}$ \\

\hline
\multicolumn{10}{|c|}{\textbf{PINN - Synchrotron}} \\
\hline
070103 & 0.018877 & 0.006314 & 0.041296 & 0.006449 & 0.001428 & 0.014024 & -65.84 & -77.38 & -66.04  \\
070517 & 0.038593 & 0.01129 & 0.139547 & 0.027535 & 0.007974 & 0.097752 & -28.65 & -29.37 & -29.95  \\
080426 & 0.03746 & 0.010364 & 0.049361 & 0.006449 & 0.001591 & 0.002561 & -82.78 & -84.65 & -94.81  \\
090727 & 0.025308 & 0.008036 & 0.136314 & 0.006922 & 0.002338 & 0.037737 & -72.65 & -70.91 & -72.32  \\
120211A & 0.041991 & 0.012822 & 0.183032 & 0.041711 & 0.017846 & 0.247526 & -0.67 & 39.18 & 35.24  \\
180620B & 0.014386 & 0.003979 & 0.054122 & 0.016198 & 0.005263 & 0.04367 & 12.6 & 32.28 & -19.31  \\

\hline
\multicolumn{10}{|c|}{\textbf{Siamese}} \\
\hline
070103 & 0.018877 & 0.006314 & 0.041296 & 0.012944 & 0.003479 & 0.031142 & -31.43 & -44.90 & -24.59 \\
070517 & 0.038593 & 0.011290 & 0.139547 & 0.018644 & 0.006036 & 0.069881 & -51.69 & -46.54 & -49.92 \\
080426 & 0.037460 & 0.010364 & 0.049361 & 0.026290 & 0.006928 & 0.039899 & -29.82 & -33.15 & -19.17 \\ 
090727 & 0.025308 & 0.008036 & 0.136314 & 0.014277 & 0.005004 & 0.074537 & -43.59 & -37.73 & -45.32 \\
120211A & 0.041991 & 0.012822 & 0.183032 & 0.035129 & 0.011317 & 0.162906 & -16.34 & -11.74 & -11.00 \\ 
180620B & 0.014386 & 0.003979 & 0.054122 & 0.011194 & 0.003639 & 0.045707 & -22.19 & -8.55 & -15.55 \\

\hline
\multicolumn{10}{|c|}{\textbf{ReFANN}} \\
\hline
070103 & 0.018877 & 0.006314 & 0.041296 & 0.013182 & 0.003490 & 0.031752 & -30.17 & -44.74 & -23.11 \\
070517 & 0.038593 & 0.011290 & 0.139547 & 0.019912 & 0.006281 & 0.071892 & -48.41 & -44.37 & -48.48 \\
080426 & 0.037460 & 0.010364 & 0.049361 & 0.023113 & 0.006165 & 0.033146 & -38.30 & -40.51 & -32.85 \\
090727 & 0.025308 & 0.008036 & 0.136314 & 0.016191 & 0.005280 & 0.080701 & -36.02 & -34.30 & -40.80 \\
120211A & 0.0420 & 0.0128 & 0.1830 & 0.0343 & 0.0116 & 0.1613 & -18.34 & -9.50 & -11.89 \\
180620B & 0.014386 & 0.003979 & 0.054122 & 0.011506 & 0.003605 & 0.044954 & -20.02 & -9.41 & -16.94 \\

\hline
\multicolumn{10}{|c|}{\textbf{Quantile}} \\
\hline
070103 & 0.018877 & 0.006314 & 0.041296 & 0.015241 & 0.004144 & 0.036931 & -19.26 & -34.37 & -10.57 \\
070517 & 0.038593 & 0.011290 & 0.139547 & 0.019948 & 0.005875 & 0.067097 & -48.31 & -47.96 & -51.92 \\
080426 & 0.037460 & 0.010364 & 0.049361 & 0.026534 & 0.007046 & 0.033772 & -29.17 & -32.02 & -31.58 \\
090727 & 0.025308 & 0.008036 & 0.136314 & 0.016404 & 0.007305 & 0.106562 & -35.18 & -9.09 & -21.83 \\
120211A & 0.041991 & 0.012822 & 0.183032 & 0.032976 & 0.009188 & 0.158321 & -21.47 & -28.34 & -13.50 \\
180620B & 0.014386 & 0.003979 & 0.054122 & 0.011232 & 0.003658 & 0.047104 & -21.93 & -8.08 & -12.97 \\
\hline

\end{tabular}
\caption{Comparison of baseline Error Fractions ($EF$), reconstructed error fractions ($EF_{\text{RC}}$), and relative percentage parameter uncertainty changes ($\%\log_{10} F_a$, $\%\log_{10} T_a$, and $\% \alpha$) for the W07 model across representative GRBs, evaluated for all PINN physics priors, Siamese, ReFANN, and PQR models. \label{tab:EF/RC}}
\end{table*}

\subsection{Gap-Aware Reconstruction \& Uncertainty Quantification} \label{Gap-Aware Reconstruction}

To ensure that light-curve reconstructions remain physically plausible within observational outages while providing statistically rigorous confidence bounds, a unified gap-aware reconstruction protocol is implemented alongside model-tailored uncertainty quantification (UQ) schemes.

\subsubsection{Adaptive Gap Insertion Strategy} \label{Adaptive Gap Insertion Strategy}

Reconstruction points are inserted exclusively within identified temporal gaps where the logarithmic time interval exceeds $\Delta \tau > 0.05$ (corresponding to a fractional time interval of $\approx 12\%$). 
This matching threshold prevents artificial over-sampling of well-observed epochs.
The number of inserted reconstruction grid points scales adaptively as a percentage of the original sample size ($N_{\mathrm{obs}}$):
\begin{itemize}
    \item $5\%$ for densely sampled light-curves ($N_{\mathrm{obs}} > 500$)
    \item $10\%$ for moderately sampled light-curves ($250 \le N_{\mathrm{obs}} \le 500$)
    \item $30\%$ for sparsely sampled light-curves ($100 \le N_{\mathrm{obs}} < 250$)
    \item $40\%$ for severely undersampled light-curves ($N_{\mathrm{obs}} < 100$), subject to a mandatory minimum baseline of 20 points across all cases.
\end{itemize}

A Boolean gap-mask array tracks point origins, recording True for newly inserted gap points and False for coordinates coincident with true observations. 
Only true gap-mask entries are written to output CSV files and rendered as yellow error bars in visualisation plots. 
This strategy resolves the systematic shift in post-reconstruction phenomenological parameters (specifically $\log_{10} T_a$ and $\log_{10} F_a$) that arises when reconstruction points are uniformly placed without accounting for the gap structure.

\subsubsection{Model-Specific Uncertainty Quantification Formulations} \label{Model-Specific Uncertainty Quantification Formulations}

Because the core architectures process uncertainty through distinct physical and probabilistic assumptions, UQ is formulated specifically for each model framework while maintaining consistent $95\%$ confidence intervals across all outputs.

\begin{table*}[htp] 
\centering
\begin{tabular}{ccc}
    \hline
    Model Framework & Primary UQ Mechanism & Input Source  \\
    \hline
    PINN & Residual + Smoothing & Post-fit residuals \& obs. errors  \\
    Siamese & MC Dropout (200 stochastic forward & Log-variance head \&  \\
     & passes at inference, dropout active) & epistemic MC spread \\
    ReFANN & MC Dropout (200 stochastic forward & Log-variance head \& \\
     & passes at inference, dropout active) & epistemic MC spread \\
    
    Polynomial Quantile Regression & Non-parametric Pinball Loss & Empirical error distributions \\
    \hline
    \end{tabular}
\caption{Overview of Uncertainty Quantification (UQ) mechanisms and underlying input sources across the four candidate machine learning light-curve reconstruction frameworks (PINN, Siamese, ReFANN, and PQR model). \label{tab:Framework}}
\end{table*}

\paragraph{PINN: Residual-Based UQ} \label{PINN Residual-Based UQ}

For the PINN framework, confidence intervals are computed from a residual-based width rather than MC-dropout uncertainty. 
The total uncertainty $\sigma_{\mathrm{total}, k}$ at a reconstruction grid point $k$ combines the model's epistemic uncertainty and the observational aleatoric uncertainty in quadrature:
$$\sigma_{\mathrm{total}, k} = \sqrt{\sigma_{\mathrm{epistemic}}^2 + \sigma_{\mathrm{aleatoric}, k}^2}$$
After training, the PINN is evaluated in deterministic evaluation mode at the observed timestamps to compute the residual vector $\hat{y}_i - y_i$. 
The epistemic uncertainty is estimated via the residual standard deviation across observed points:
$$\sigma_{\mathrm{epistemic}} = \text{std}(\hat{y}_i - y_i)$$
The localised aleatoric uncertainty $\sigma_{\mathrm{aleatoric}, k}$ at a reconstruction grid point $\tau_k$ is obtained by interpolating the observational flux uncertainties $\sigma_\mathrm{obs}$ from observed timestamps $\tau_\mathrm{obs}$ to grid points $\tau_k$, followed by Gaussian smoothing (kernel size $\sigma = 8$ grid points) to eliminate interpolation jitter:
$$\sigma_{\mathrm{aleatoric}, k} = \text{Smooth}\Big(\text{interp}(\tau_k;\, \tau_\mathrm{obs},\, \sigma_\mathrm{obs})\Big)$$
The resulting $95\%$ confidence interval for a reconstructed logarithmic flux point $\hat{y}(\tau_k)$ is then given by $\hat{y}(\tau_k) \pm 1.96 \, \sigma_{\mathrm{total}, k}$.

MC-dropout uncertainty (200 forward passes with dropout active) was explored but rejected: GELU activations with dropout collapse toward zero in extrapolation regions \citep{Gal2016, Gal2017}, pulling the MC-mean below the deterministic prediction and producing asymmetric confidence intervals even when the formula is symmetric. 
The residual-based approach yields symmetric bands that are physically interpretable: wider when the PINN fit is poor, narrower when the PINN closely tracks the data.

\paragraph{ReFANN and Siamese Architectures: MC-Dropout UQ} \label{ReFANN and Siamese: MC-Dropout UQ}

The UQ for the ReFANN and Siamese NN follows a different approach from other models. 
Both models use MC Dropout: after training, 200 stochastic forward passes are performed with dropout kept active, producing 200 stochastic reconstructions for each time point. 
From these passes, two sources of uncertainty are estimated: the aleatoric uncertainty, obtained from a dedicated output head that directly predicts the log-variance of the data, and the epistemic uncertainty, given by the variance across the 200 predicted means. 

The two are then combined in quadrature to obtain the total predictive standard deviation, $\sigma_{\mathrm{total}} = \sqrt{\sigma_{\mathrm{epistemic}}^2 + \sigma_{\mathrm{aleatoric}}^2}$.
Unlike the percentile-based confidence intervals used for the DGP, QSS, BNN, and TCN models, the $95\%$ confidence interval here is computed as $\hat{y} \pm 1.96\,\sigma_{\mathrm{total}}$, where $\hat{y}$ is the mean prediction across the 200 MC samples. 
Hence, unlike the other models, the predictive uncertainties in ReFANN and the Siamese network are obtained entirely from the probabilistic NN framework (heteroscedastic regression combined with MC dropout), without directly propagating observational measurement errors or estimating uncertainty from post-fit residuals.

\paragraph{Polynomial Quantile Regression: Log-Scale Transformation \& Distribution-Free Bounds} \label{Quantile UQ}

For the PQR model framework, data pre-processing consists of a direct base-10 logarithmic transformation of observed temporal and flux values ( $\log_{10} T$ and $\log_{10} F$). 
This log-space conversion effectively compresses the extreme physical dynamic scales inherent to GRB events, establishing the necessary numerical stability for robust polynomial fitting.

Uncertainty bounds are estimated non-parametrically across the target quantiles ($q \in \{0.1, 0.5, 0.9\}$) defined in §~\ref{Quantile} (Eq.~\ref{eq:quantile_degree}), using MC error sampling drawn directly from normal observational error distributions.
Confidence intervals are rendered as $\hat{y} \pm 1.96 \sigma$, where $\sigma$ represents the 68th percentile of absolute observed logarithmic flux uncertainties.

\section{Results} \label{section:results}
\begin{figure*}[!htp]
\centering

    \includegraphics[width=.24\textwidth, height=.20\textwidth]{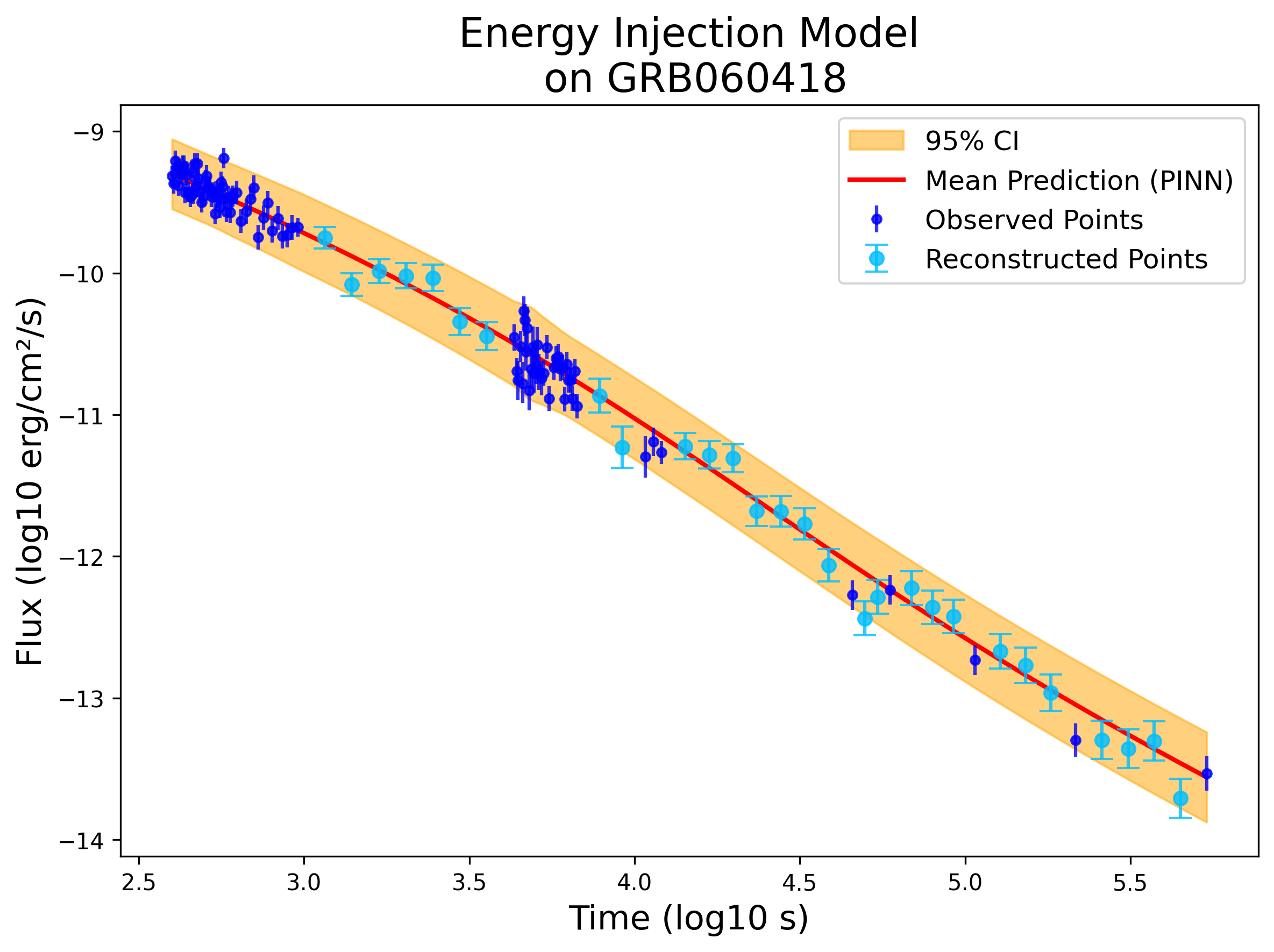}
    \includegraphics[width=.24\textwidth, height=.20\textwidth]{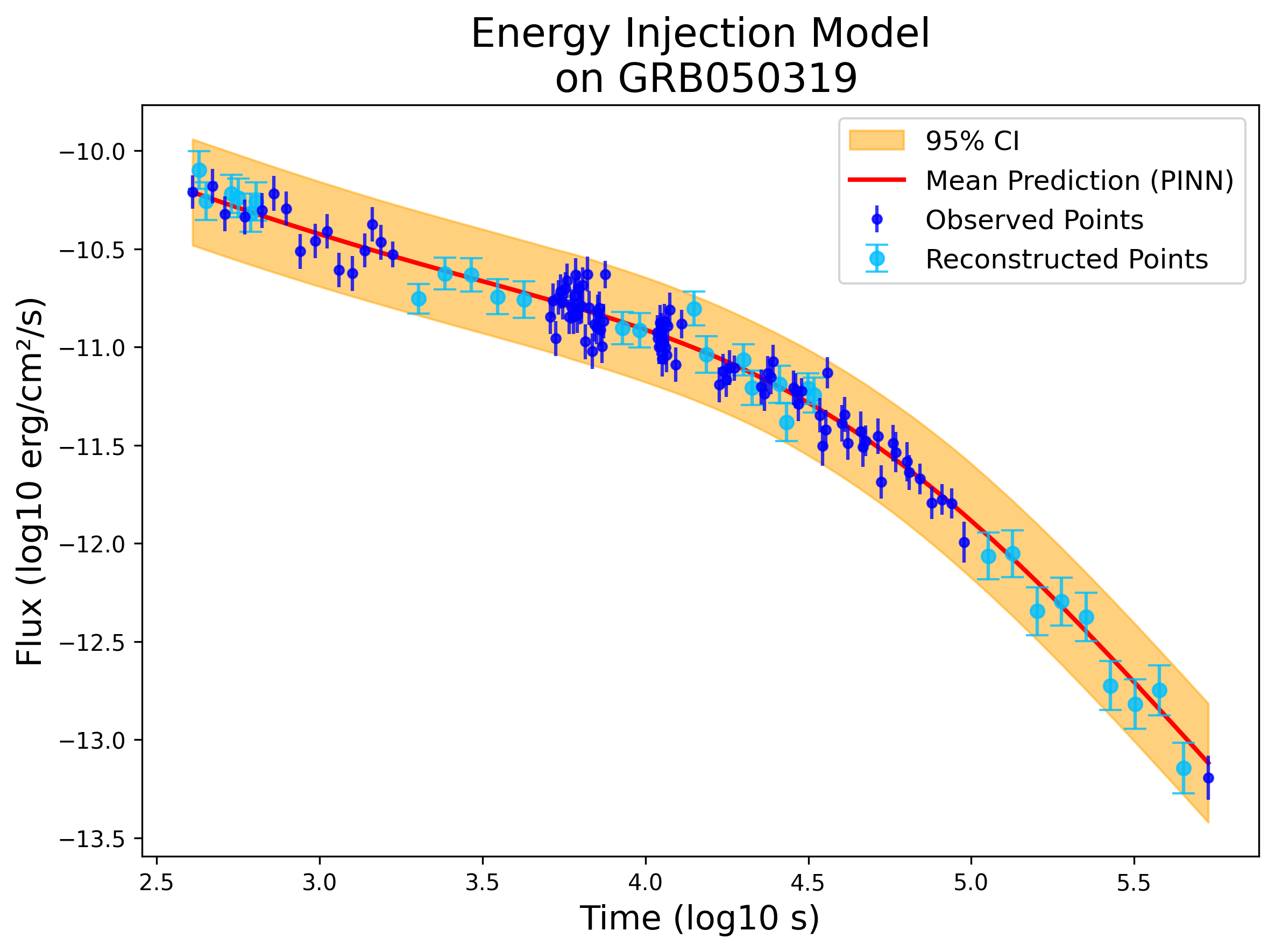}
    \includegraphics[width=.24\textwidth, height=.20\textwidth]{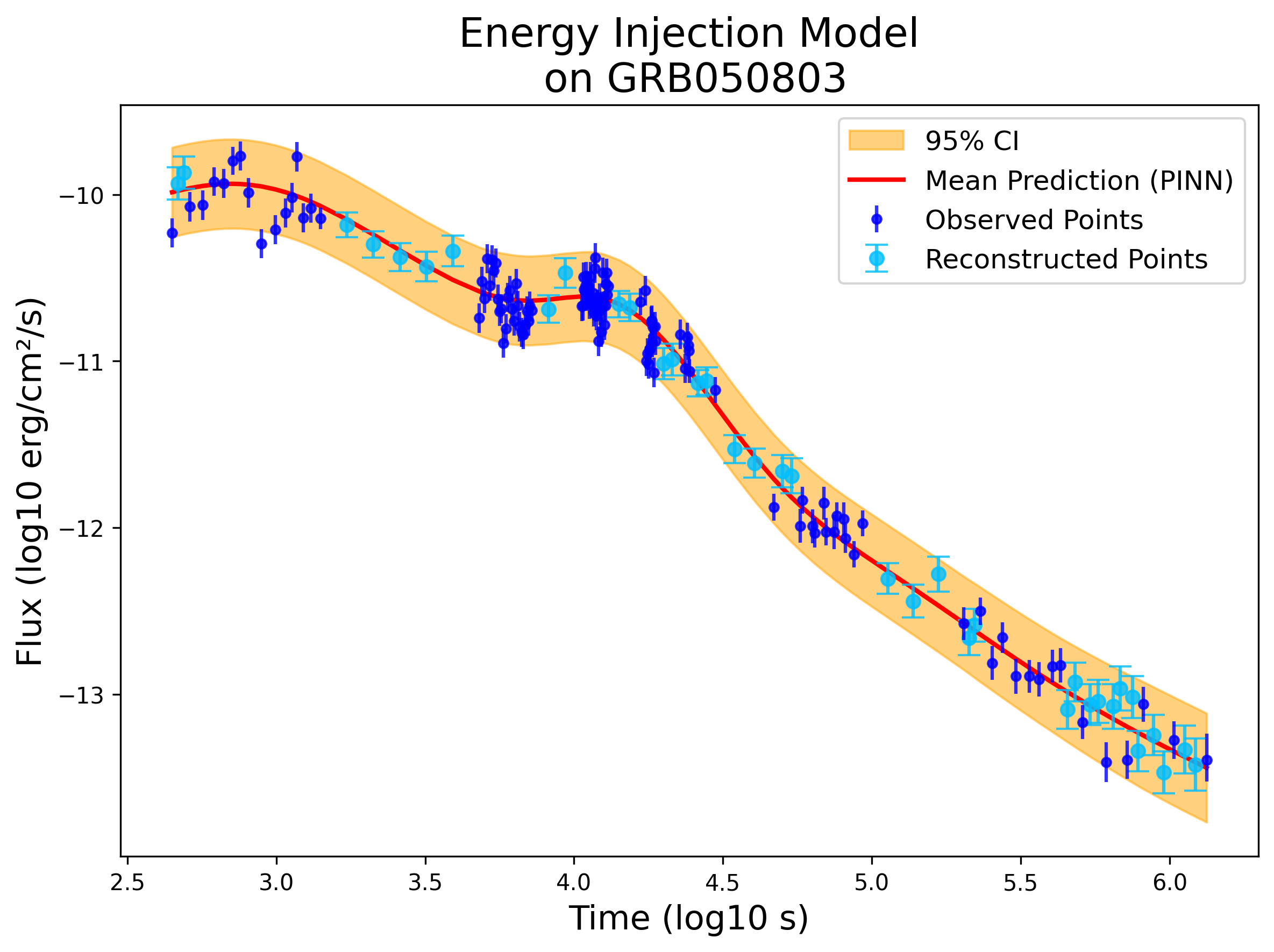}
    \includegraphics[width=.24\textwidth, height=.20\textwidth]{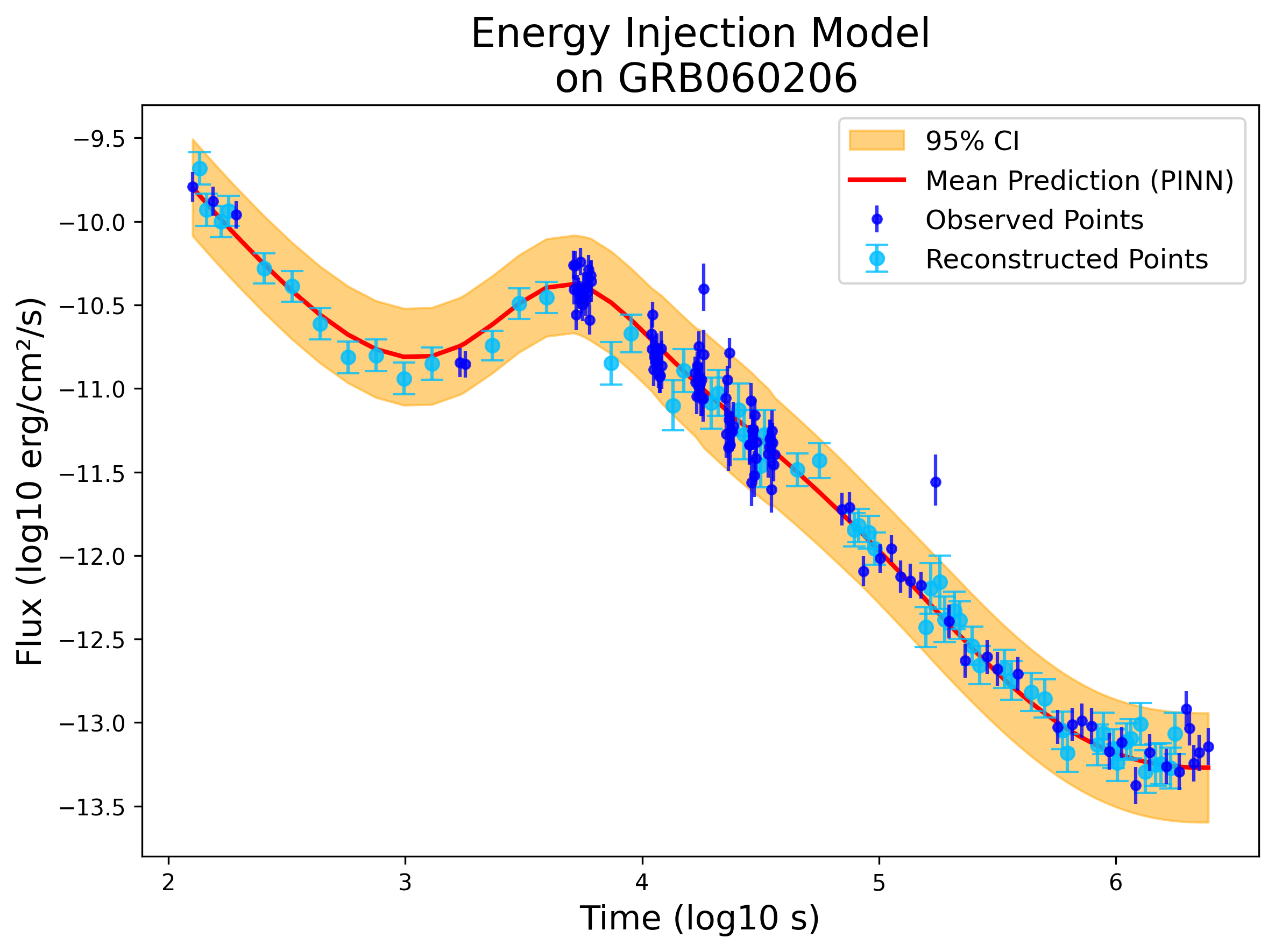}  
    \includegraphics[width=.24\textwidth, height=.20\textwidth]{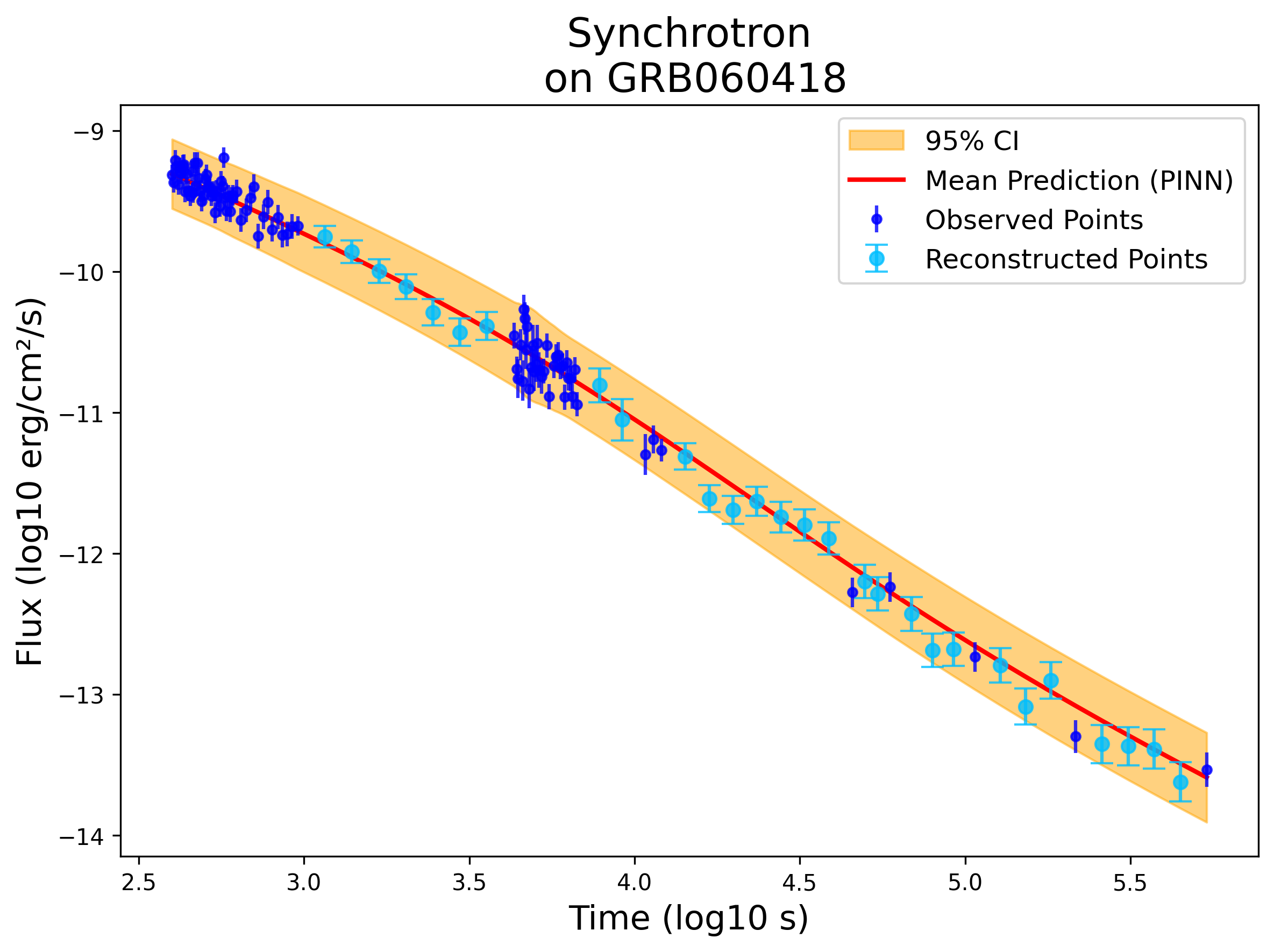}
    \includegraphics[width=.24\textwidth, height=.20\textwidth]{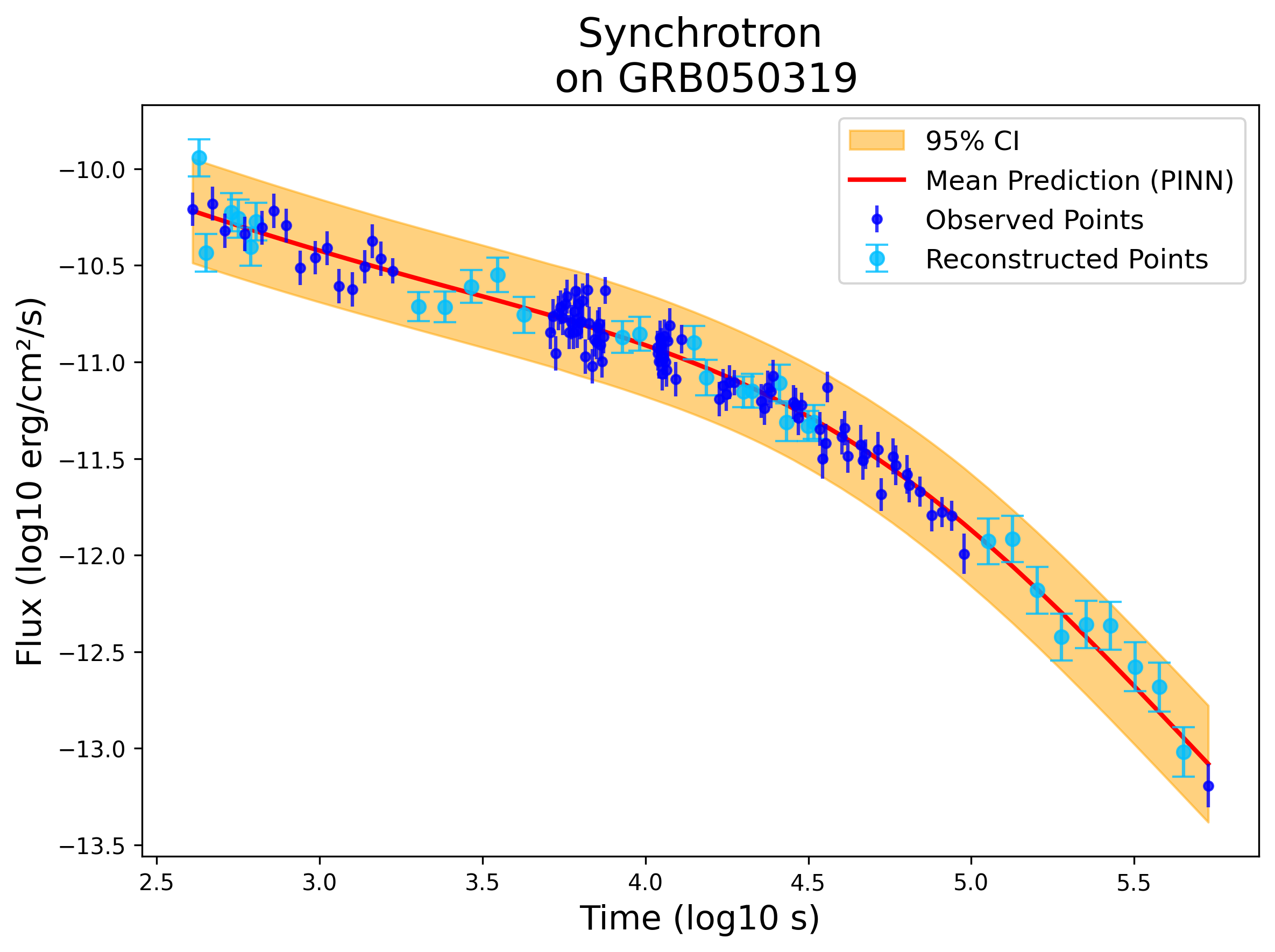}
    \includegraphics[width=.24\textwidth, height=.20\textwidth]{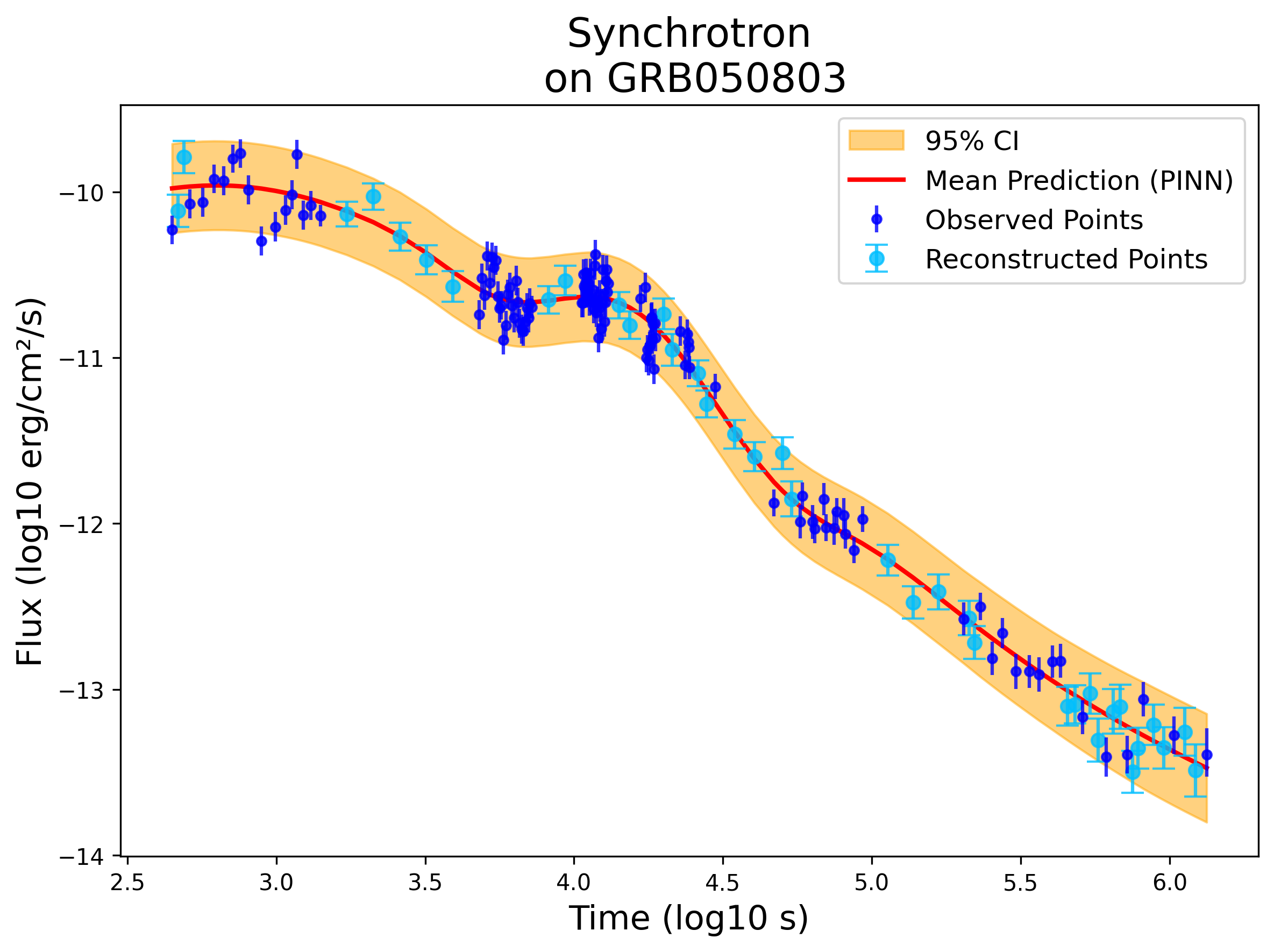}
    \includegraphics[width=.24\textwidth, height=.20\textwidth]{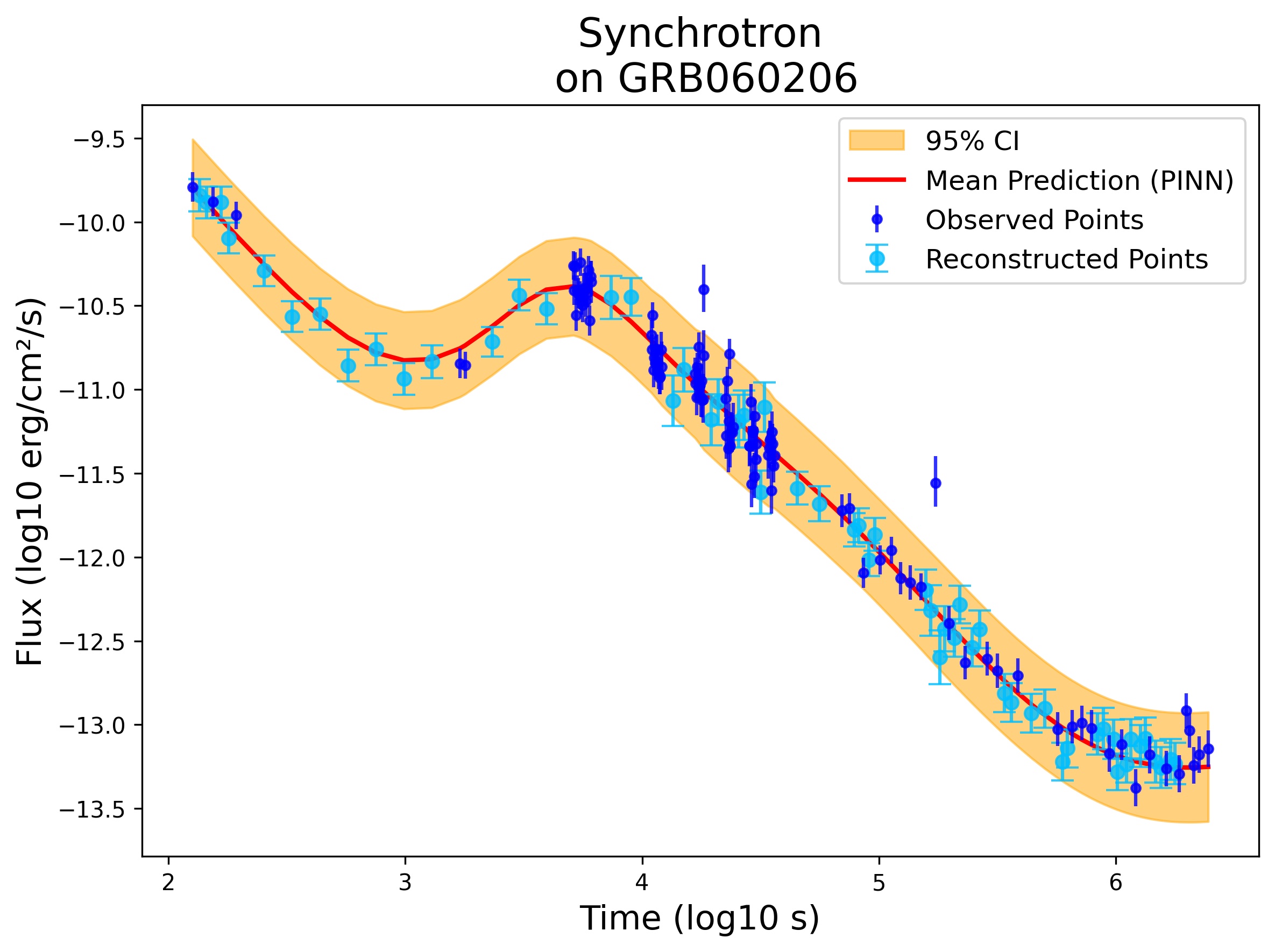}  

    \includegraphics[width=.24\textwidth, height=.20\textwidth]{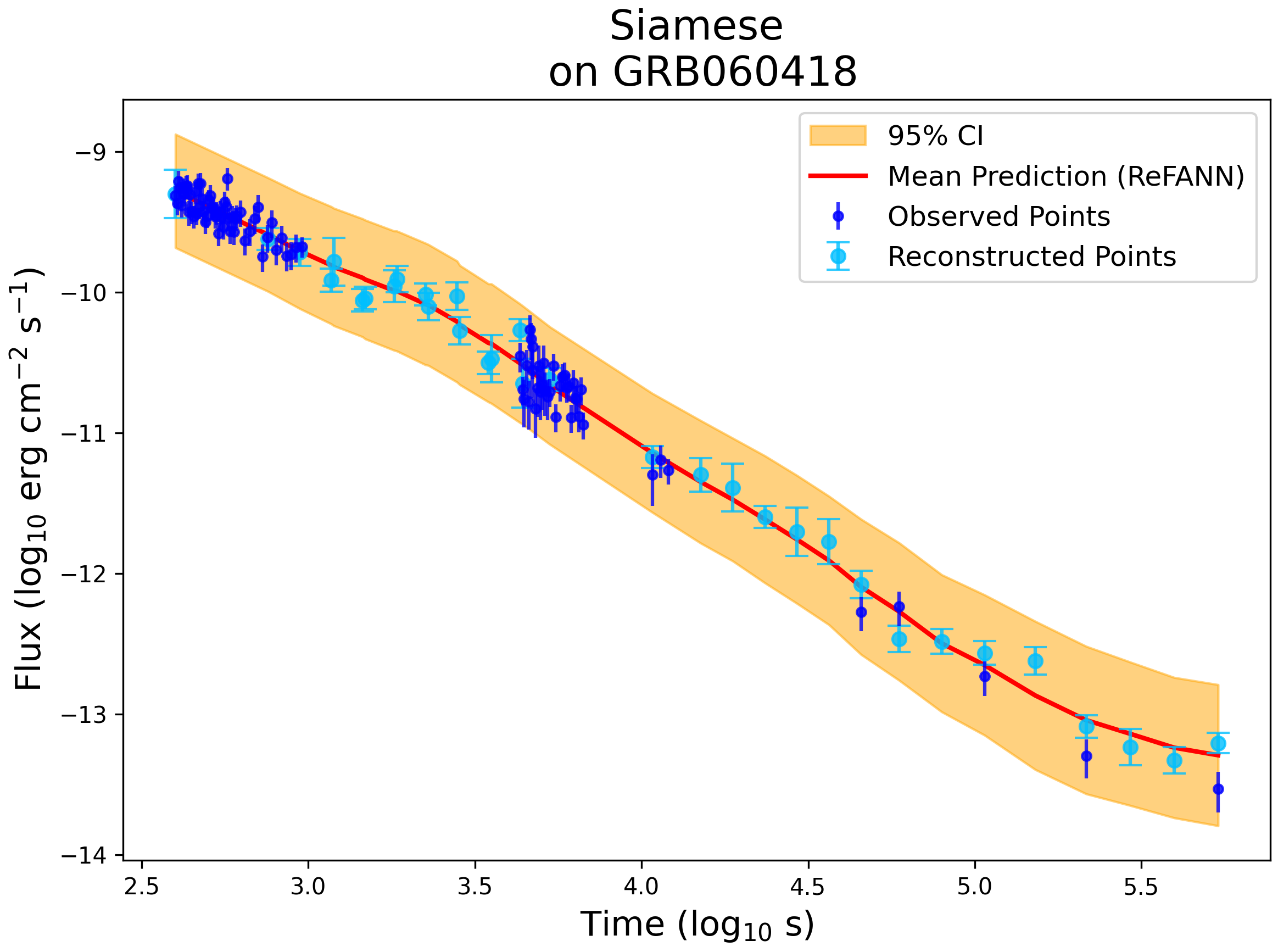}
    \includegraphics[width=.24\textwidth, height=.20\textwidth]{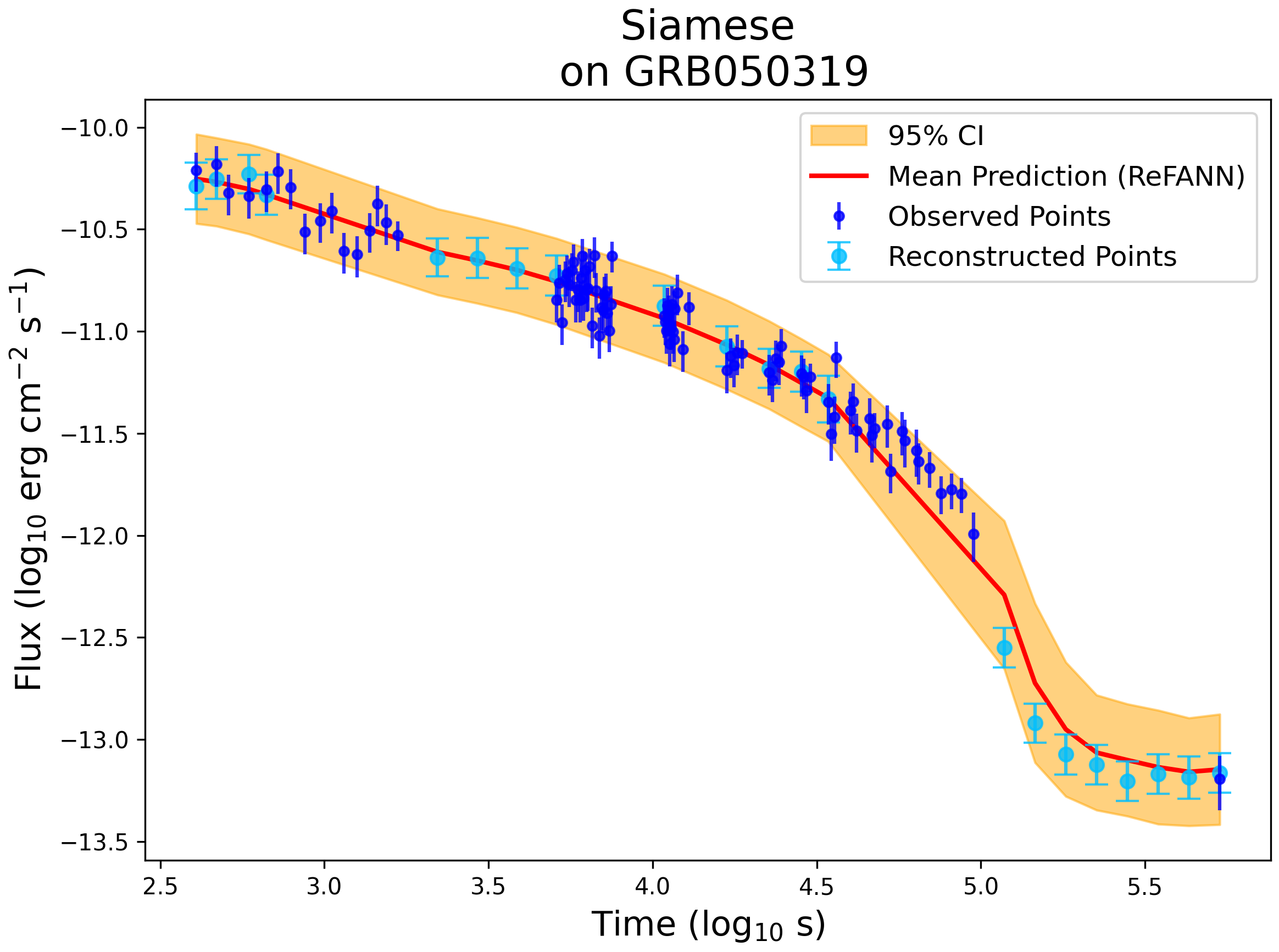}
    \includegraphics[width=.24\textwidth, height=.20\textwidth]{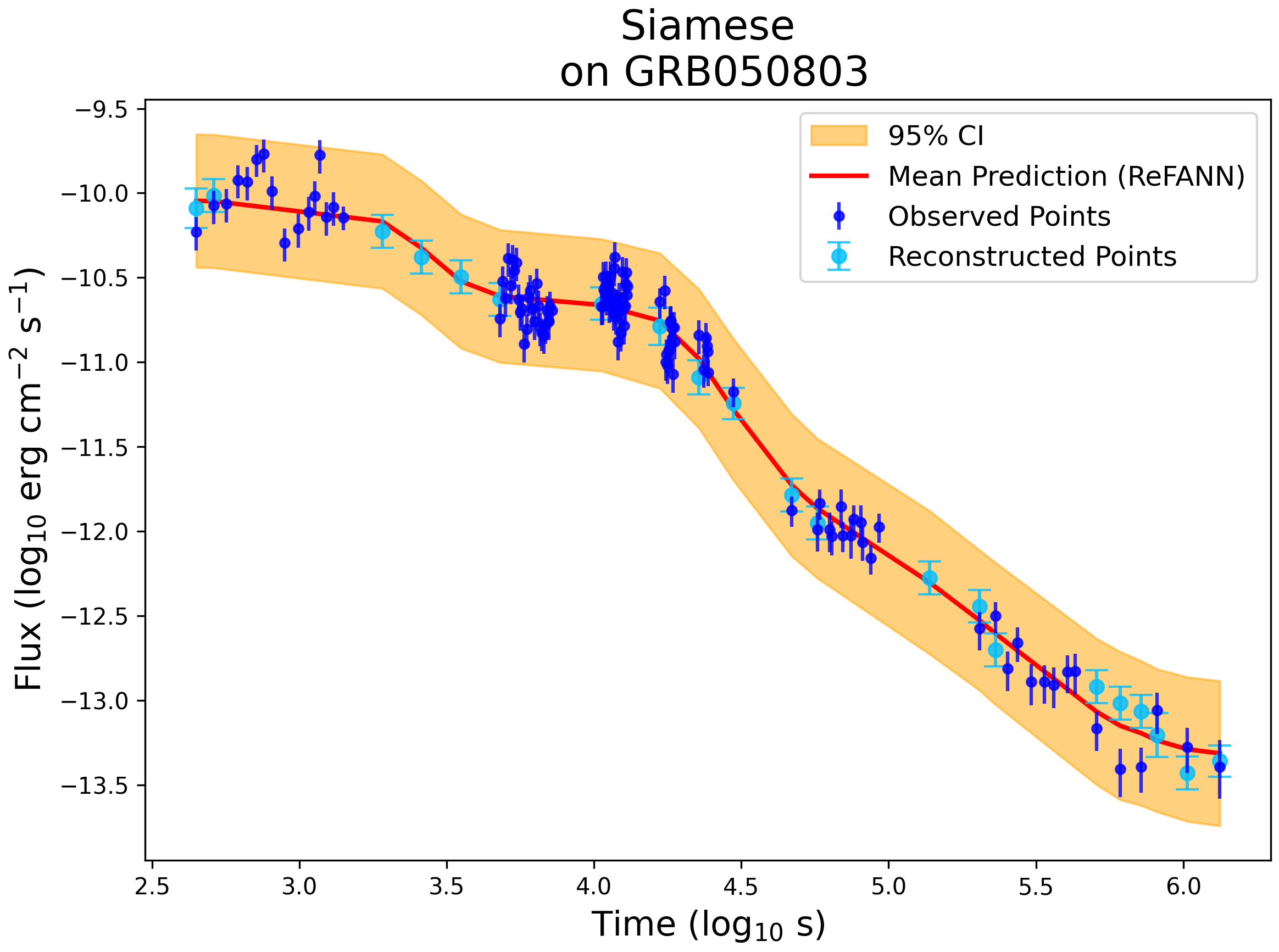}
    \includegraphics[width=.24\textwidth, height=.20\textwidth]{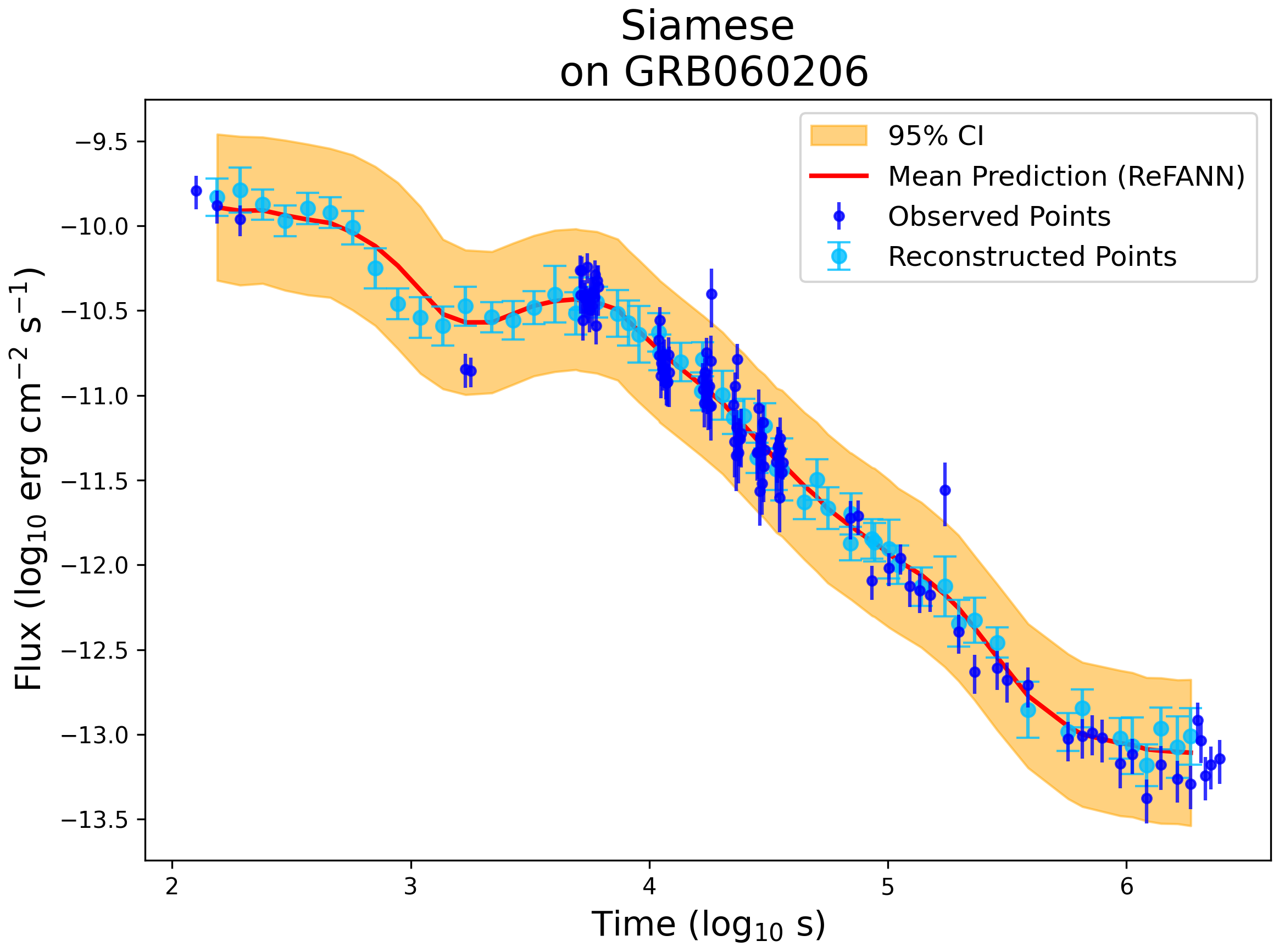}

    \includegraphics[width=.24\textwidth, height=.20\textwidth]{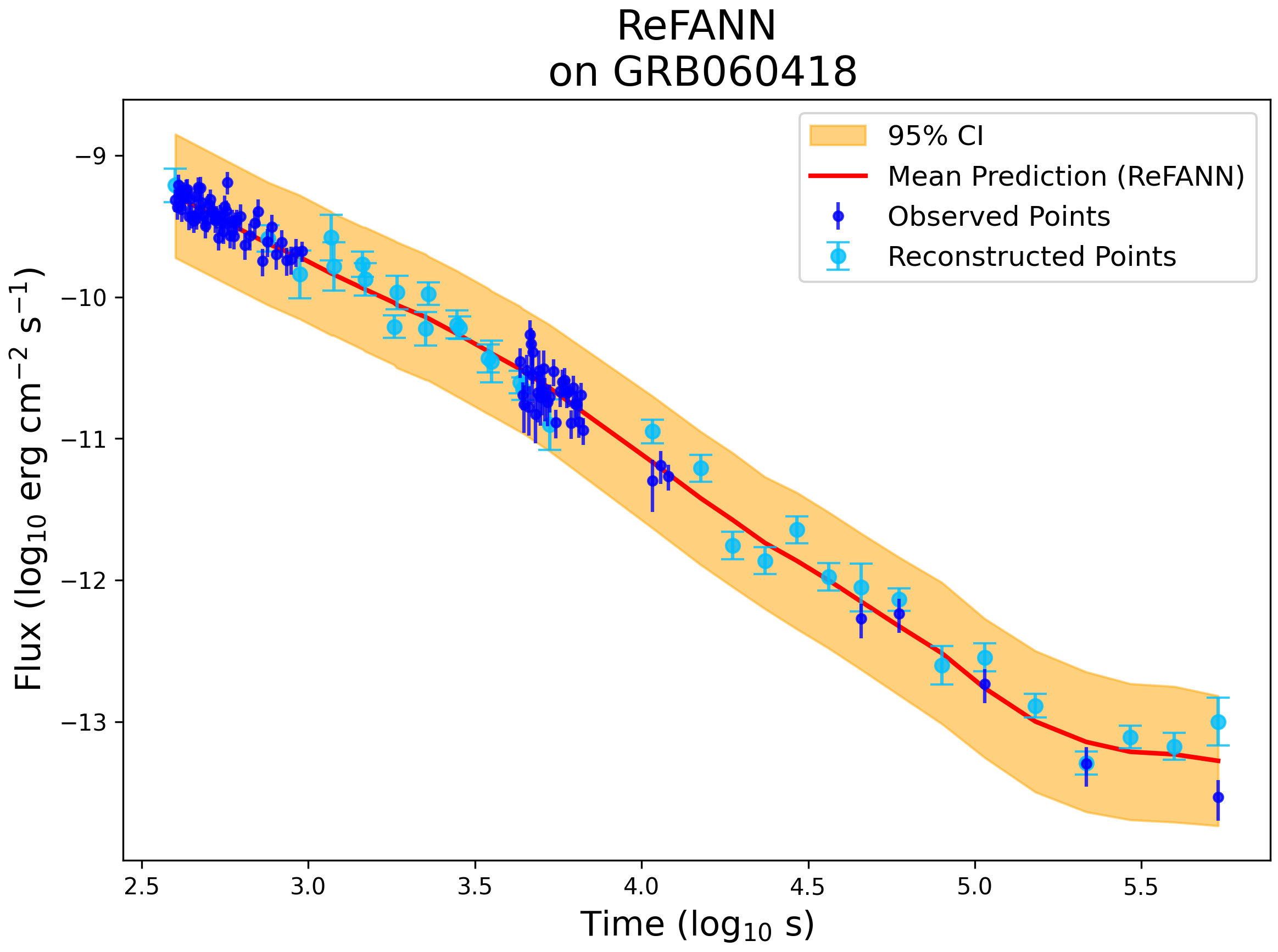}
    \includegraphics[width=.24\textwidth, height=.20\textwidth]{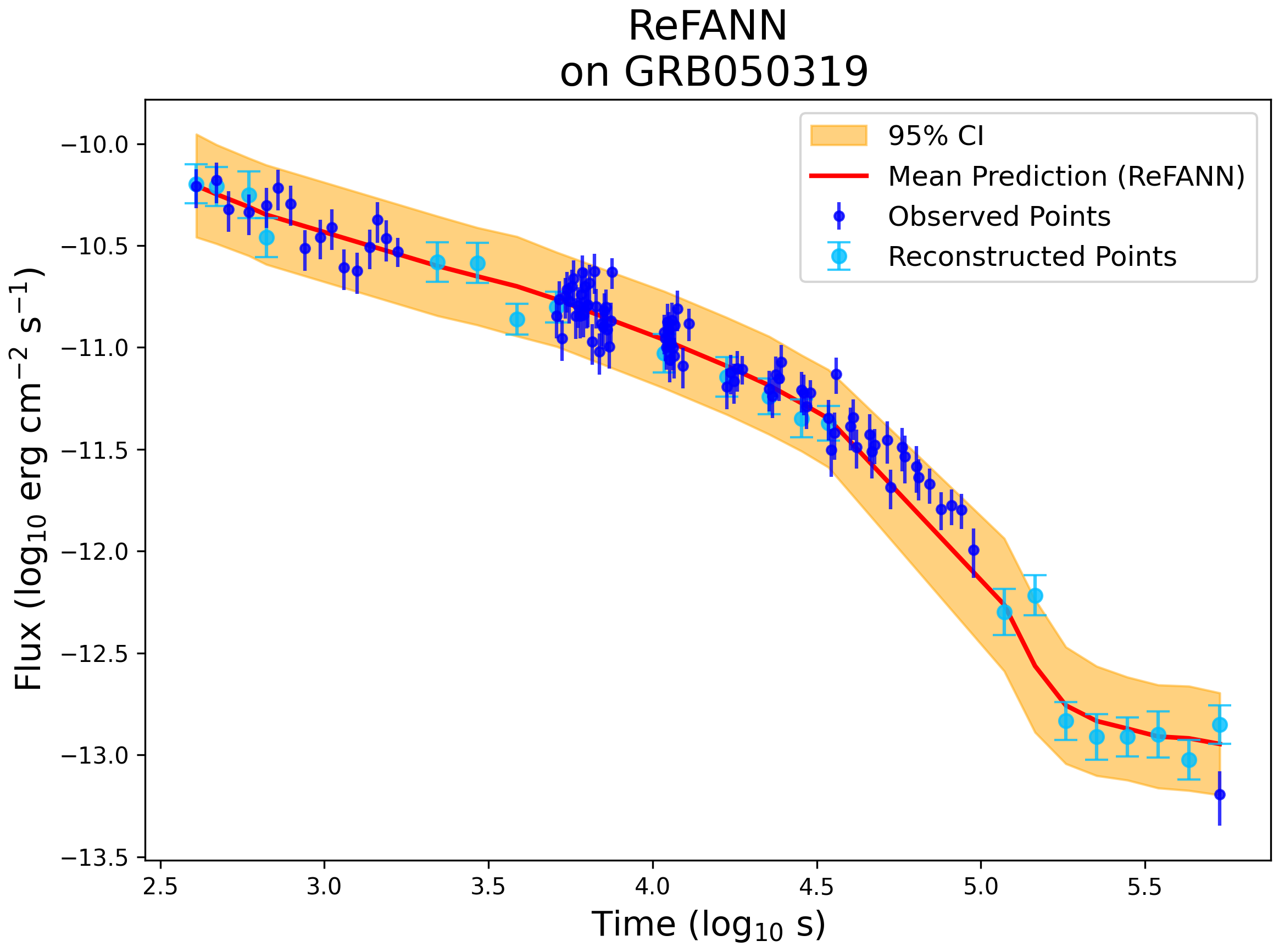}
    \includegraphics[width=.24\textwidth, height=.20\textwidth]{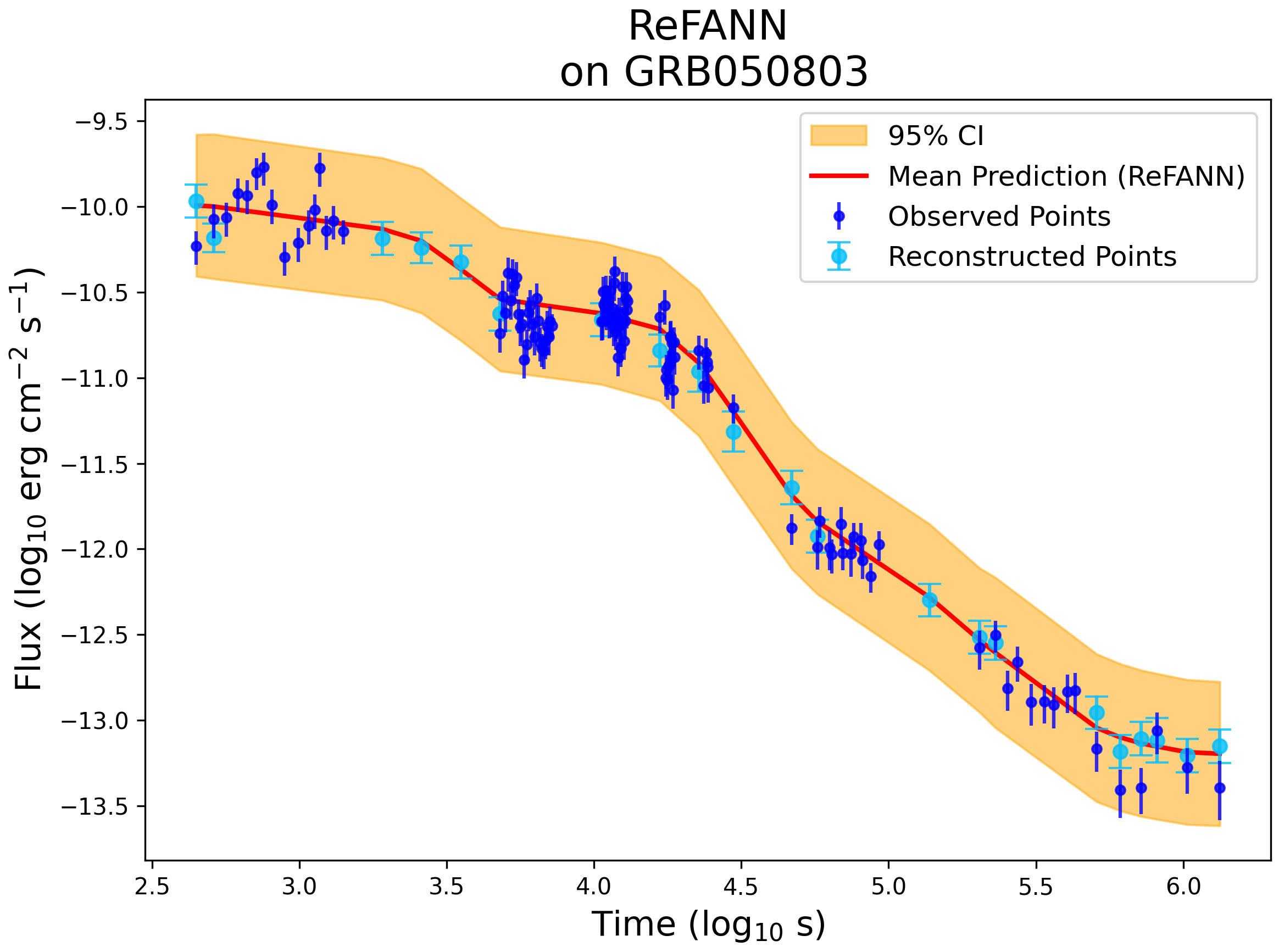}
    \includegraphics[width=.24\textwidth, height=.20\textwidth]{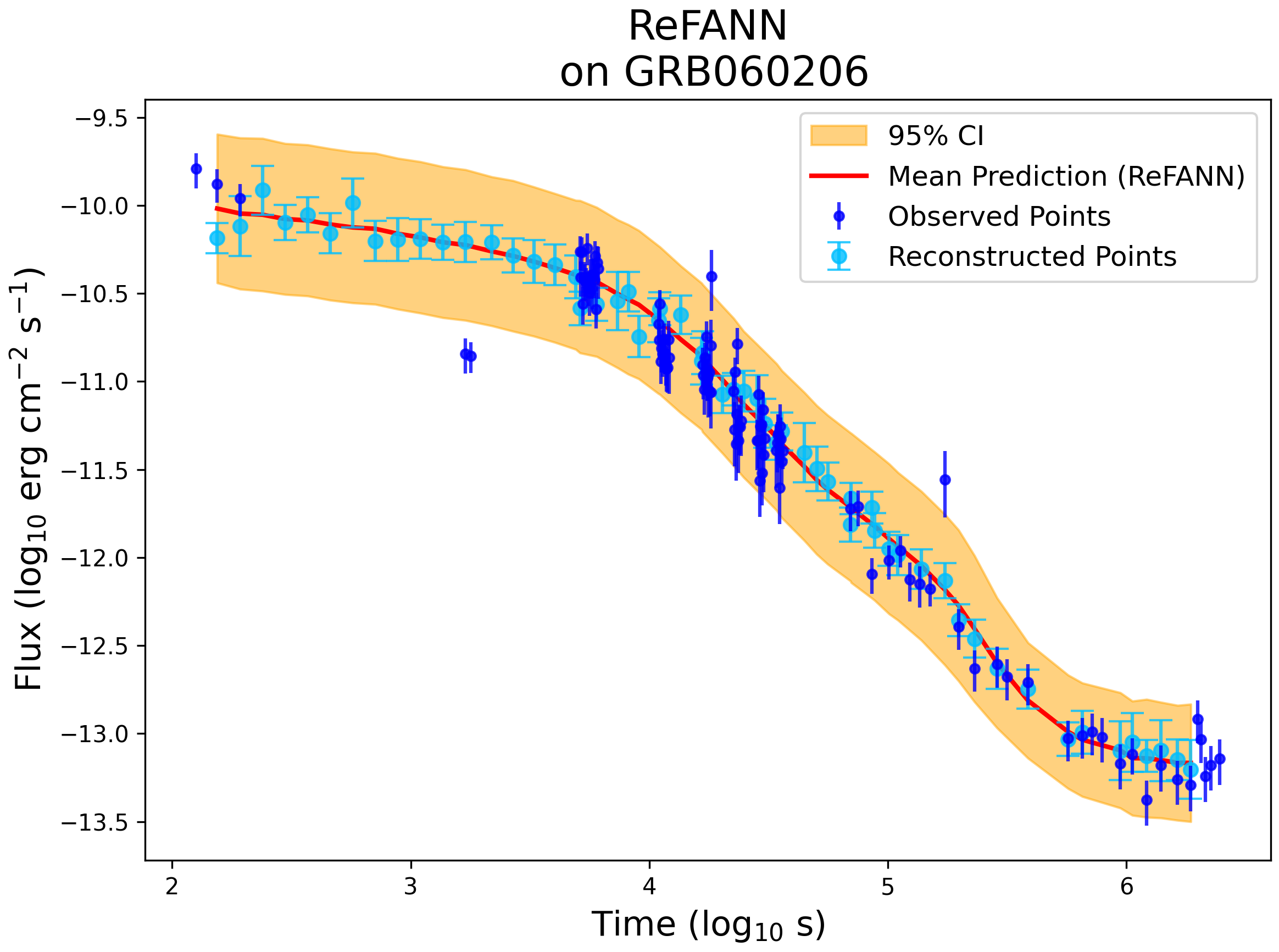}

    \includegraphics[width=.24\textwidth, height=.20\textwidth]{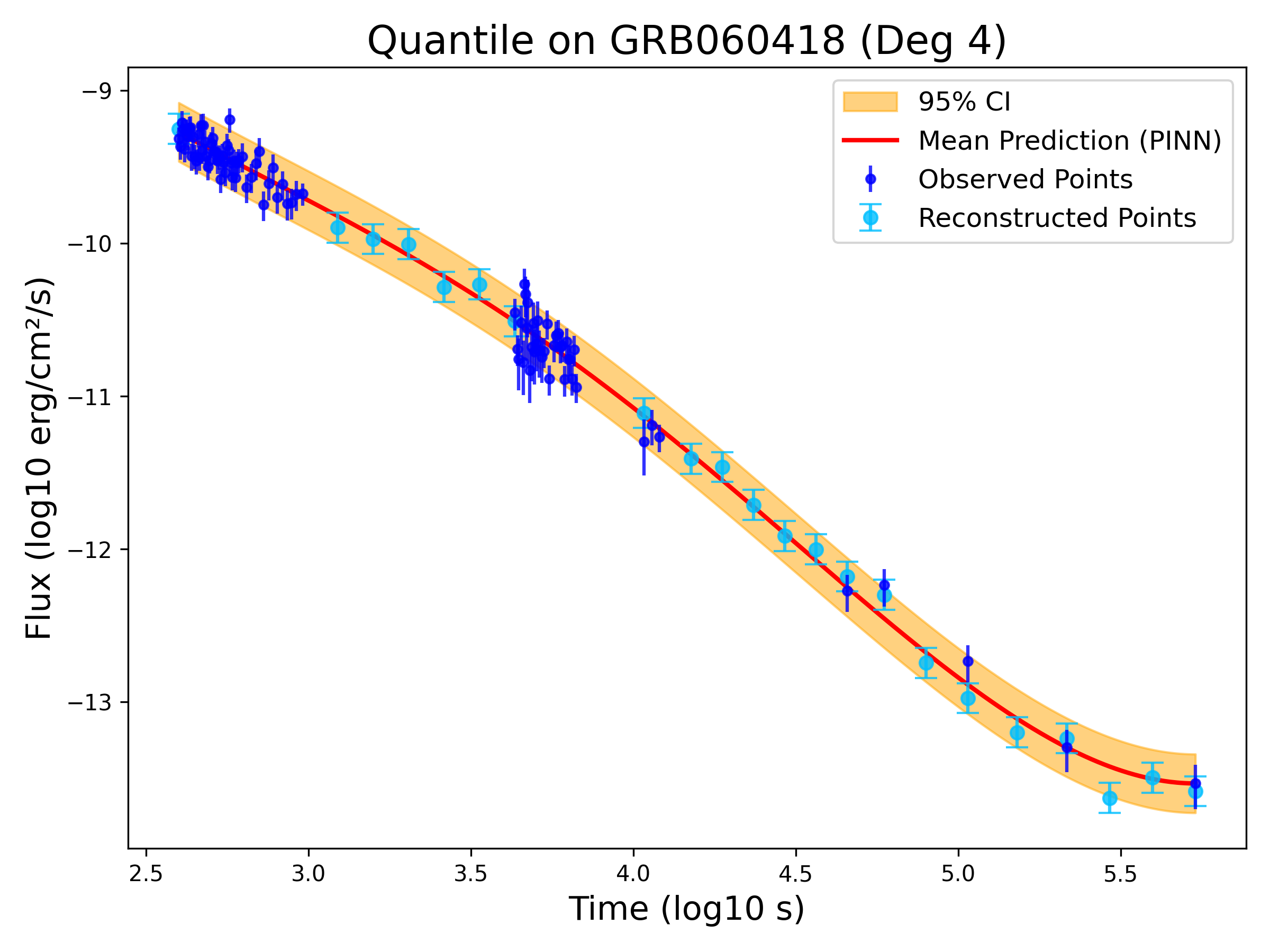}
    \includegraphics[width=.24\textwidth, height=.20\textwidth]{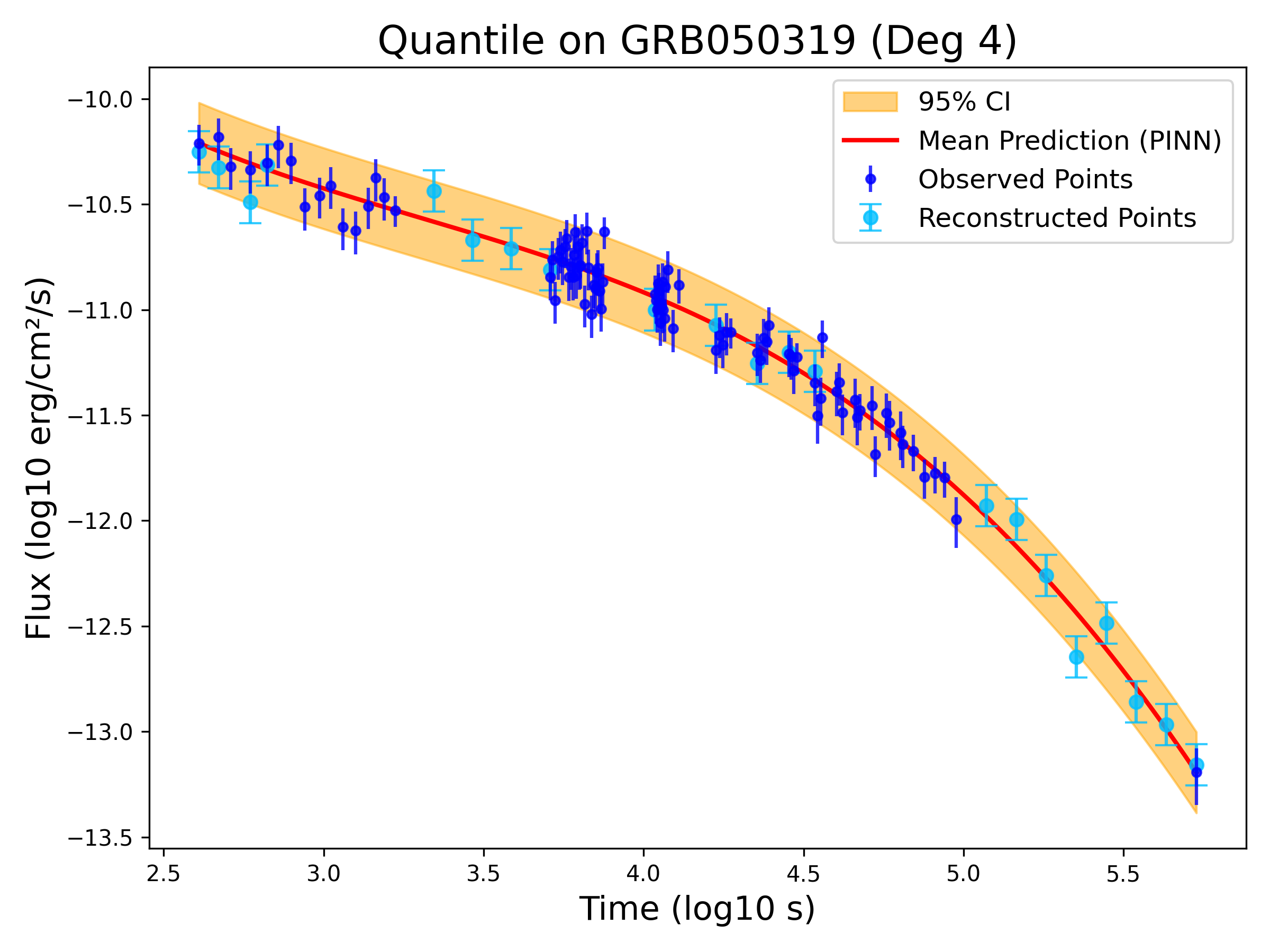}
    \includegraphics[width=.24\textwidth, height=.20\textwidth]{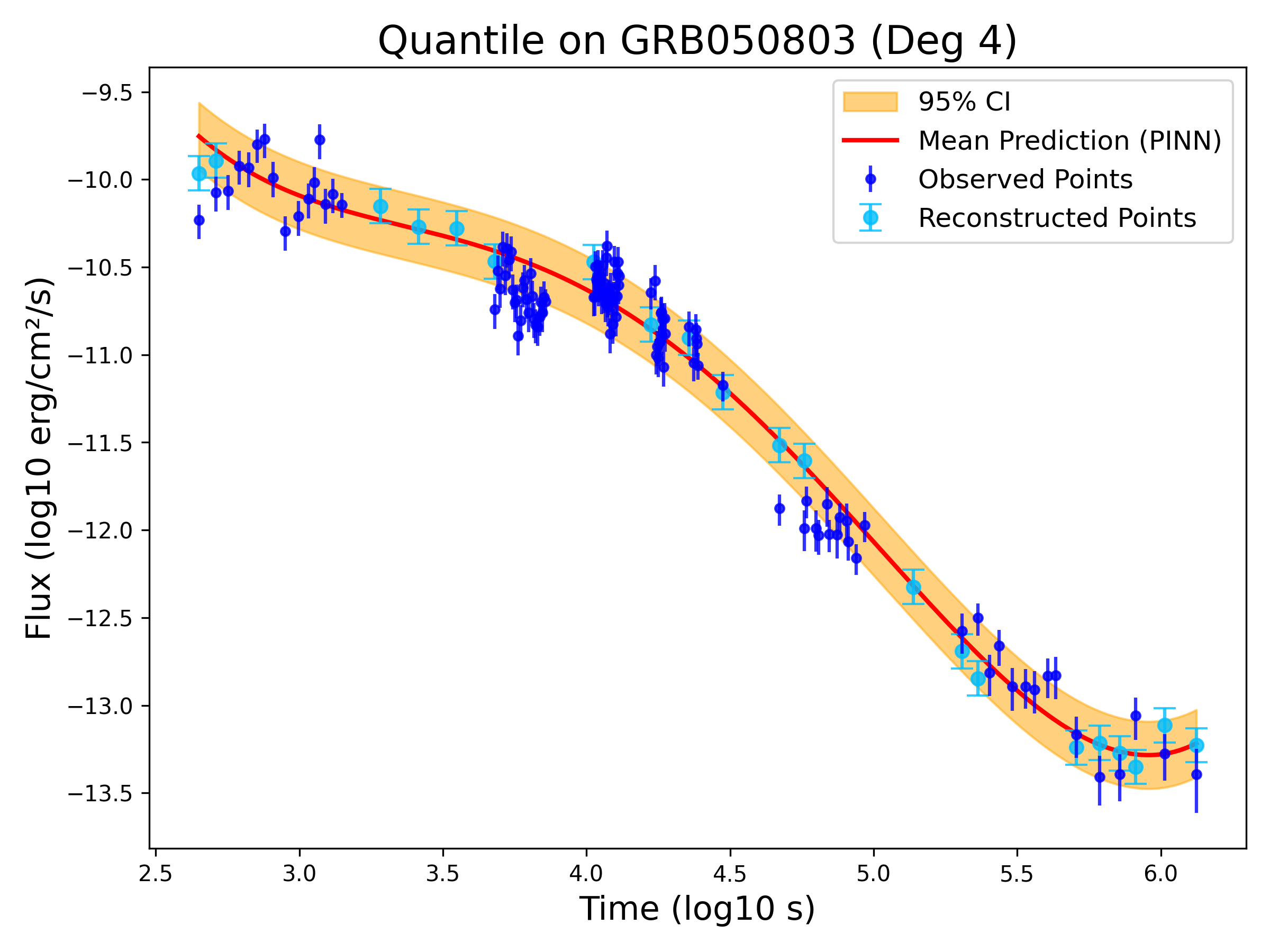}
    \includegraphics[width=.24\textwidth, height=.20\textwidth]{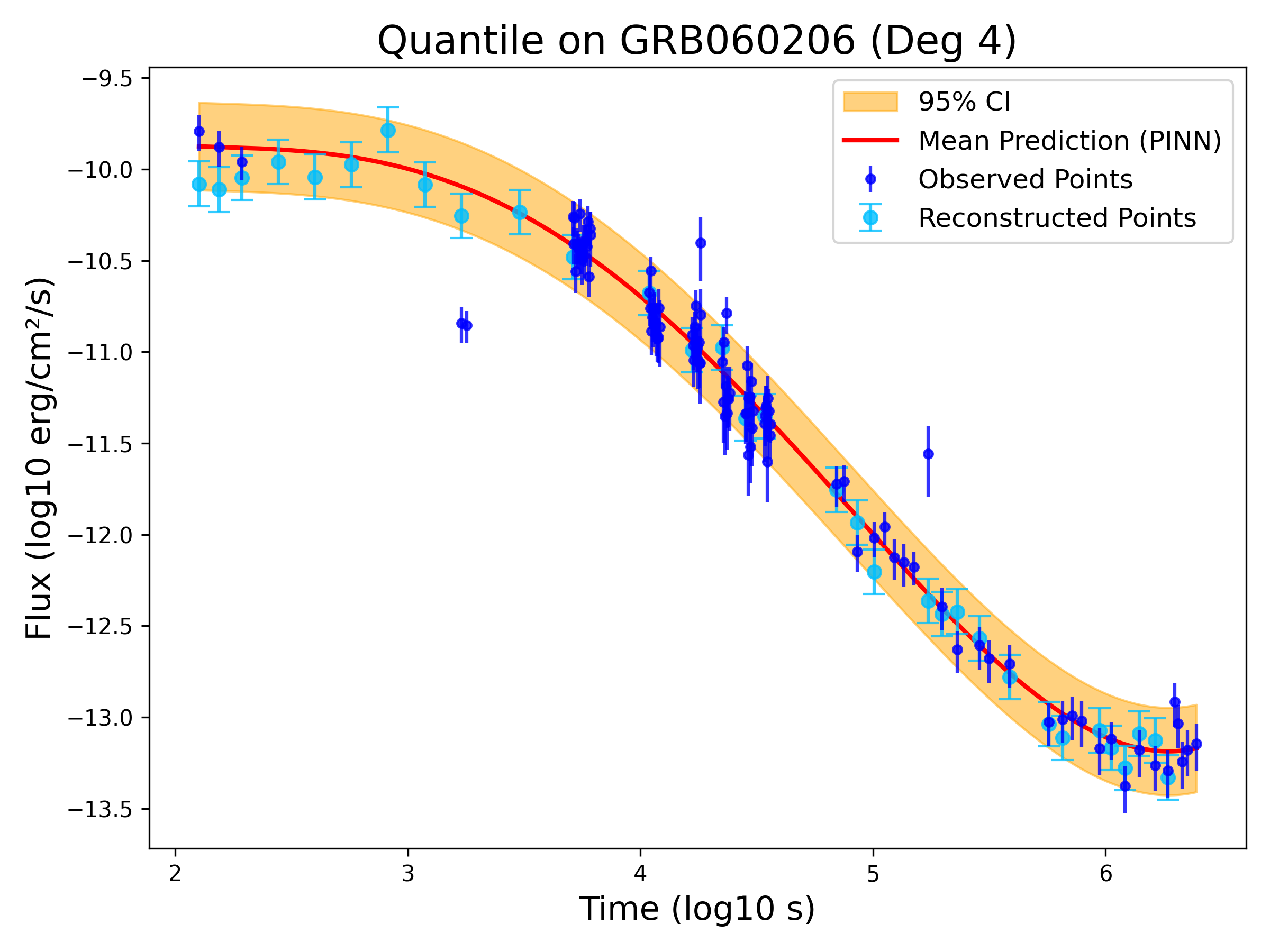}  
    
\caption{Reconstructions of the LCs for four GRB varieties are shown from left to right: i) good GRBs (Column 1); ii) GRB LCs with a break toward the end (Column 2); iii) GRB LCs with flares or bumps in the afterglow (Column 3); and iv) GRB LCs with flares or bumps and a double break toward the end (Column 4). The models are arranged from top to bottom as follows: i) PINN (Row 1); ii) ReFANN (Row 2); iii) Siamese Model (Row 3); and iv) Quantile Model (Row 4). \label{fig:Reconstrusted_grbs}}
\end{figure*}

\begin{figure*}[!htp]
\centering
\begin{adjustbox}{width=\textwidth}
\begin{tabular}{c}

\includegraphics[width=.25\textwidth, height=.17\textwidth]{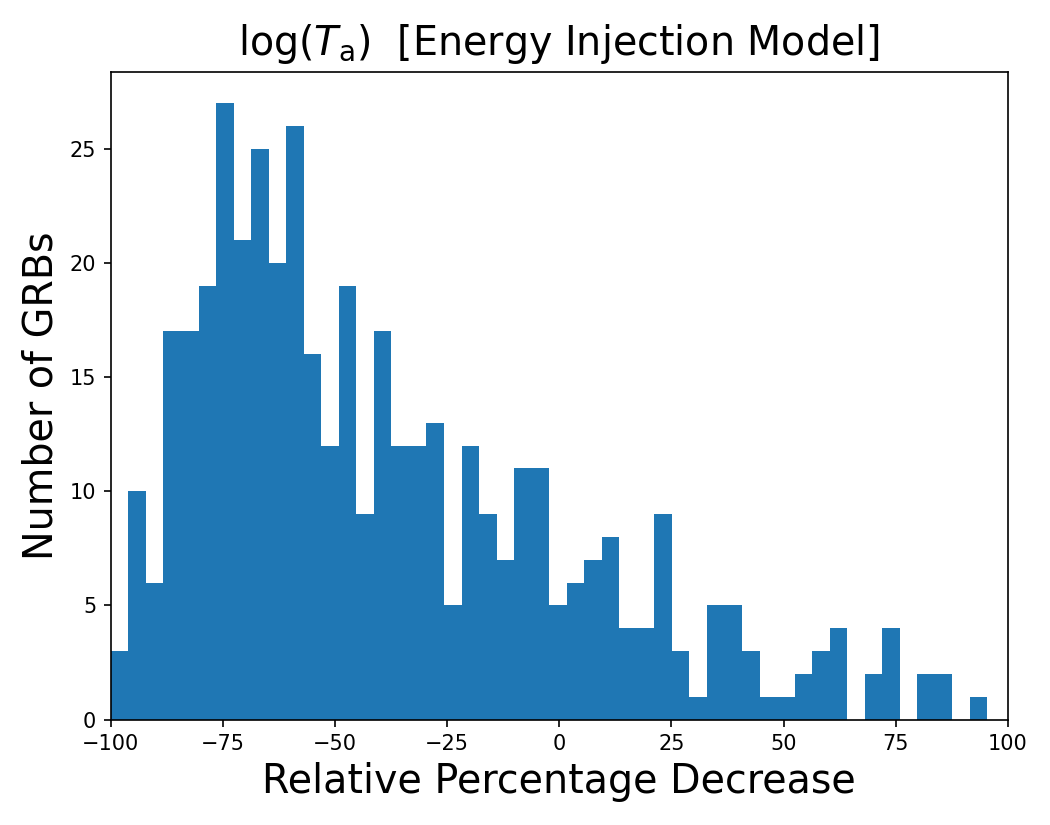}
\includegraphics[width=.25\textwidth, height=.17\textwidth]{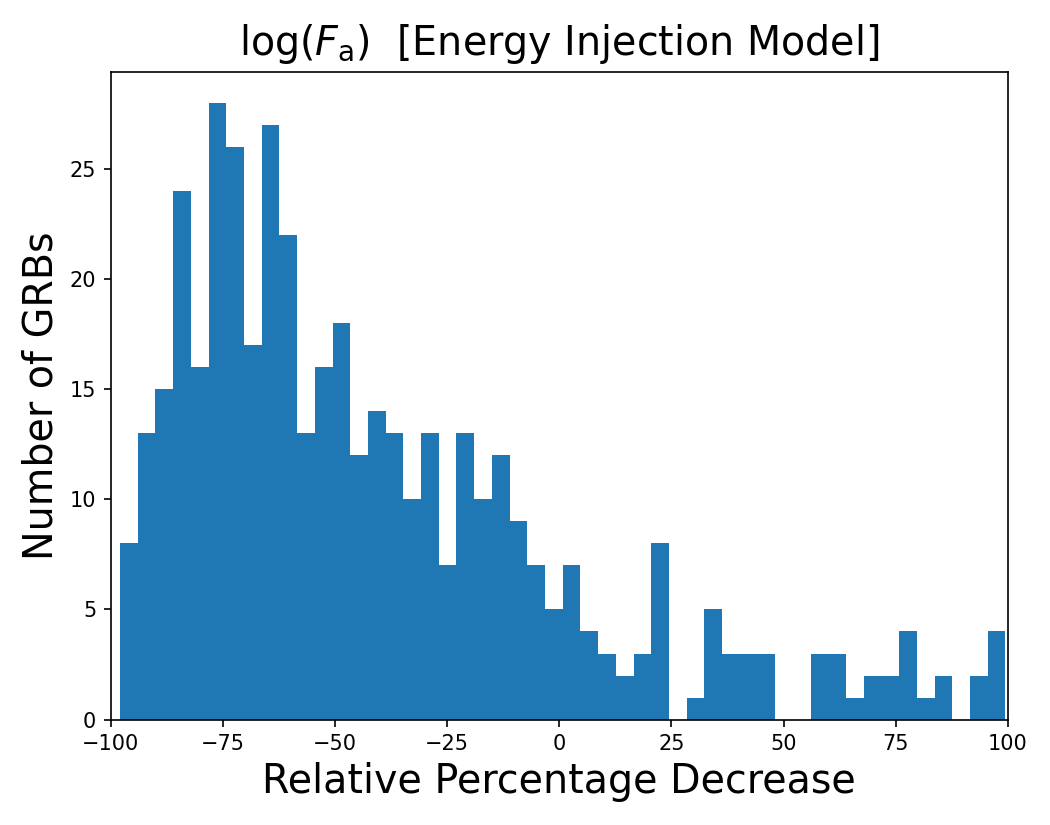}
\includegraphics[width=.25\textwidth, height=.17\textwidth]{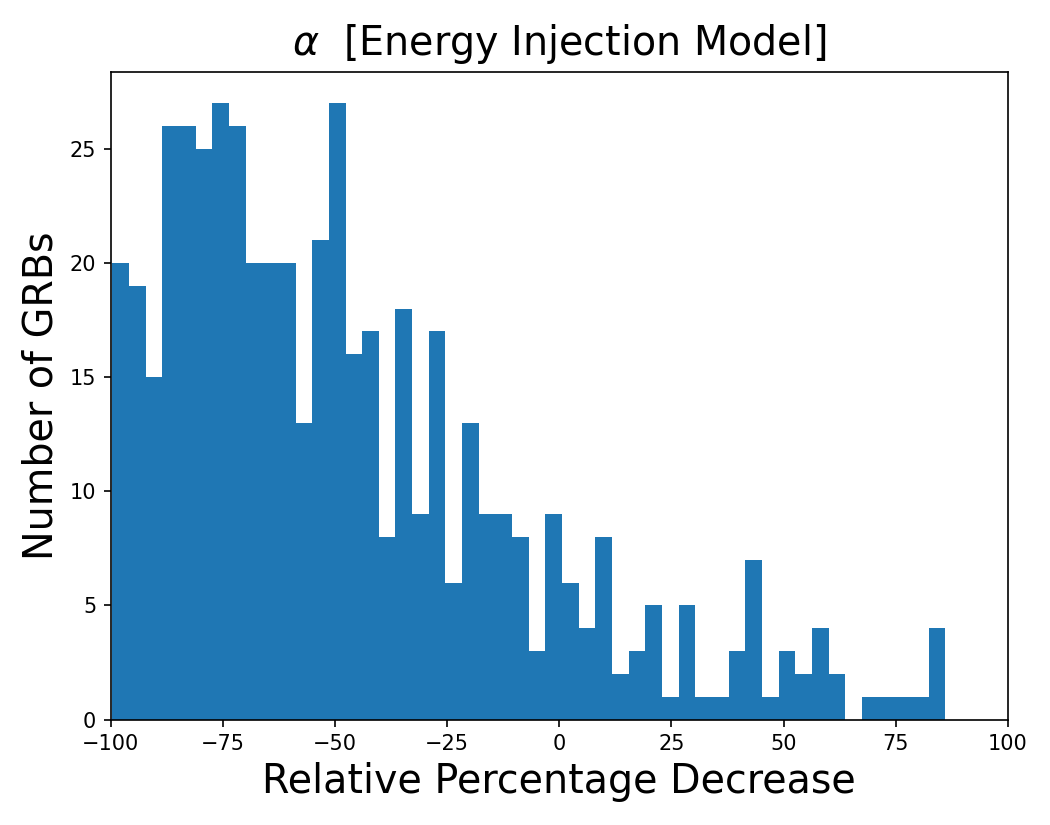} \\

\includegraphics[width=.25\textwidth, height=.17\textwidth]{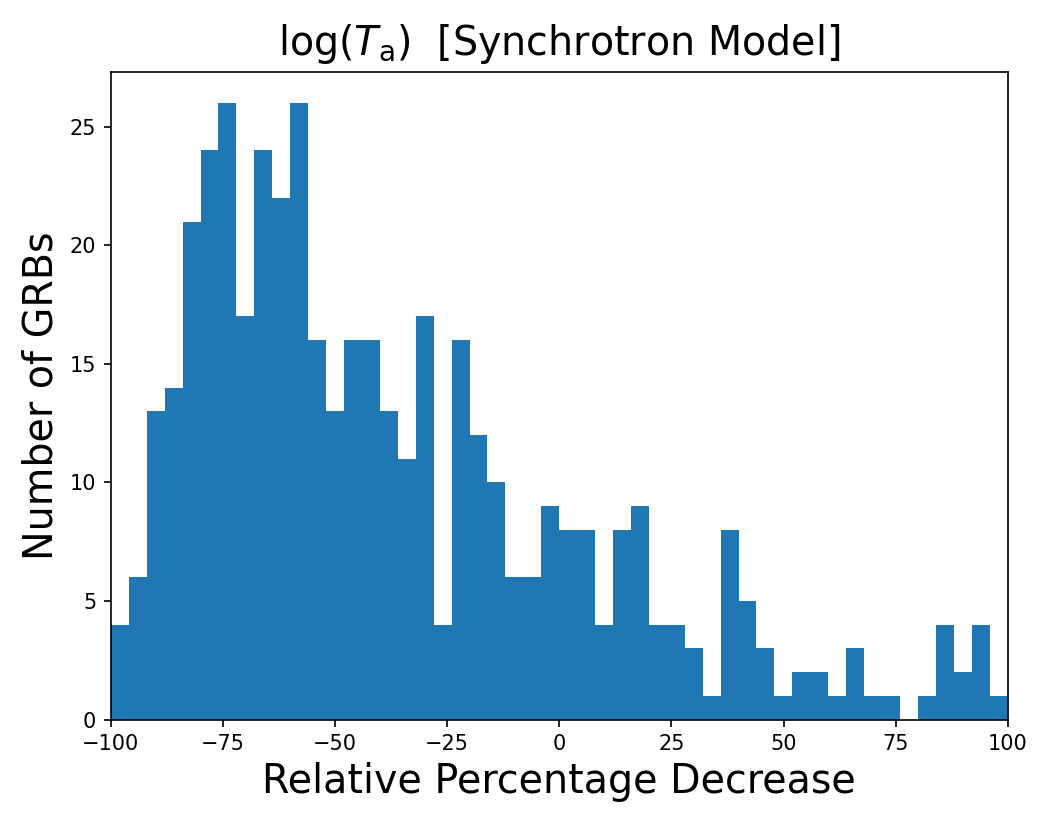}
\includegraphics[width=.25\textwidth, height=.17\textwidth]{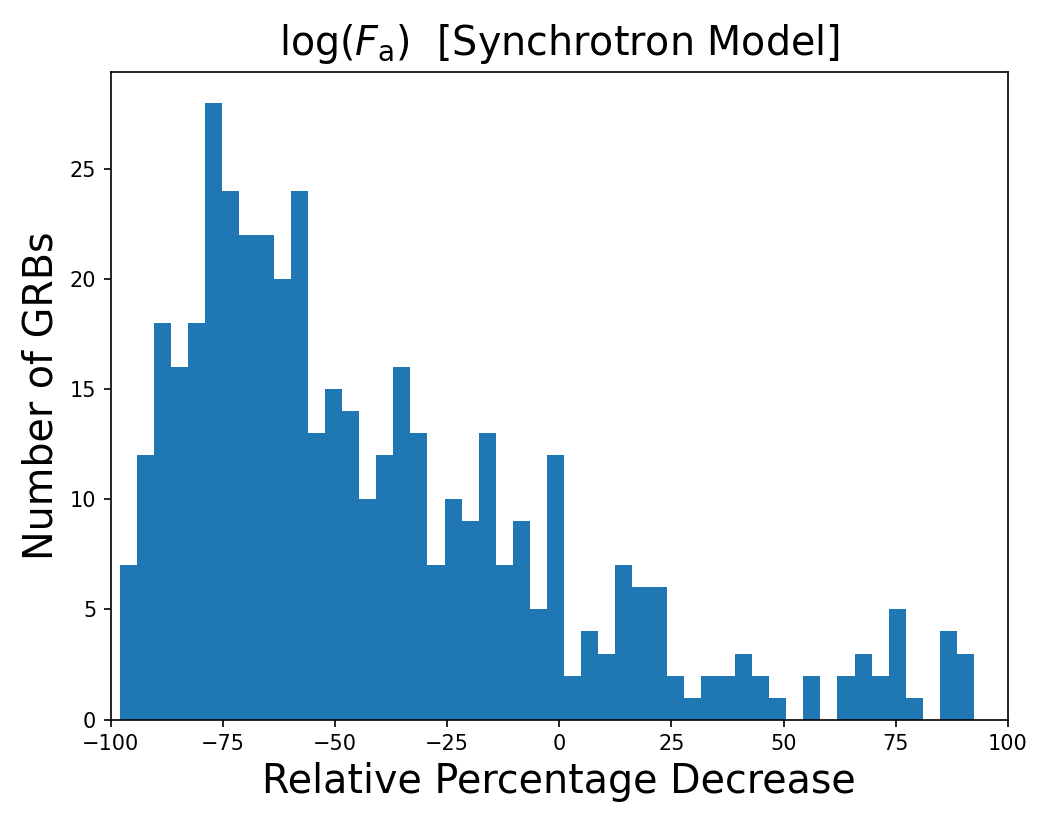}
\includegraphics[width=.25\textwidth, height=.17\textwidth]{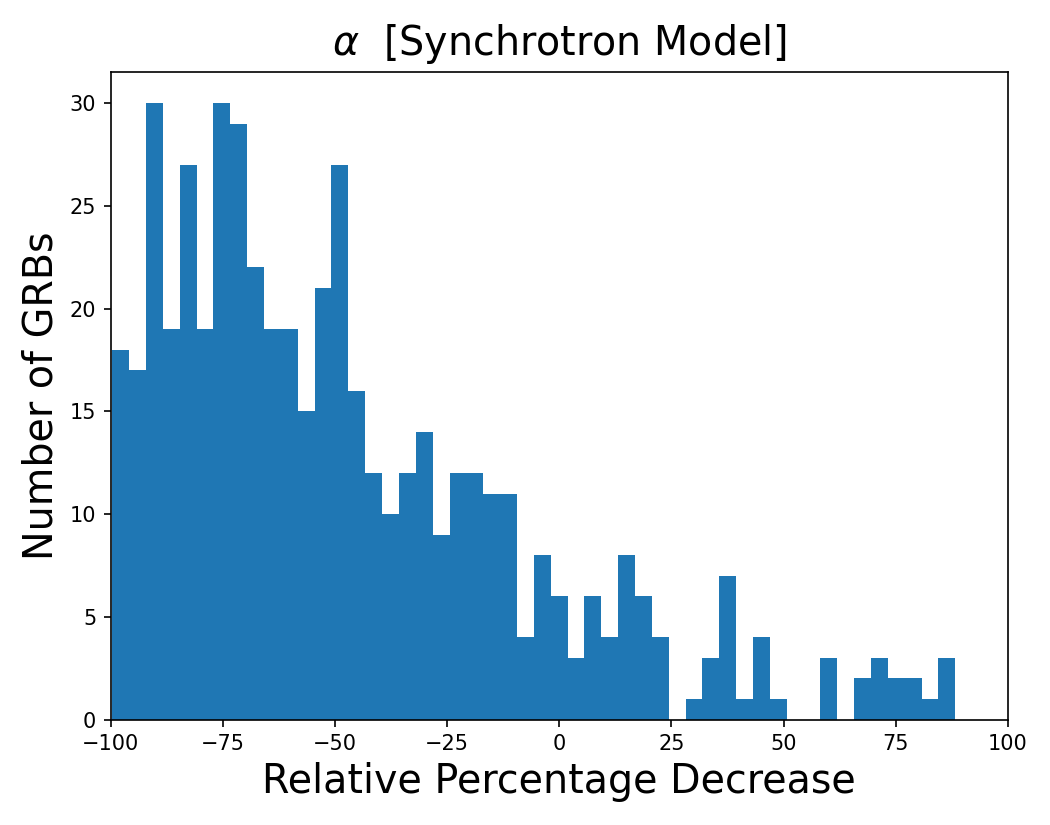} \\

\includegraphics[width=.25\textwidth, height=.17\textwidth]{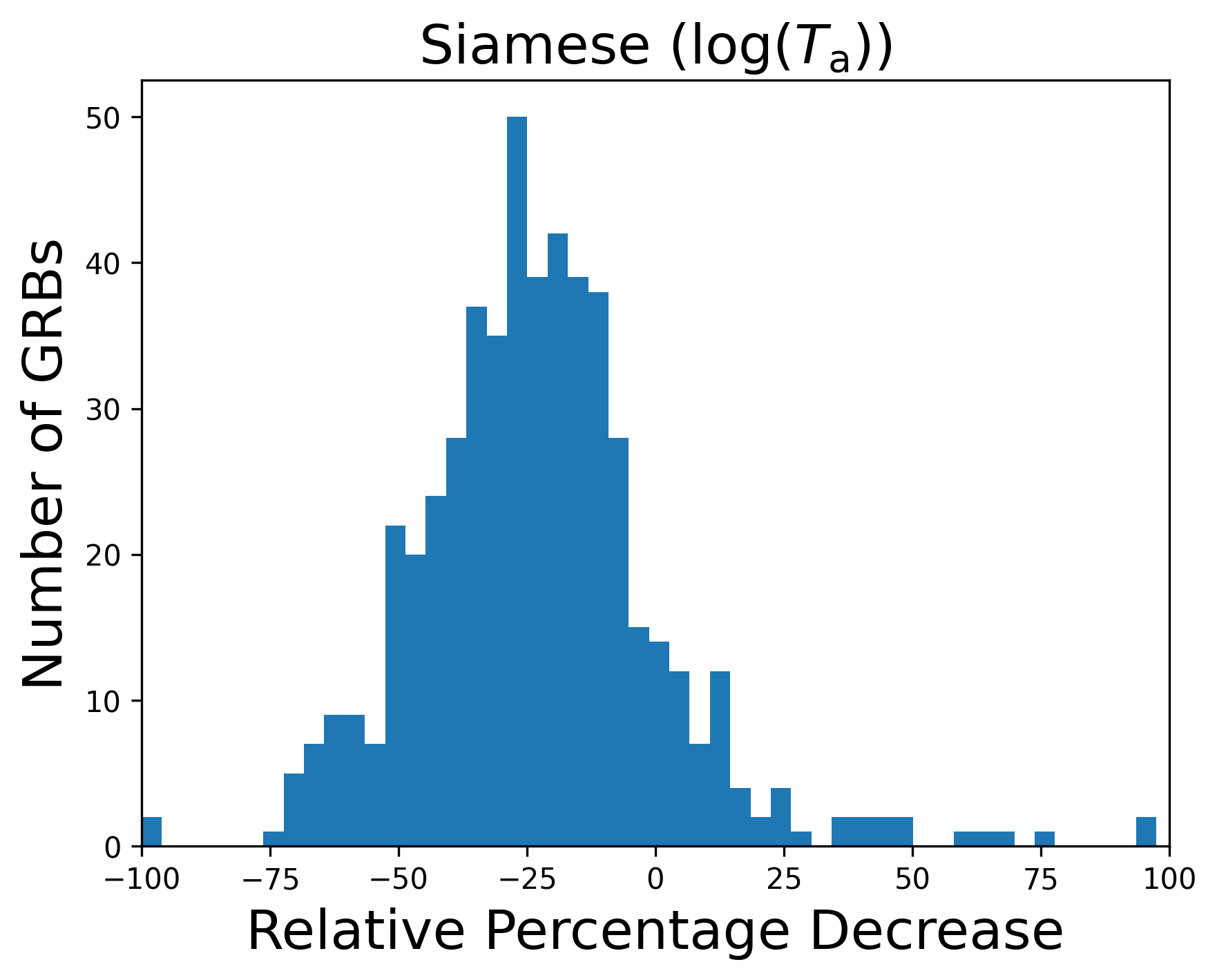}
\includegraphics[width=.25\textwidth, height=.17\textwidth]{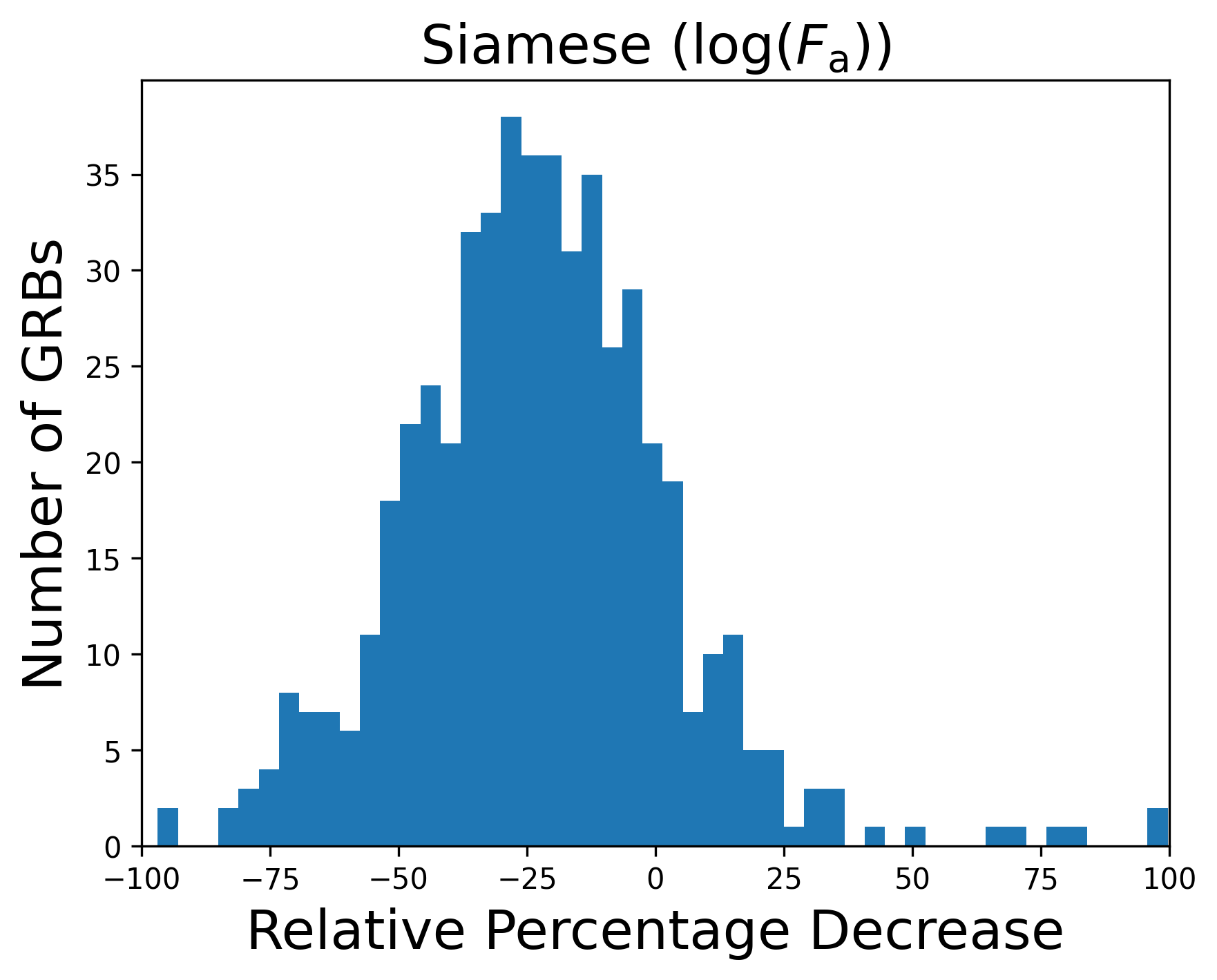}
\includegraphics[width=.25\textwidth, height=.17\textwidth]{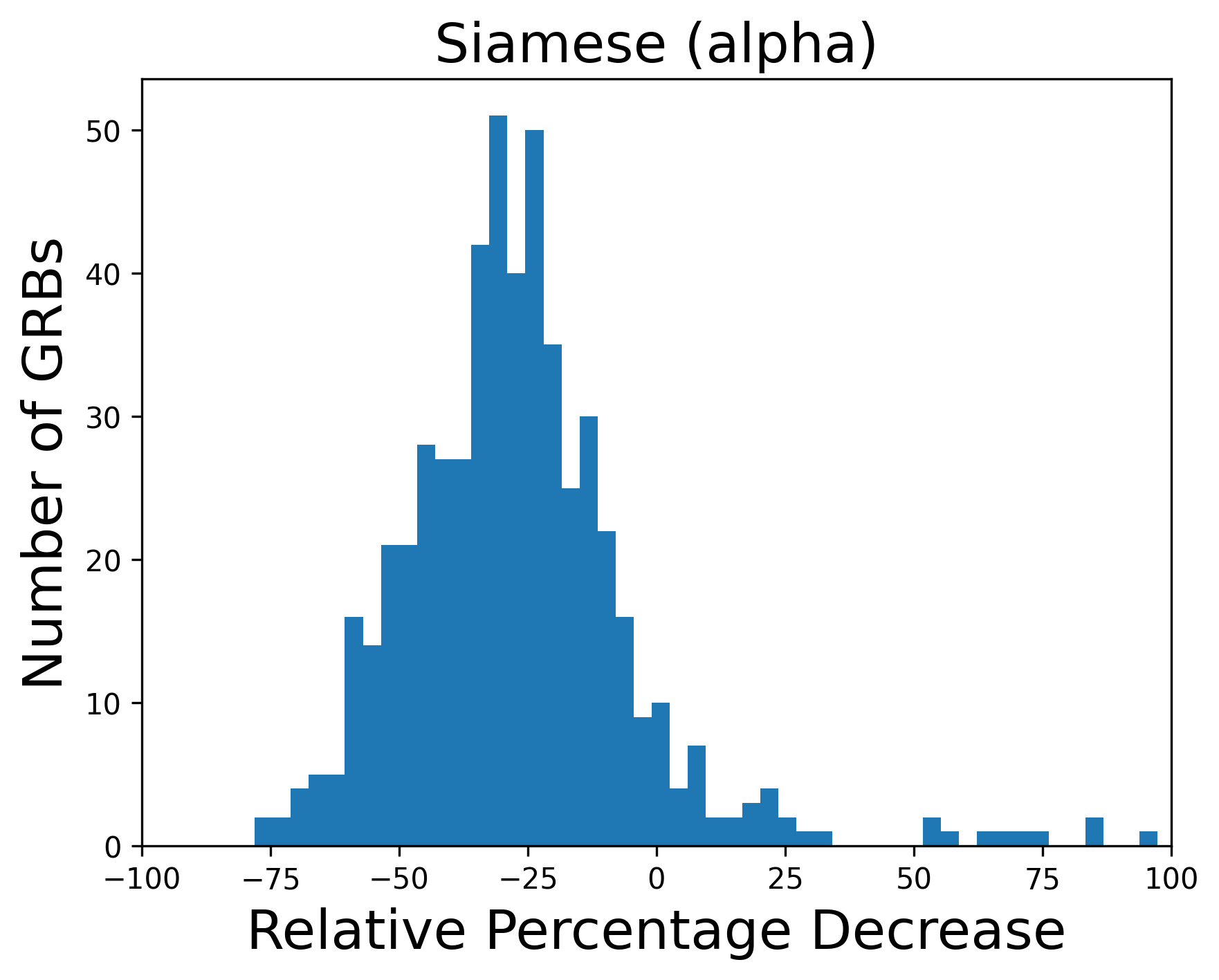} \\

\includegraphics[width=.25\textwidth, height=.17\textwidth]{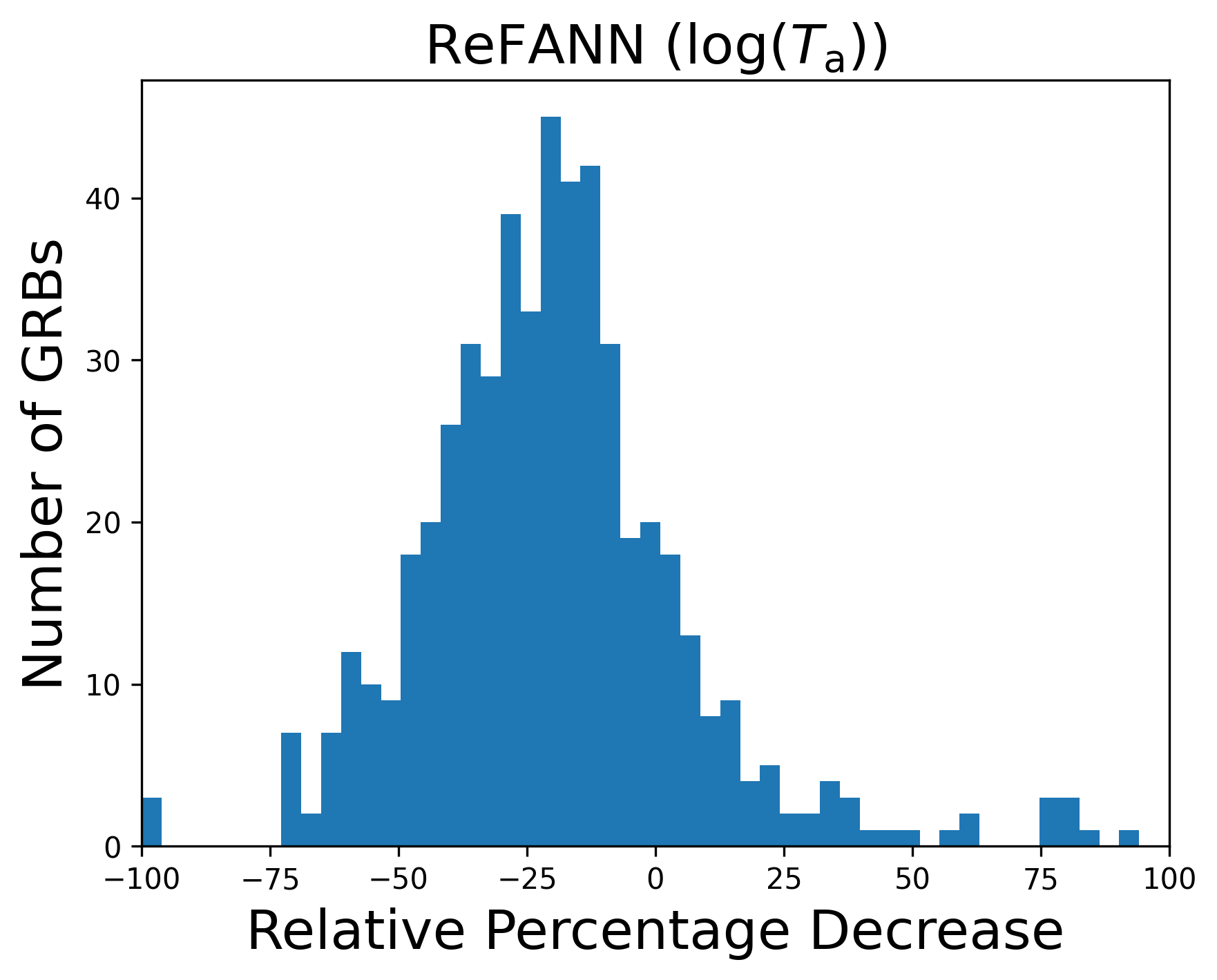}
\includegraphics[width=.25\textwidth, height=.17\textwidth]{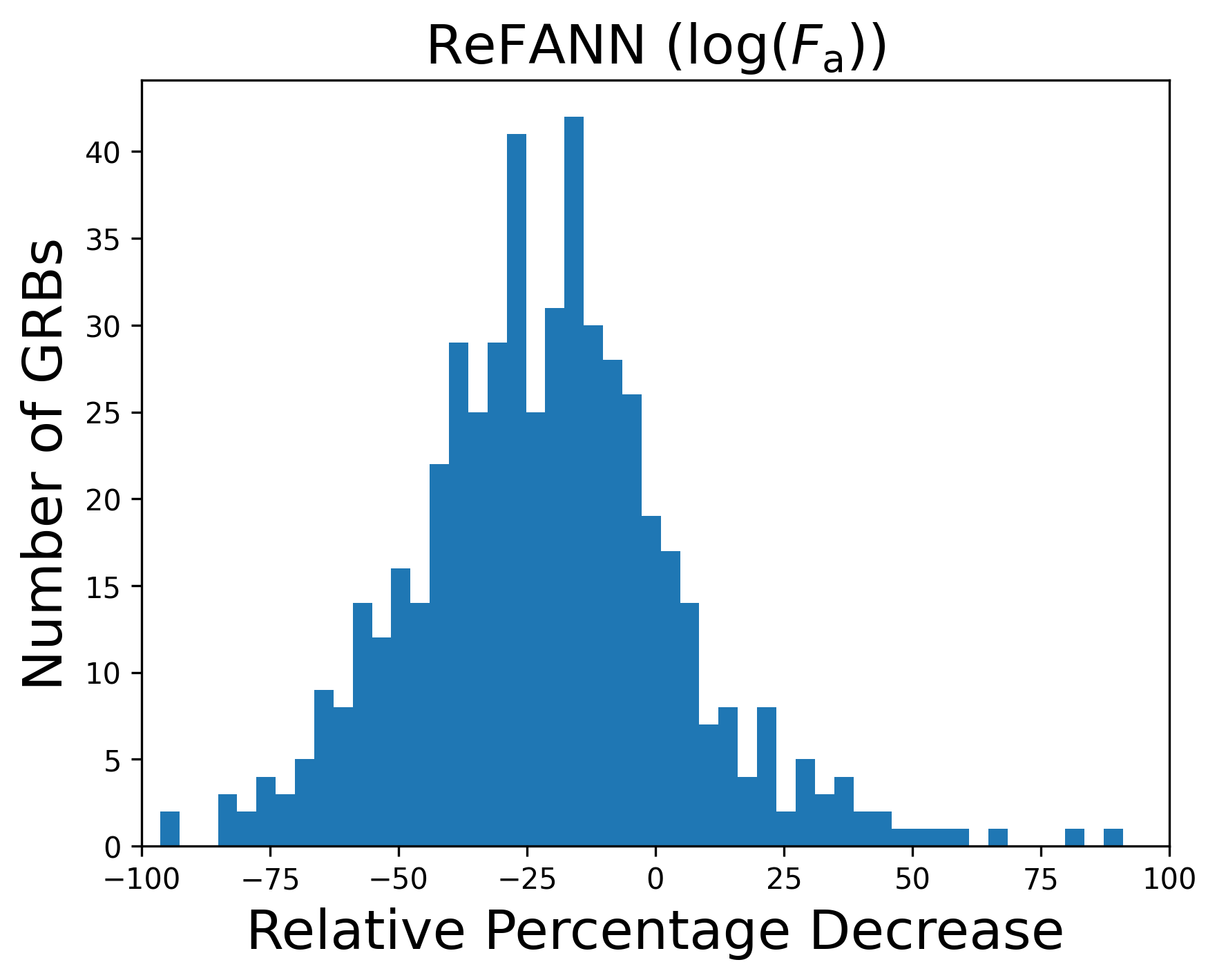}
\includegraphics[width=.25\textwidth, height=.17\textwidth]{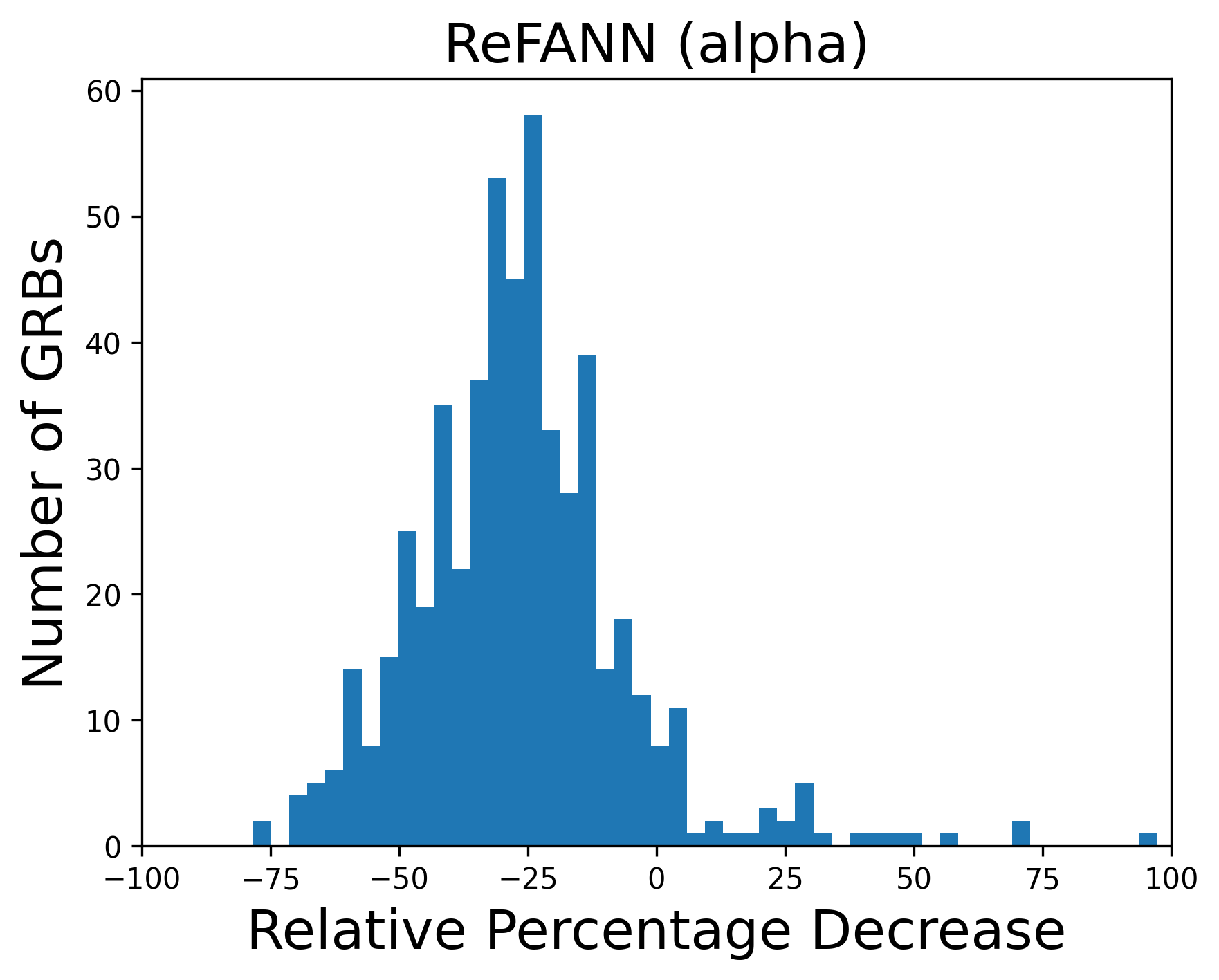} \\

\includegraphics[width=.25\textwidth, height=.17\textwidth]{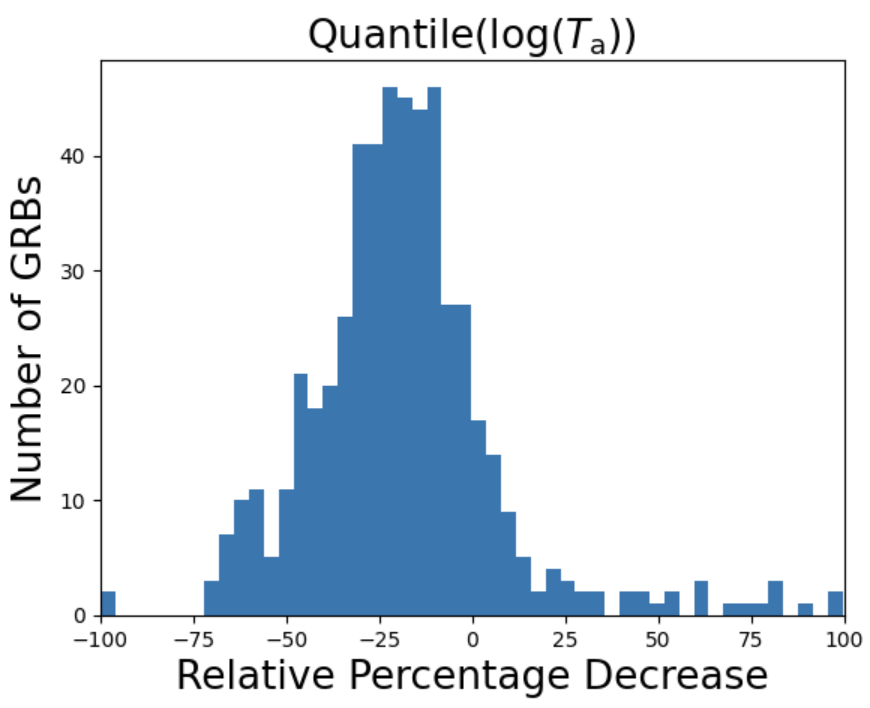}
\includegraphics[width=.25\textwidth, height=.17\textwidth]{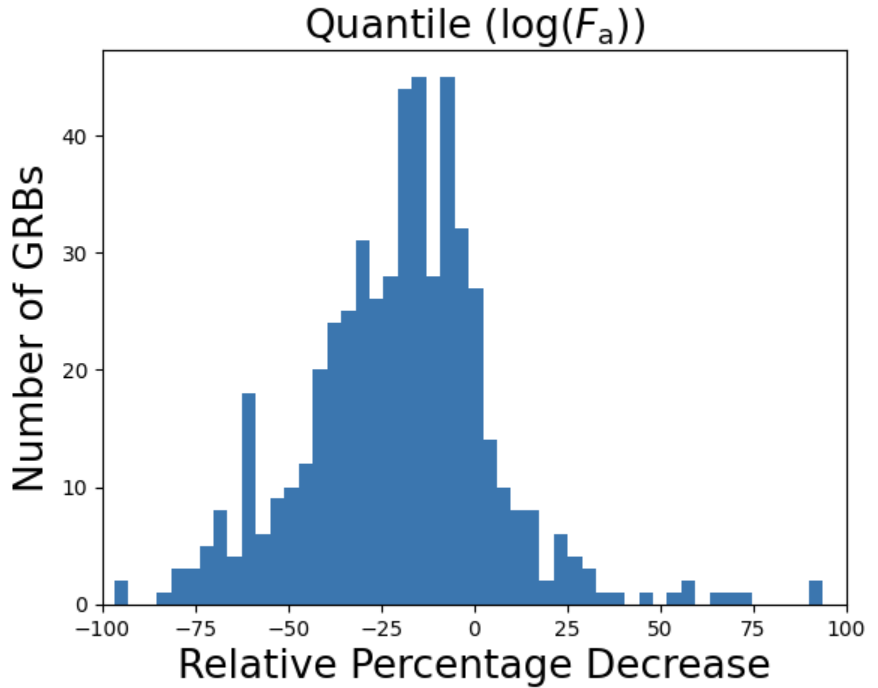}
\includegraphics[width=.25\textwidth, height=.17\textwidth]{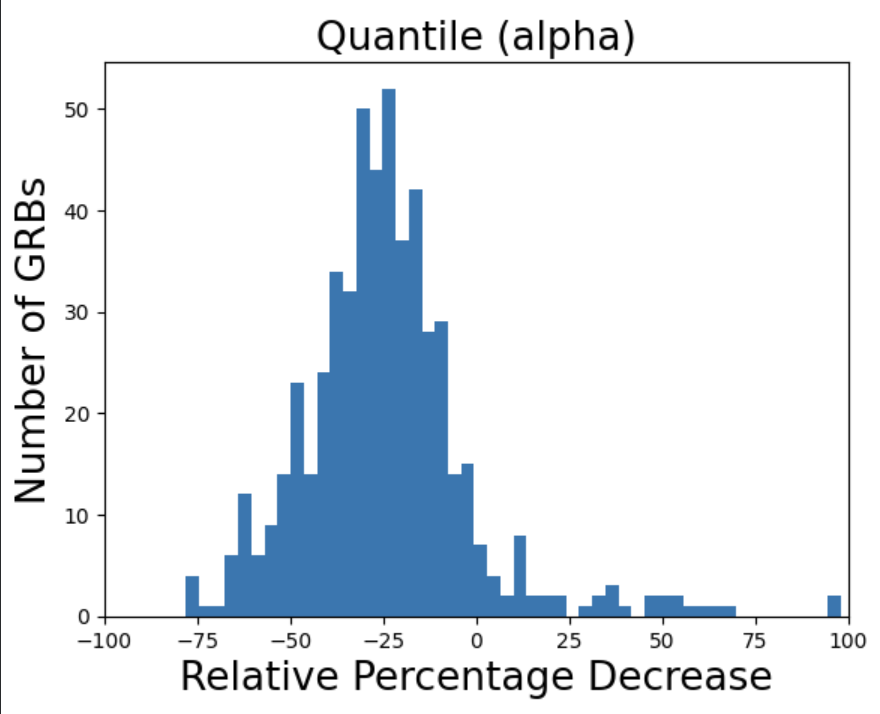} \\

\end{tabular}
\end{adjustbox}
\caption{Distribution plots of the three W07 parameters are arranged in a grid, with the parameters shown from left to right: i) $\log T\_a$ (Column 1); ii) $\log F\_a$ (Column 2); and iii) $\alpha$ (Column 3). The models are arranged from top to bottom: i) PINN (Row 1); ii) ReFANN (Row 2); iii) Siamese Model (Row 3); and iv) Quantile (Row 4). \label{fig:percentage-dec}}
\end{figure*}

The reconstruction results for each GRB category and all models are presented in Fig. \ref{fig:Reconstrusted_grbs}. For ease of comparison with the previous work, the same morphological classes of GRBs are shown. 
The rows differ only in the reconstruction method used. The histogram distributions of the relative percentage decrease for the three parameters are shown in Fig. \ref{fig:percentage-dec}.

\begin{figure*}[b!]
\centering
    \includegraphics[width=.24\textwidth, height=.20\textwidth]{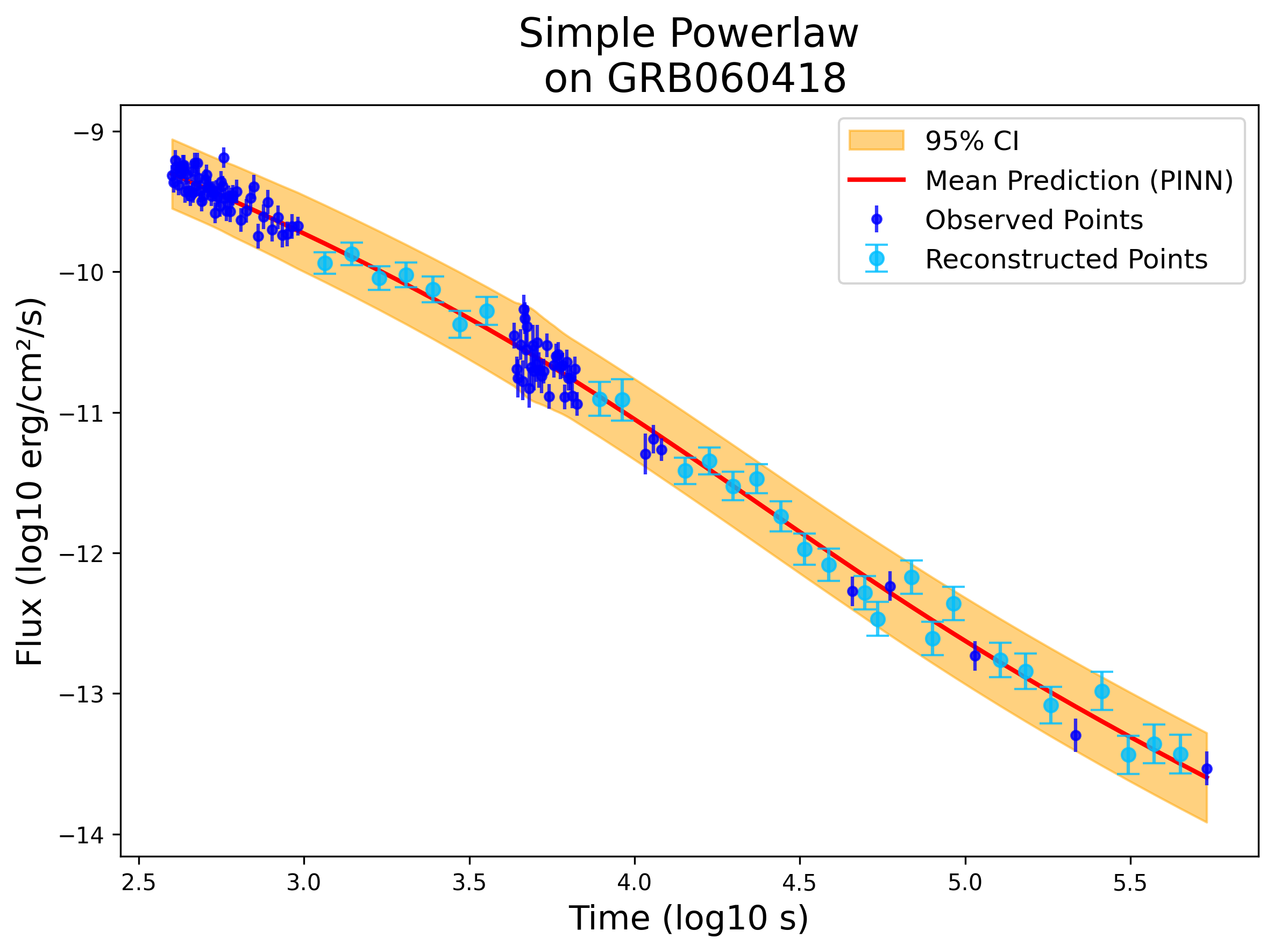}
    \includegraphics[width=.24\textwidth, height=.20\textwidth]{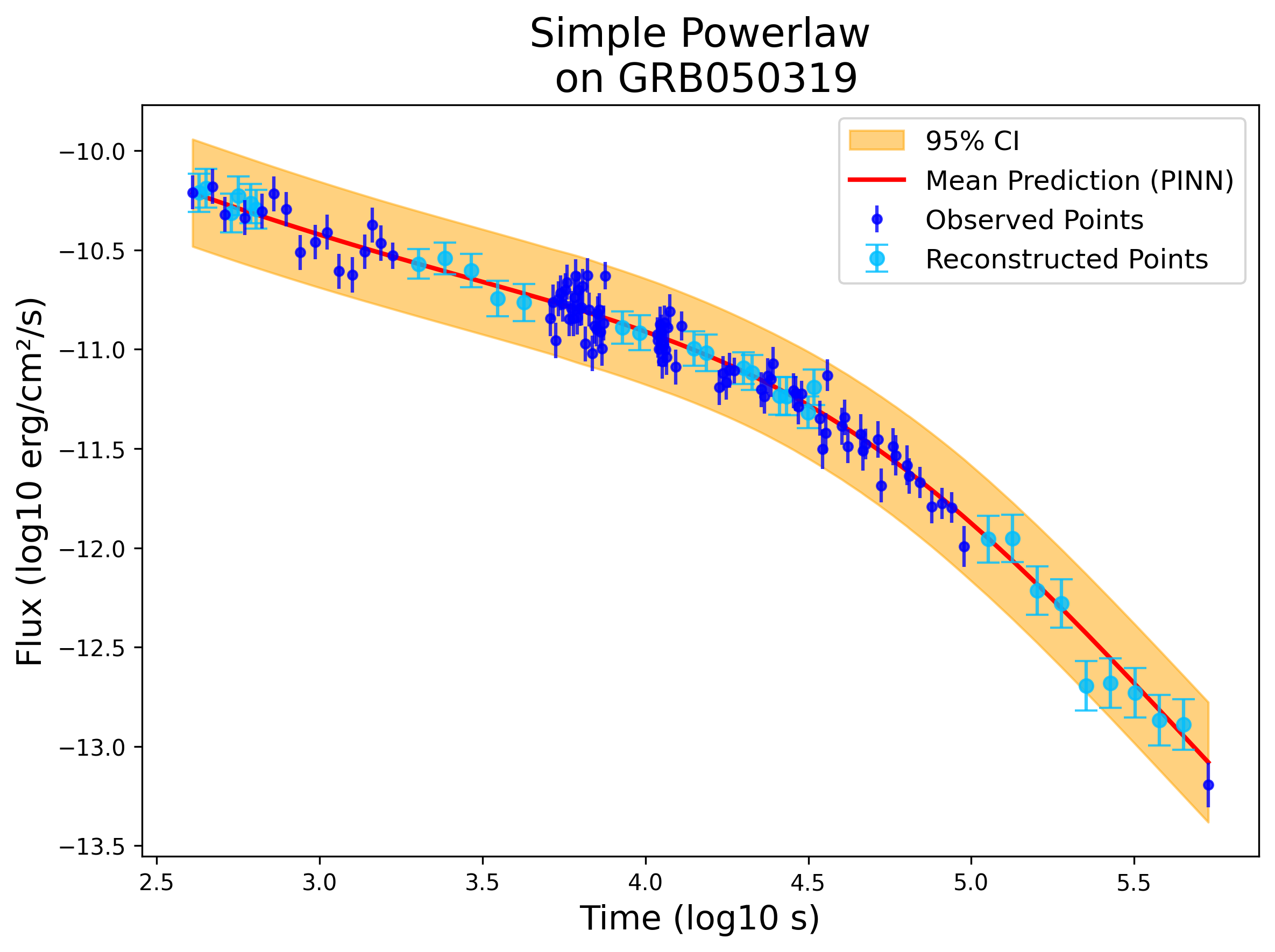}
    \includegraphics[width=.24\textwidth, height=.20\textwidth]{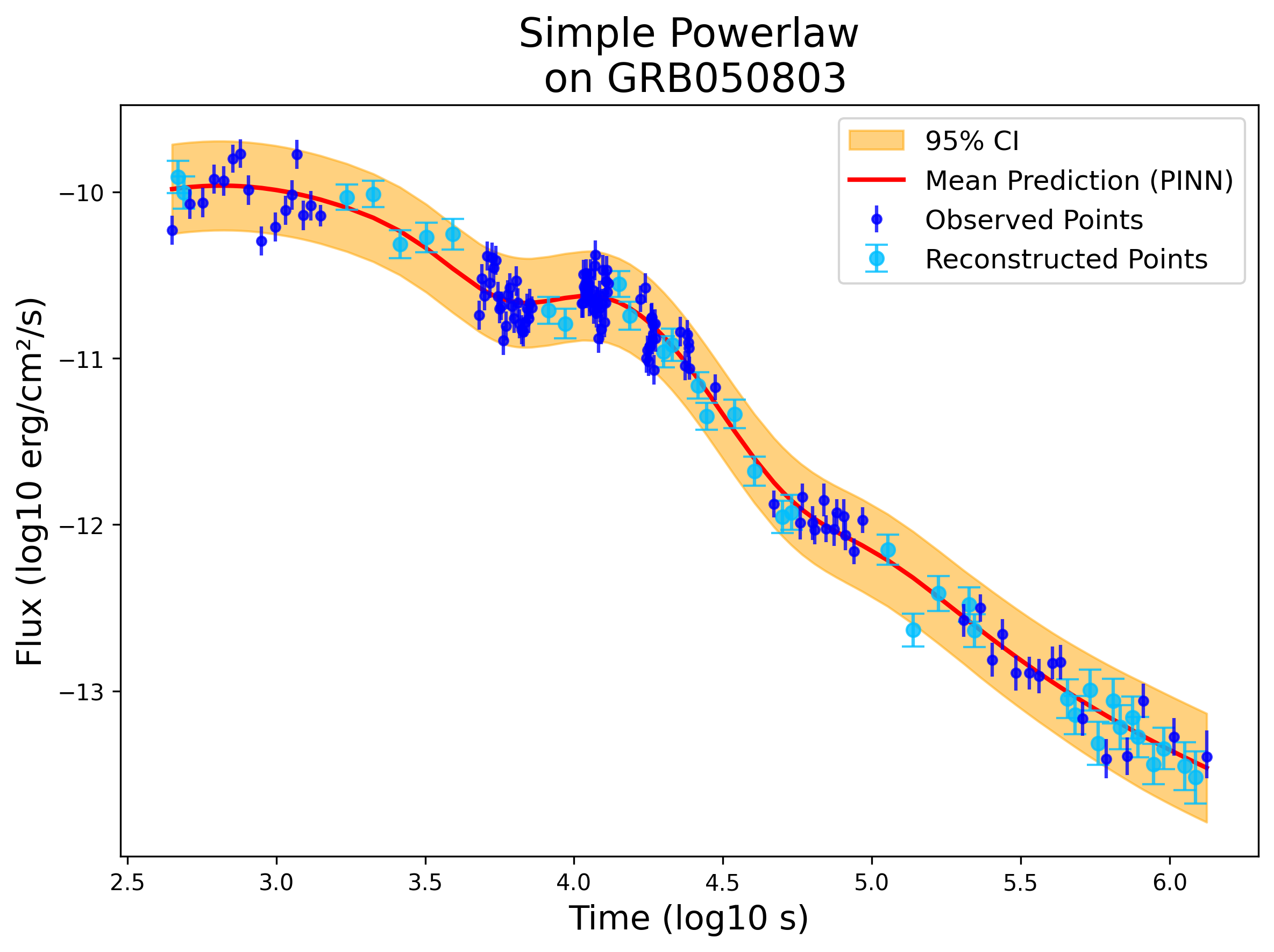}
    \includegraphics[width=.24\textwidth, height=.20\textwidth]{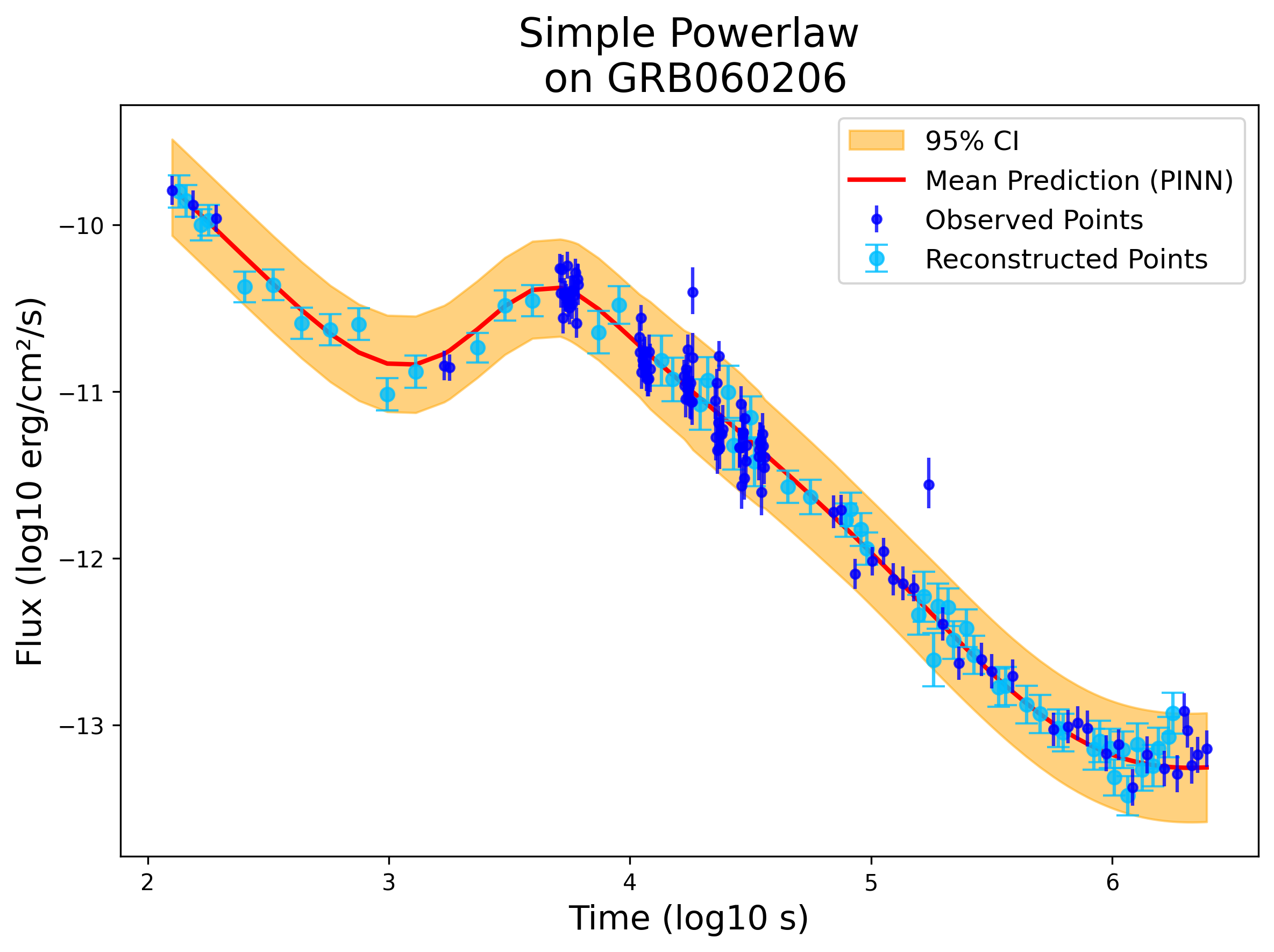}  

    \includegraphics[width=.24\textwidth, height=.20\textwidth]{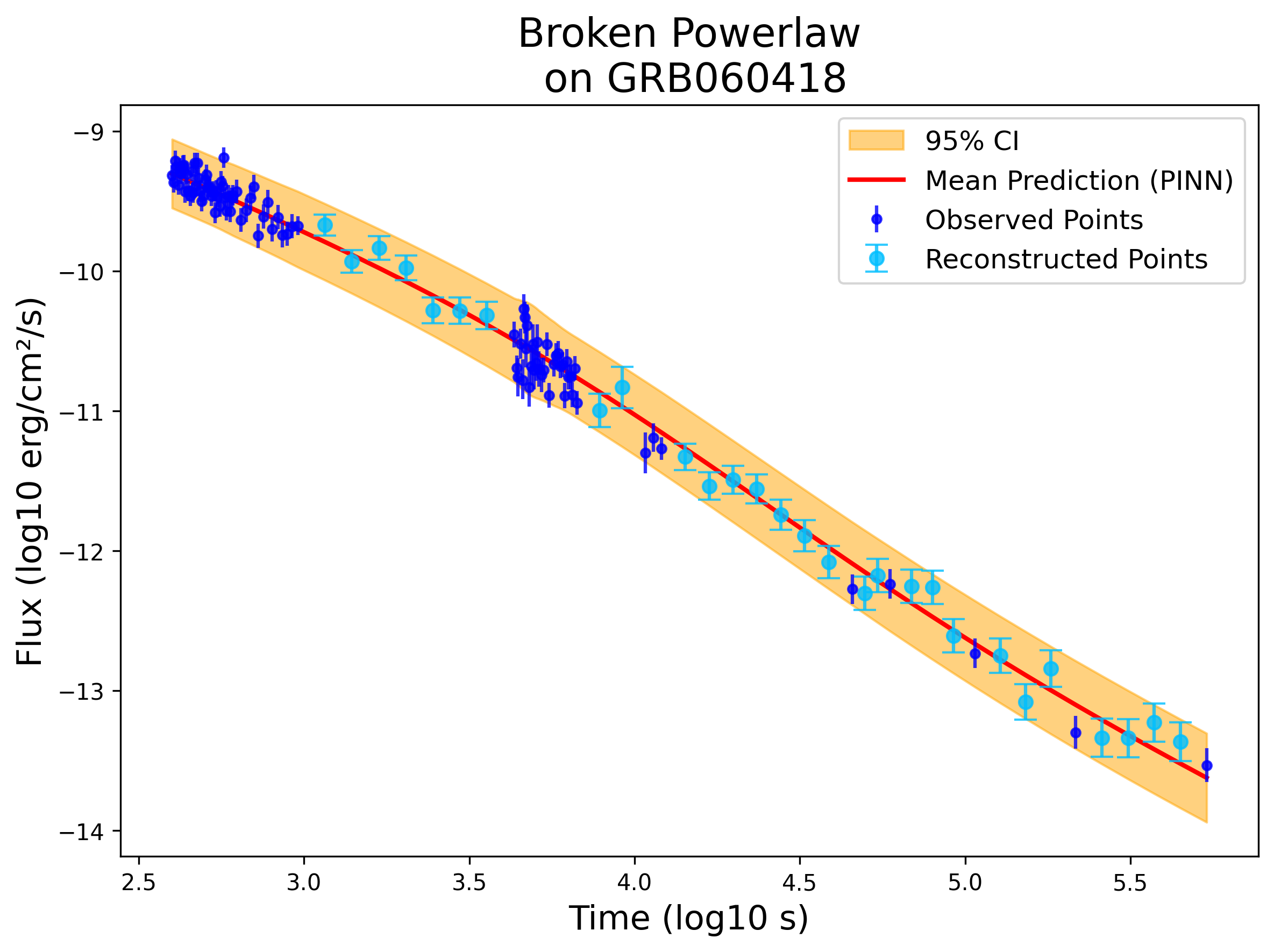}
    \includegraphics[width=.24\textwidth, height=.20\textwidth]{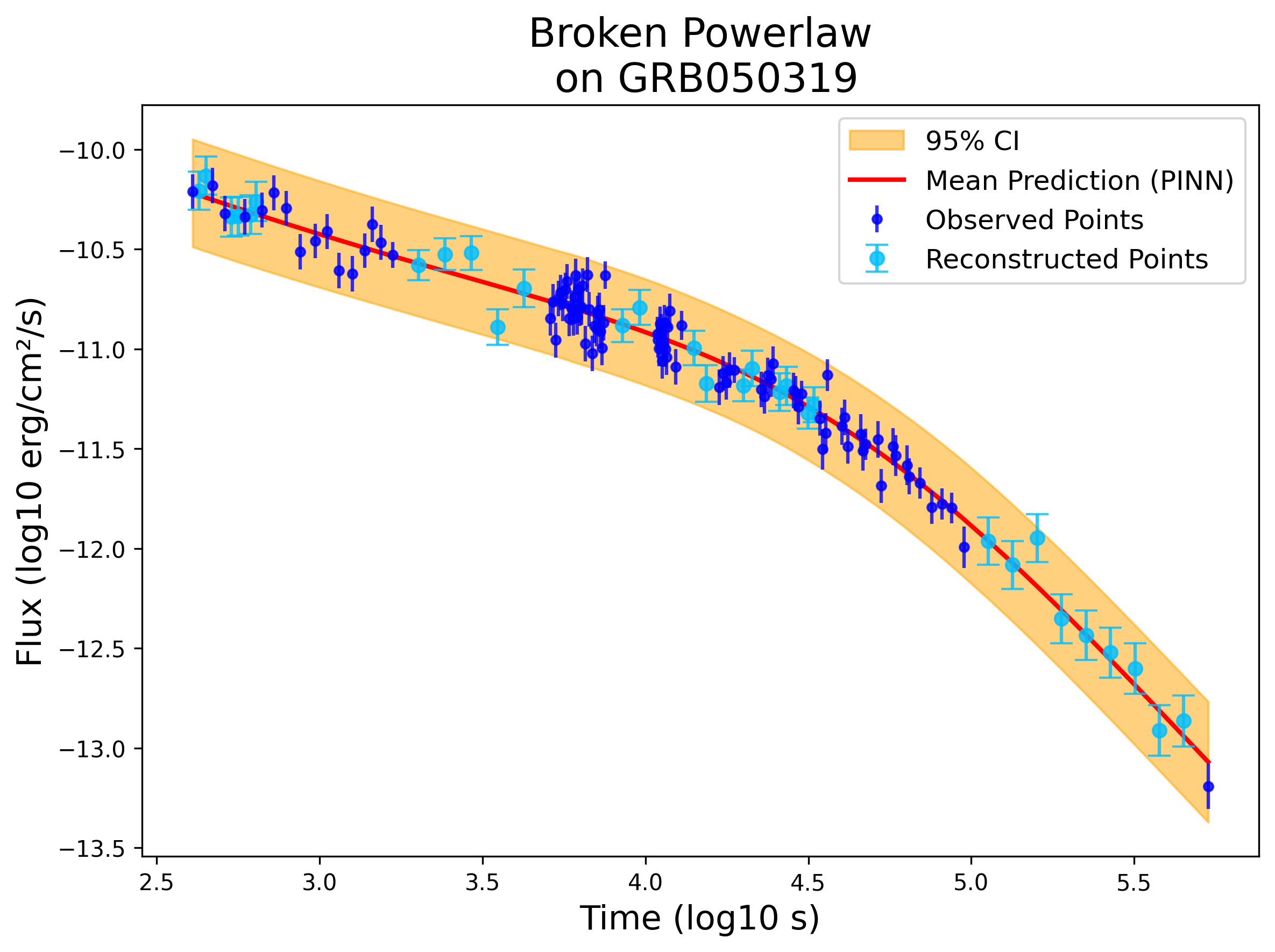}
    \includegraphics[width=.24\textwidth, height=.20\textwidth]{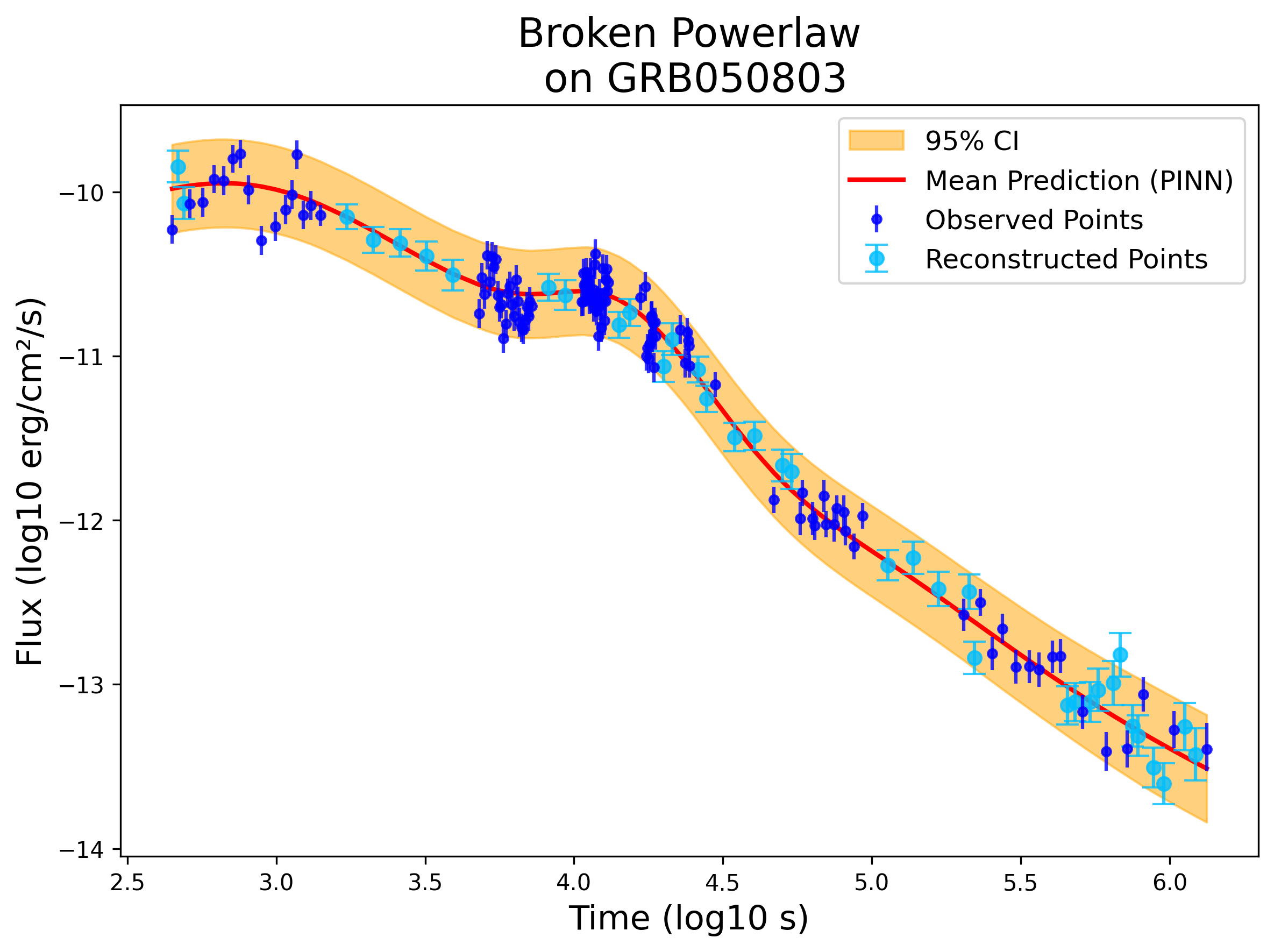}
    \includegraphics[width=.24\textwidth, height=.20\textwidth]{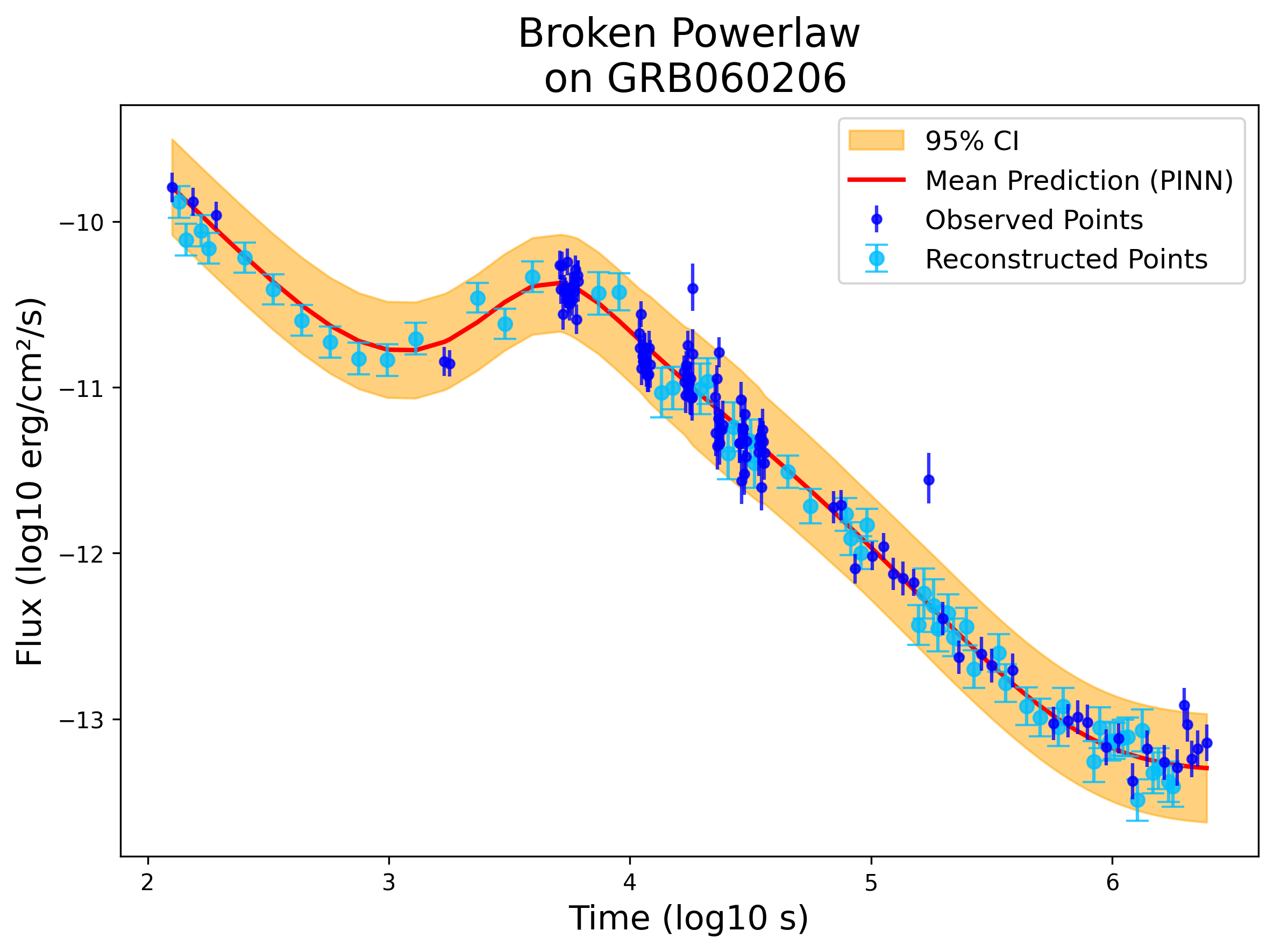}  

    \includegraphics[width=.24\textwidth, height=.20\textwidth]{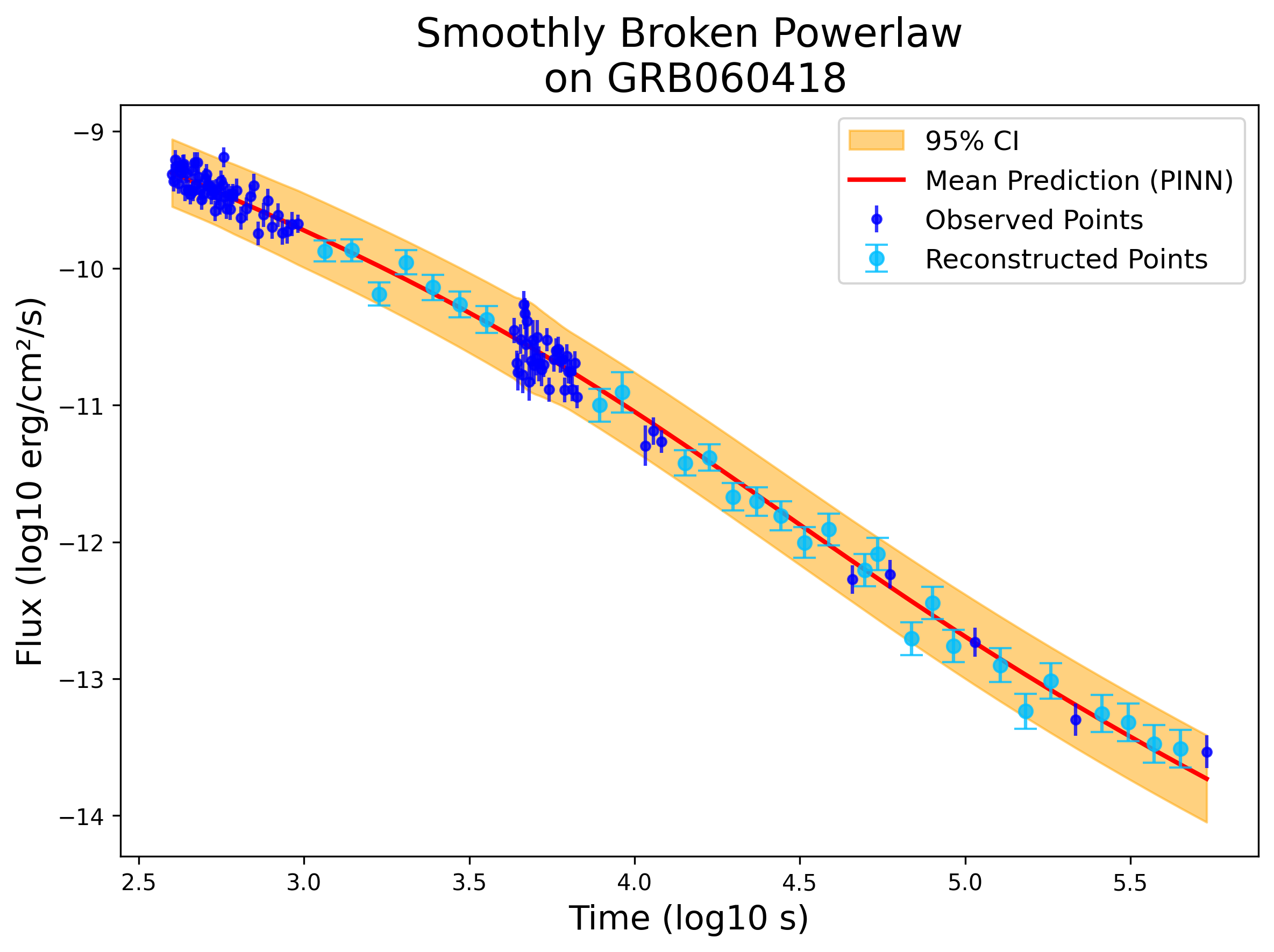}
    \includegraphics[width=.24\textwidth, height=.20\textwidth]{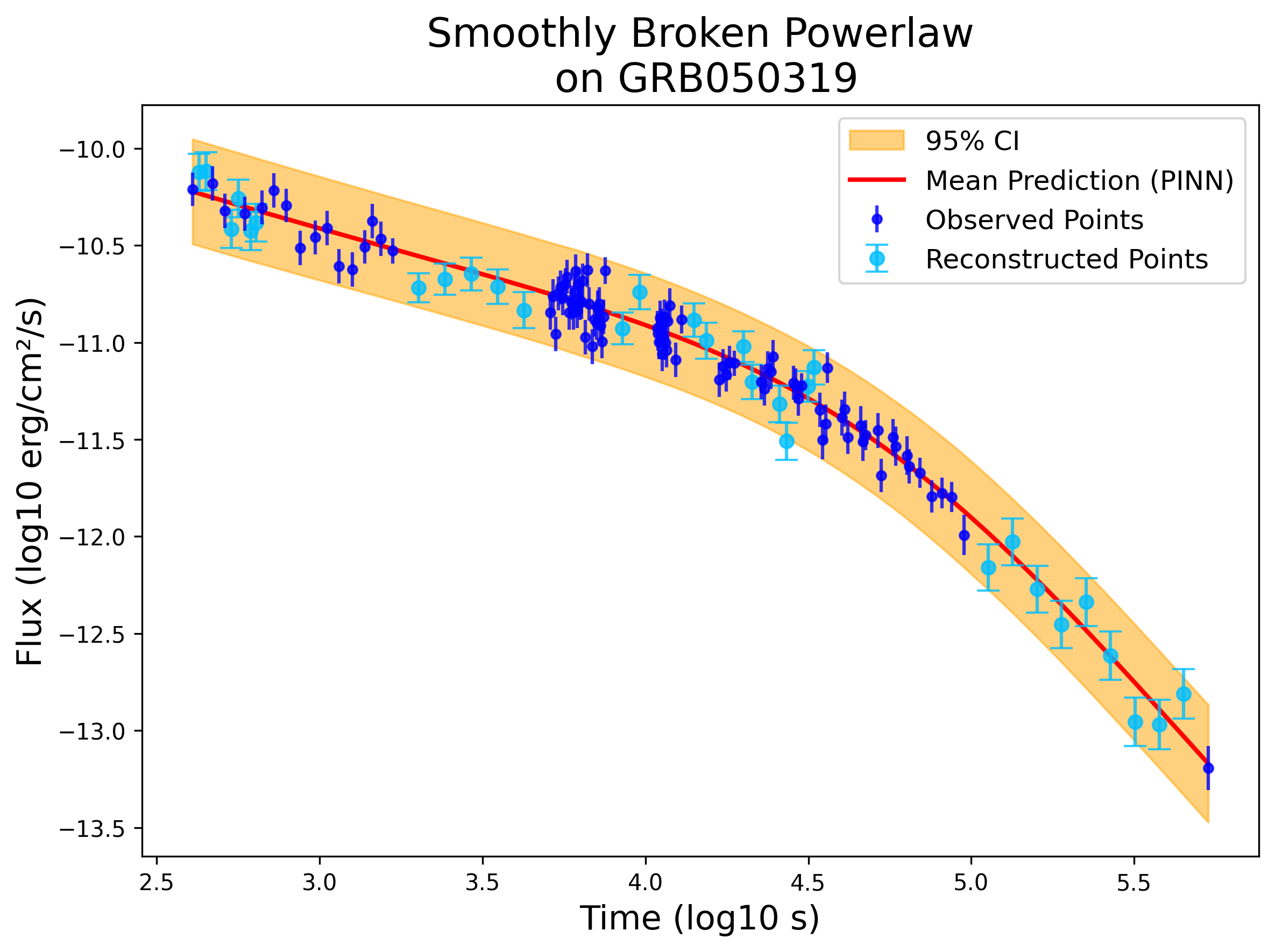}
    \includegraphics[width=.24\textwidth, height=.20\textwidth]{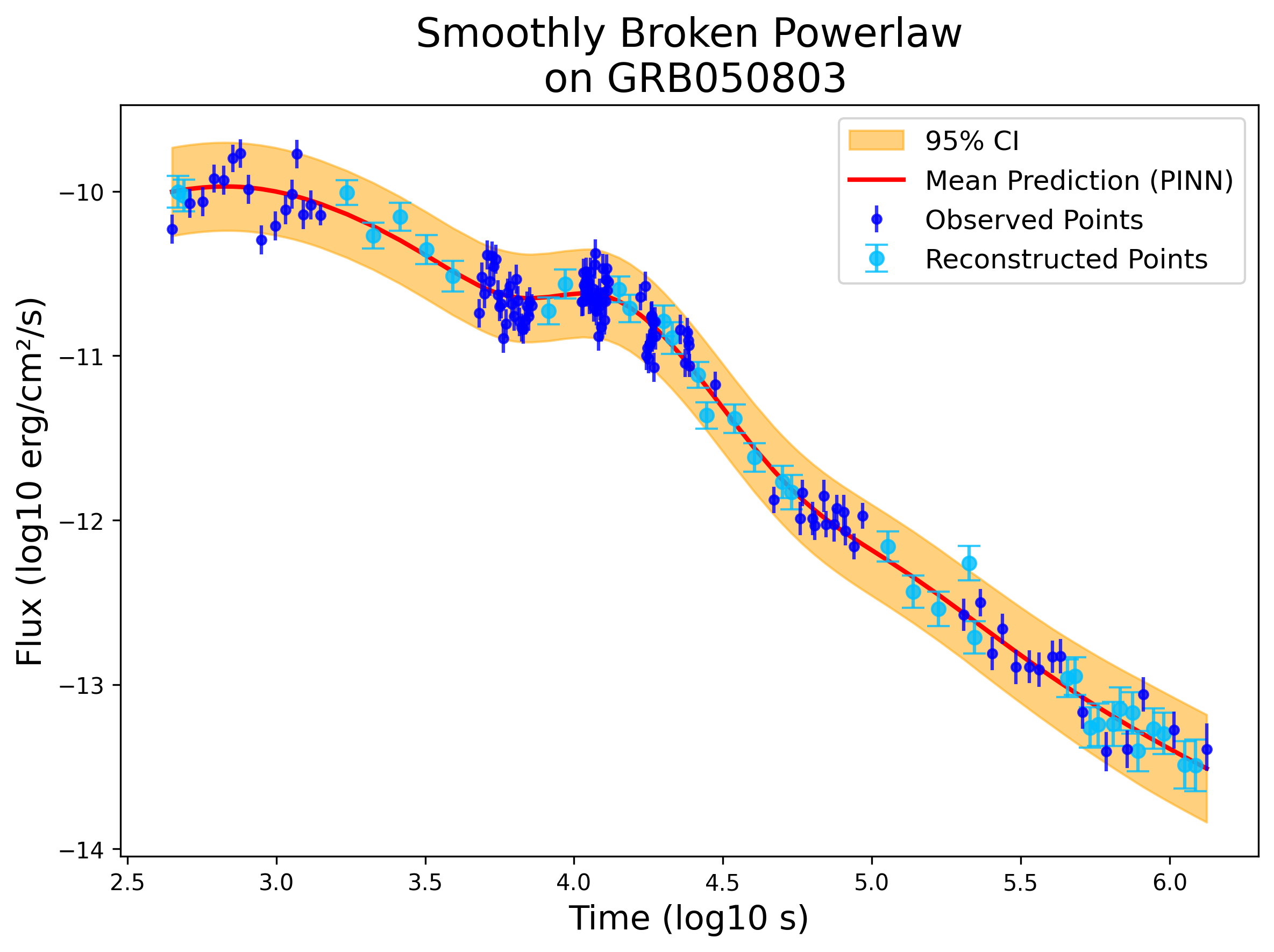}
    \includegraphics[width=.24\textwidth, height=.20\textwidth]{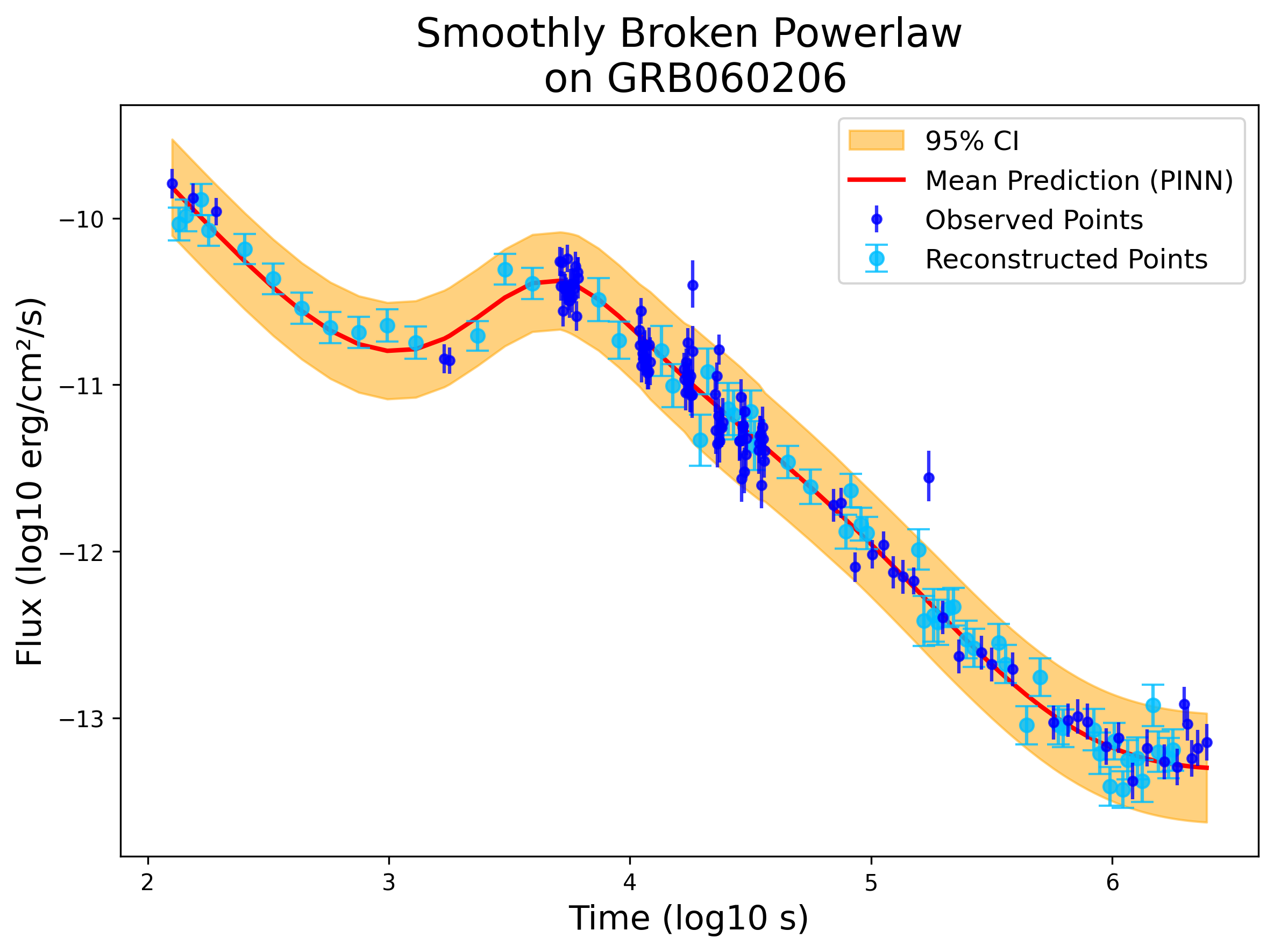}  

    \includegraphics[width=.24\textwidth, height=.20\textwidth]{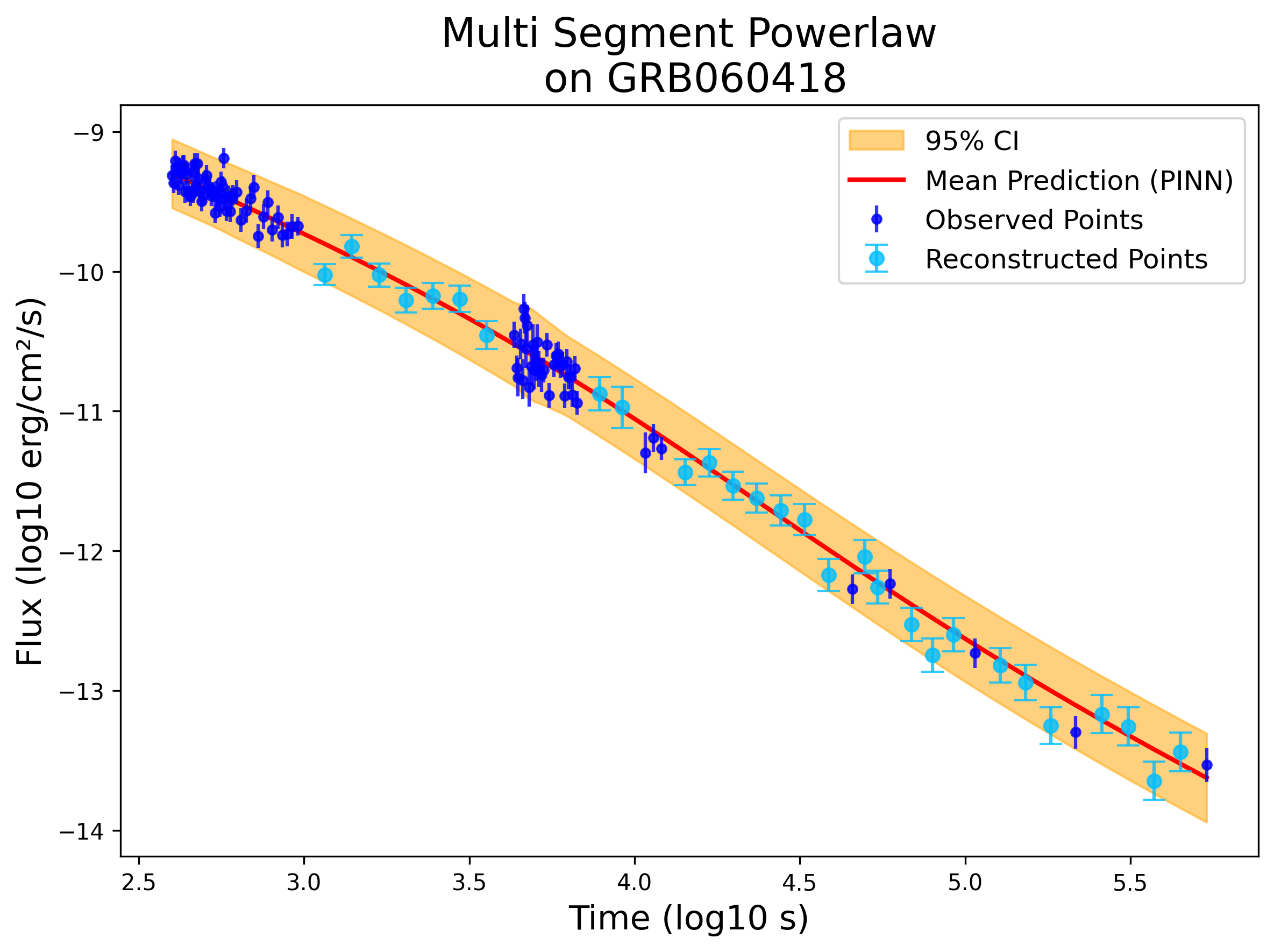}
    \includegraphics[width=.24\textwidth, height=.20\textwidth]{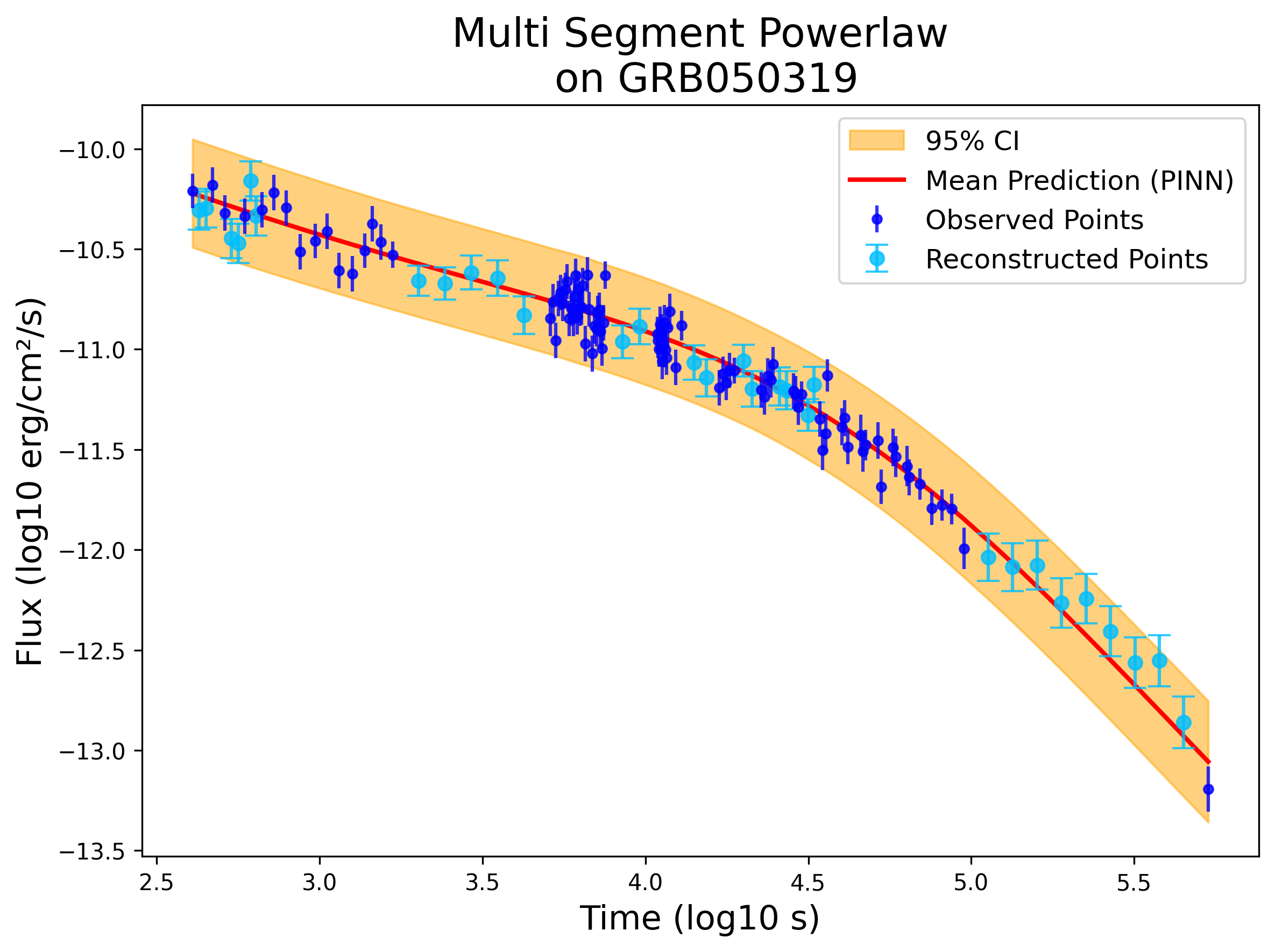}
    \includegraphics[width=.24\textwidth, height=.20\textwidth]{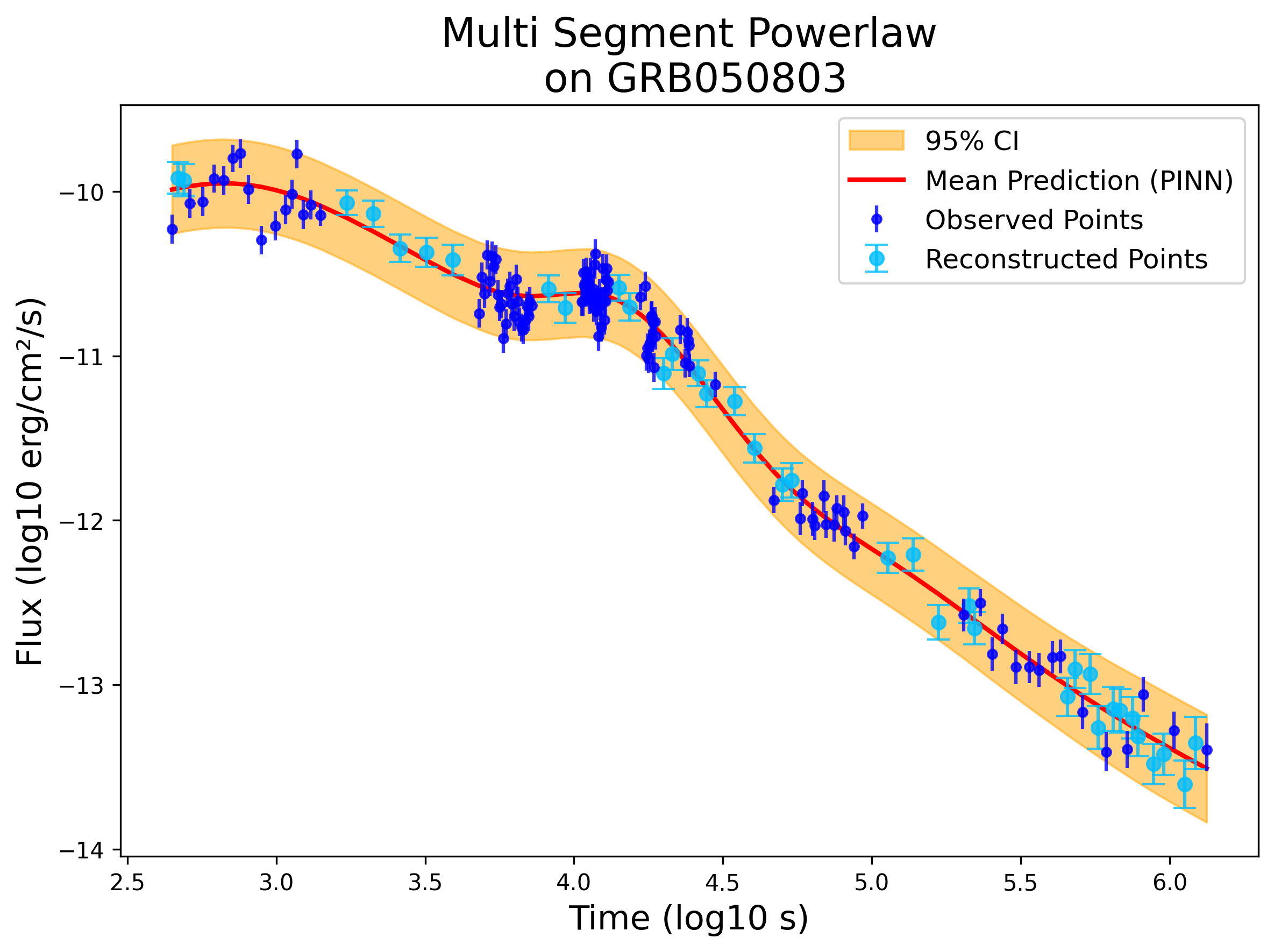}
    \includegraphics[width=.24\textwidth, height=.20\textwidth]{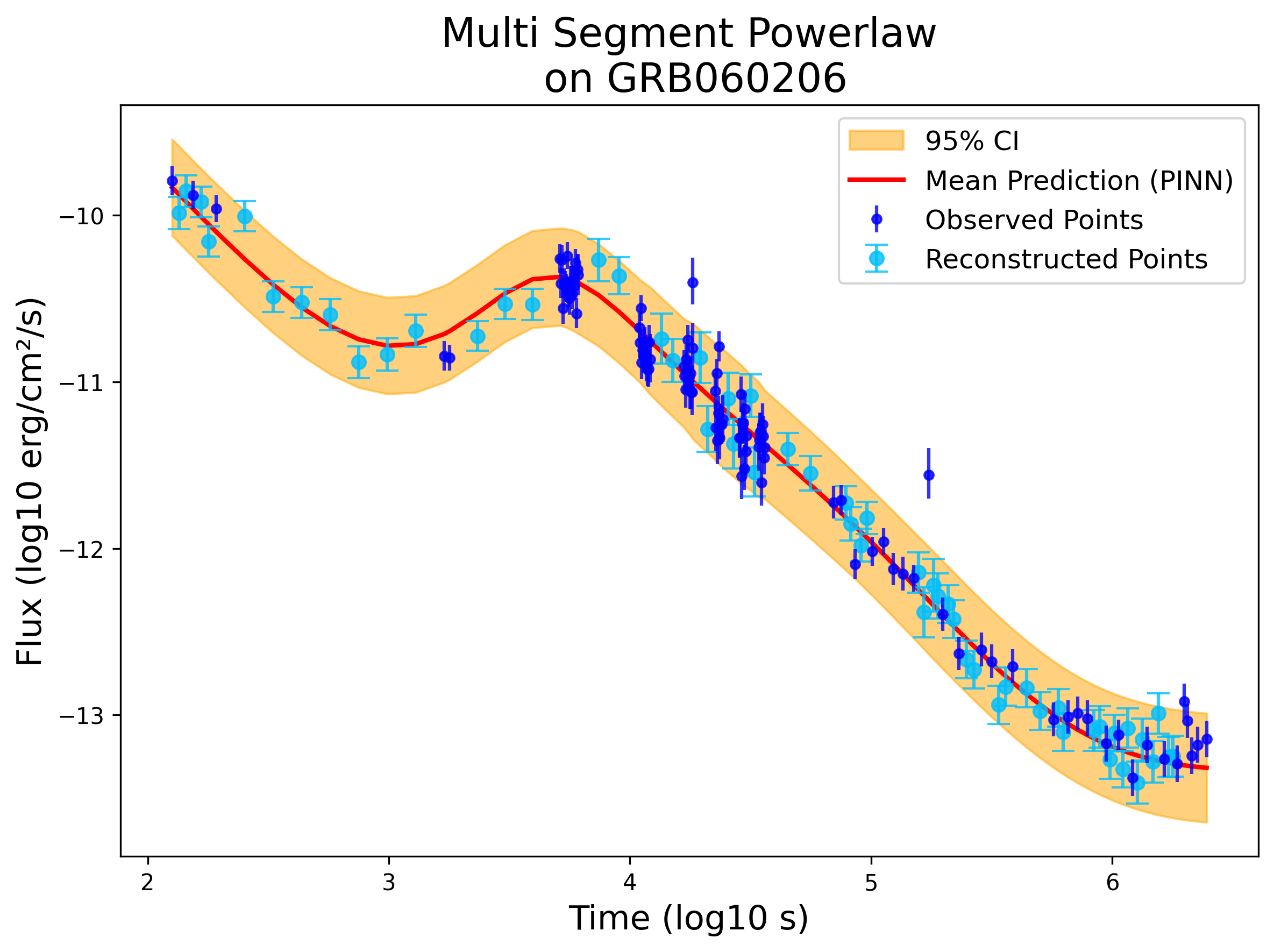}  

    \includegraphics[width=.24\textwidth, height=.20\textwidth]{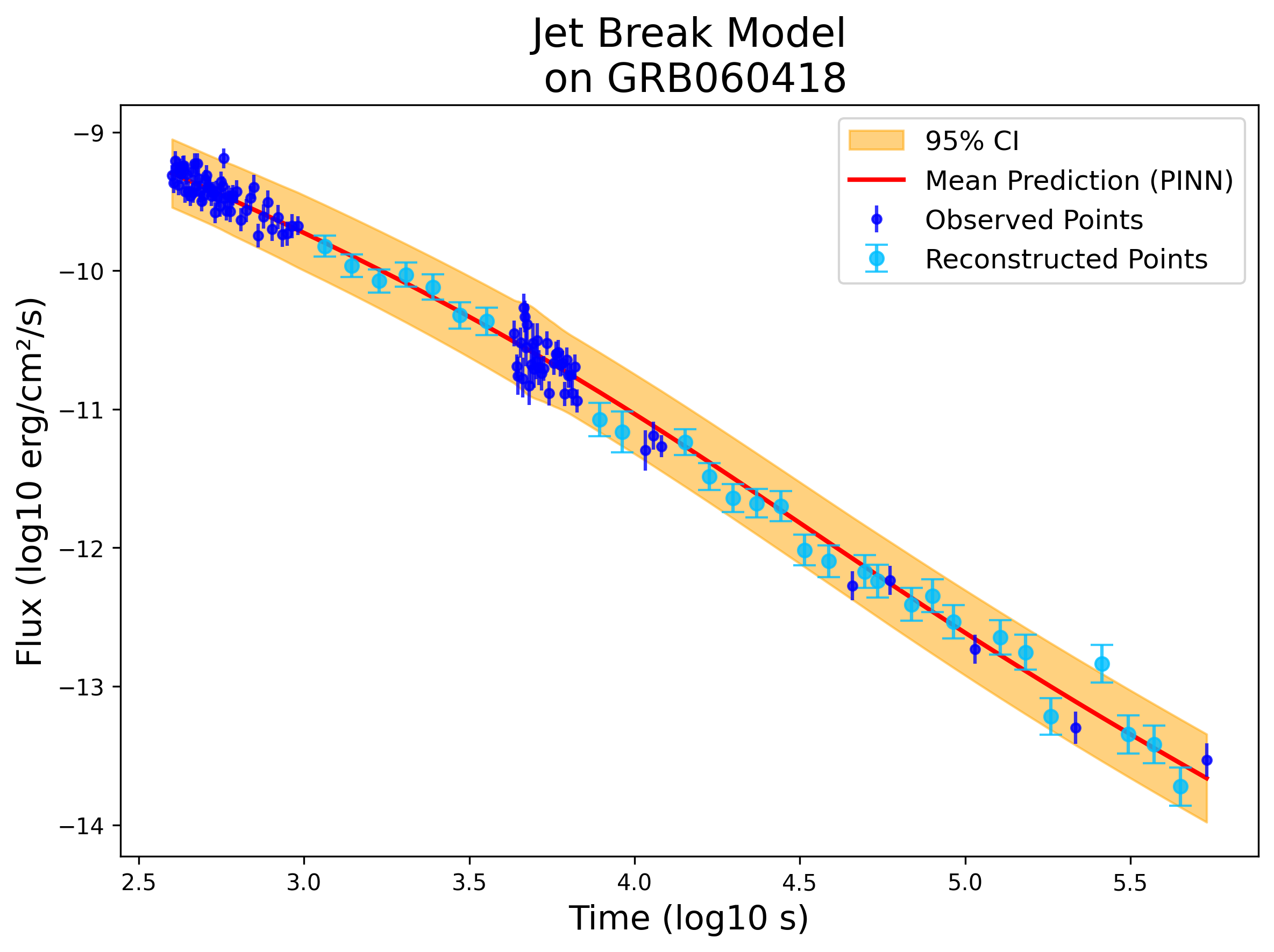}
    \includegraphics[width=.24\textwidth, height=.20\textwidth]{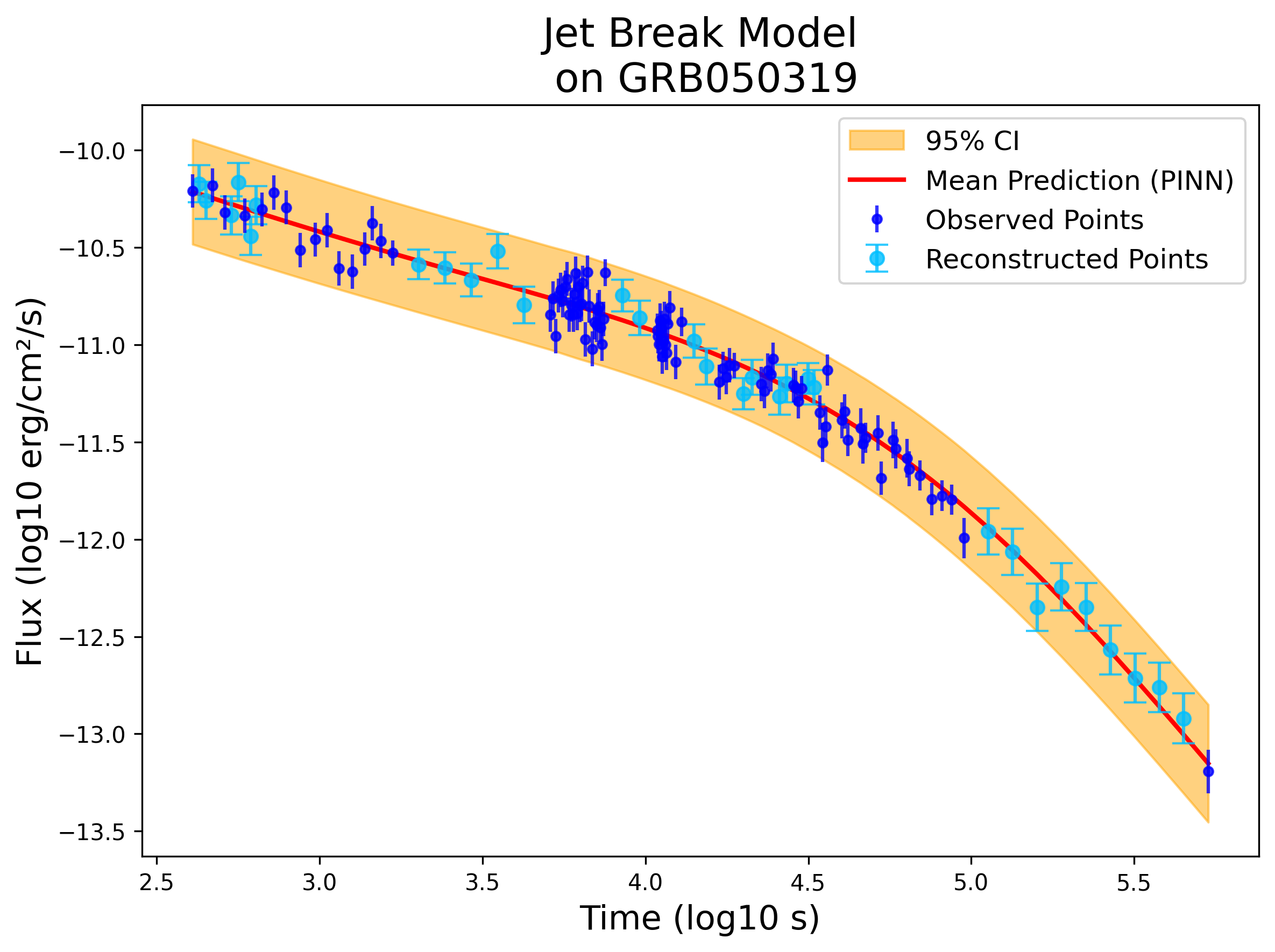}
    \includegraphics[width=.24\textwidth, height=.20\textwidth]{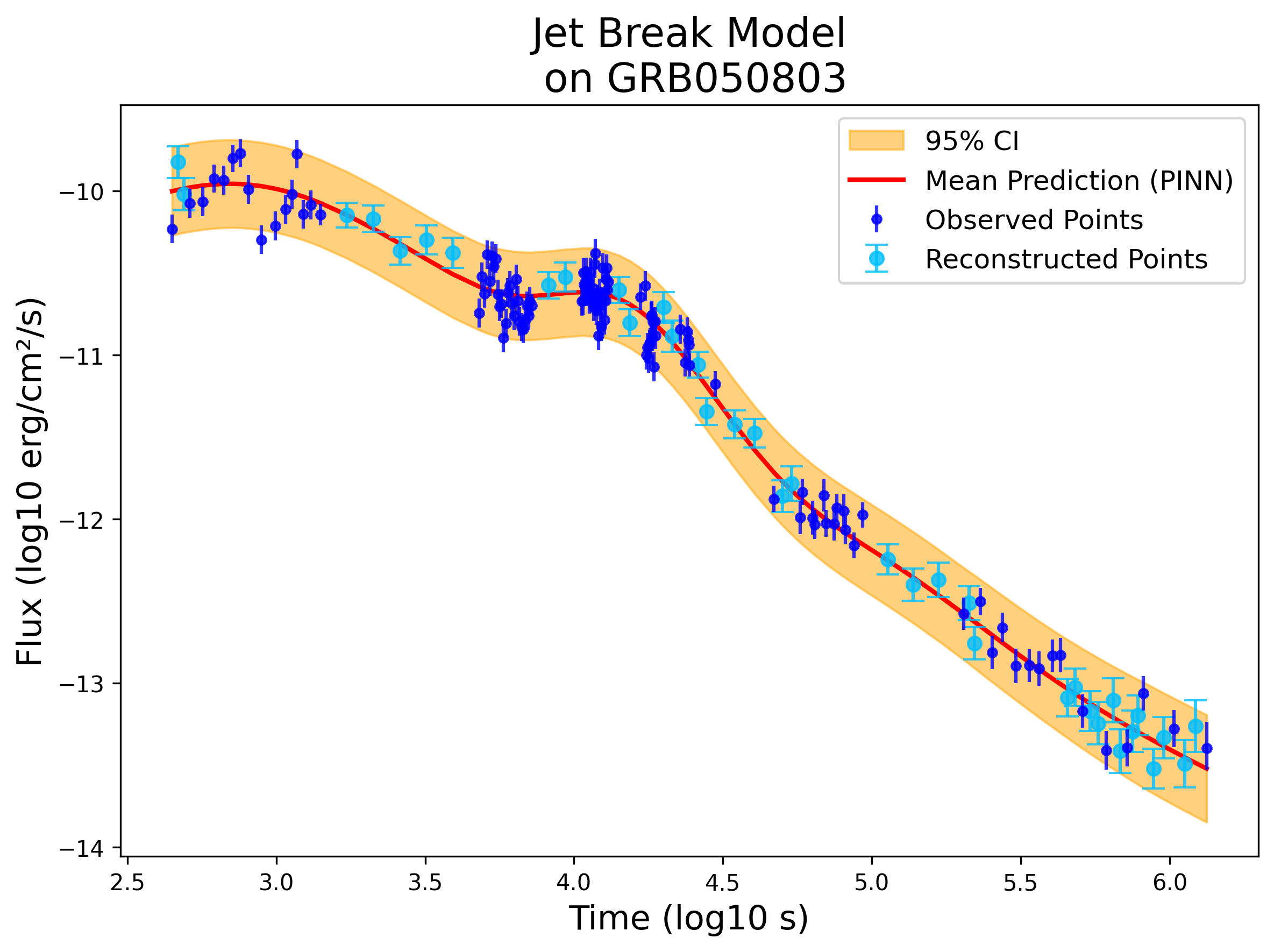}
    \includegraphics[width=.24\textwidth, height=.20\textwidth]{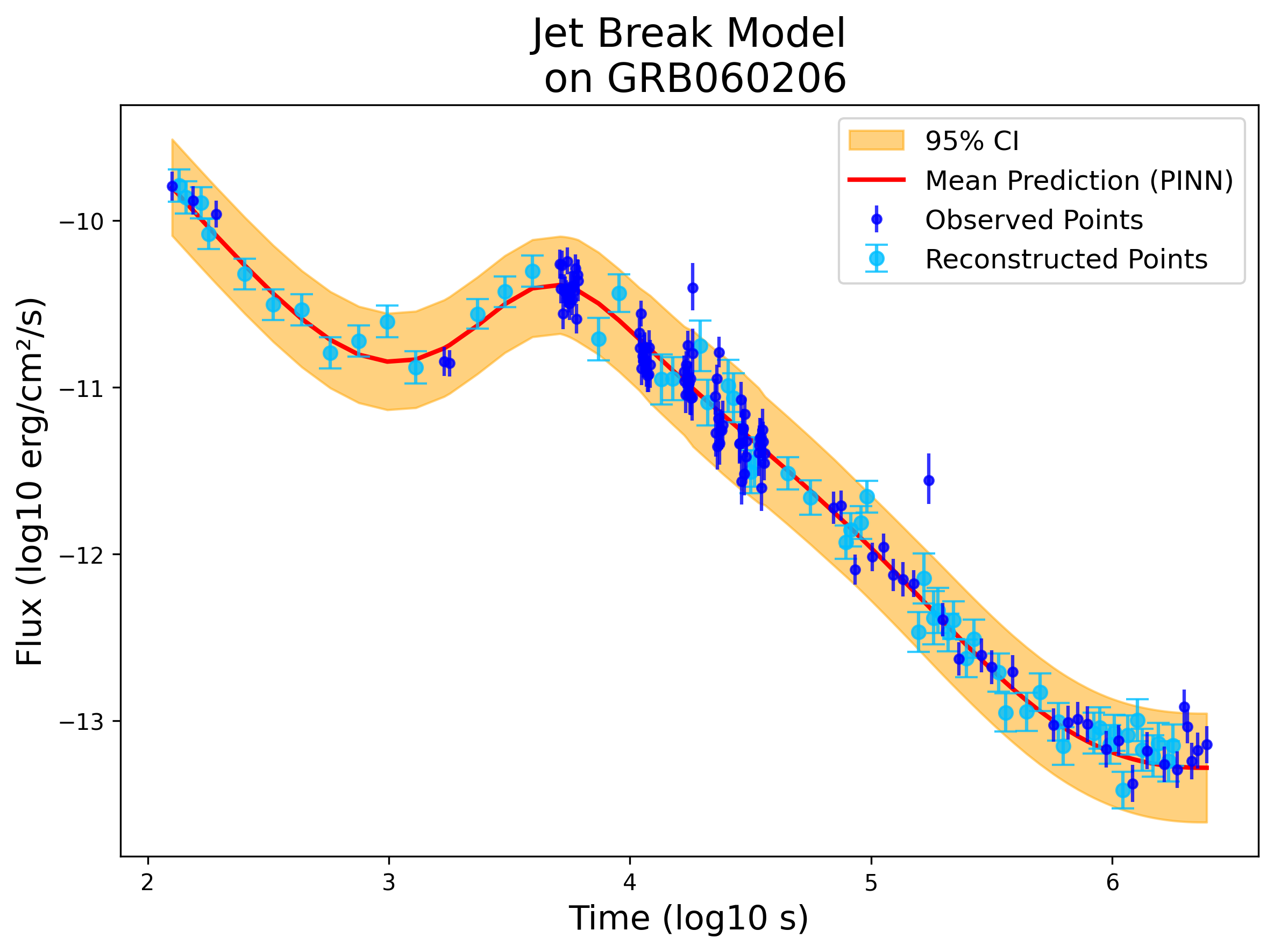}  

\end{figure*}

Table \ref{tab:reconstruction_comparison} shows how effectively the models' reconstructed data perform in comparison to the initial observed data for all GRBs. 
Although the models perform well for most GRB reconstructions, a few outliers are observed in the W07 parameters. 
We classify outliers as instances where the relative percentage increase exceeds $100\%$. Table \ref{tab:model_attributes} shows the performance strengths and weaknesses of each model.

The PINN framework was evaluated independently for each of its seven physics priors (Section~\ref{Physics Prior: Seven Analytical Models}), since each prior regularises the reconstruction toward a different afterglow morphology and is expected to perform differently depending on which morphology a given GRB actually follows. 
Of the 545 GRBs for which a reconstruction was attempted under every prior before Table~\ref{tab:reconstruction_comparison}, we have eliminated non-physical W07 fits, negative $F_a$, or failed covariance estimates,
to yield a convergent W07 fit on the augmented light-curve; Table~\ref{tab:reconstruction_comparison} lists the resulting mean percentage decrease in the  $\log_{10} T_a$, $\log_{10} F_a$, and $\alpha$ error fractions together with the outlier rate for each prior.

\begin{figure*}[!htbp]
\centering
\tabcolsep=0pt
\renewcommand{\arraystretch}{1.2} 
\begin{adjustbox}{width=\textwidth}
\begin{tabular}{c}
\includegraphics[width=.25\textwidth, height=.17\textwidth]{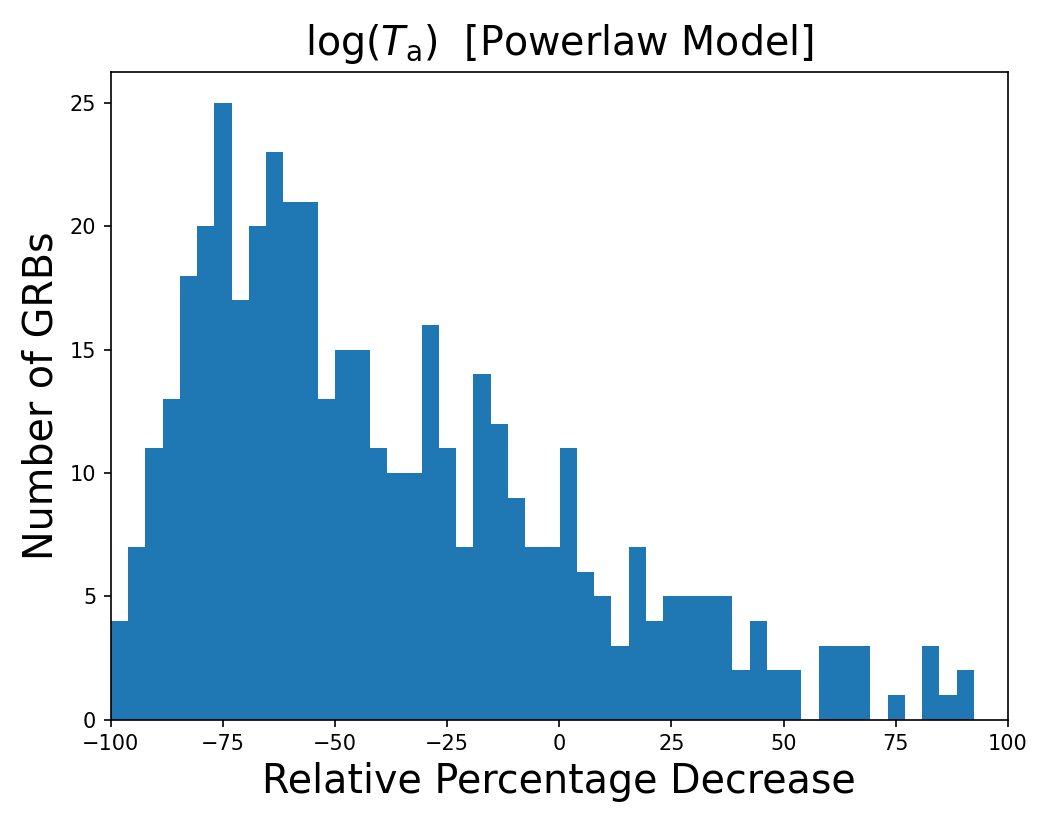}
\includegraphics[width=.25\textwidth, height=.17\textwidth]{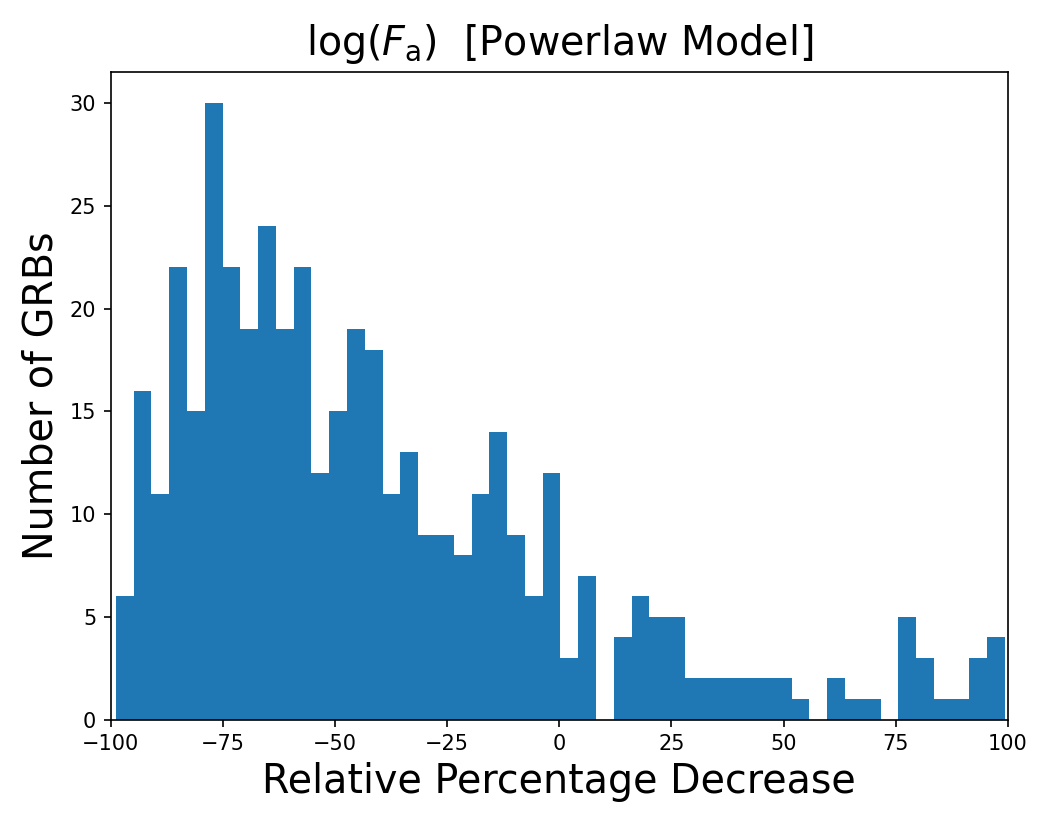}
\includegraphics[width=.25\textwidth, height=.17\textwidth]{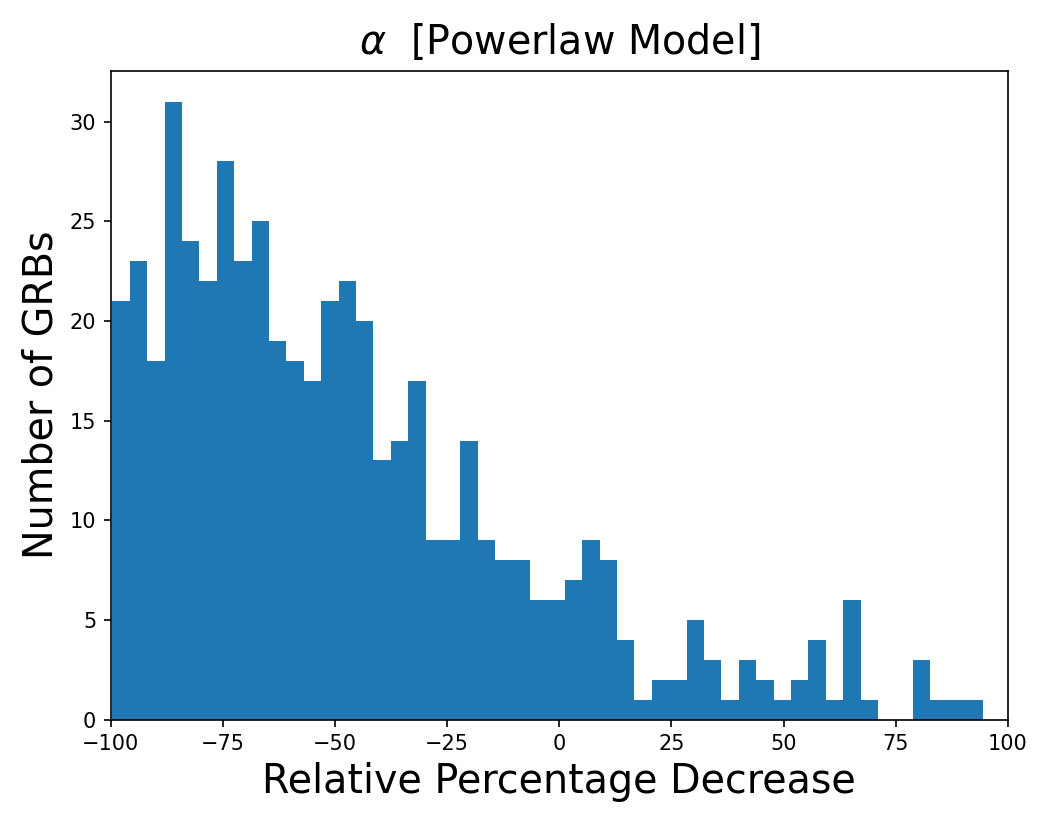} \\

\includegraphics[width=.25\textwidth, height=.17\textwidth]{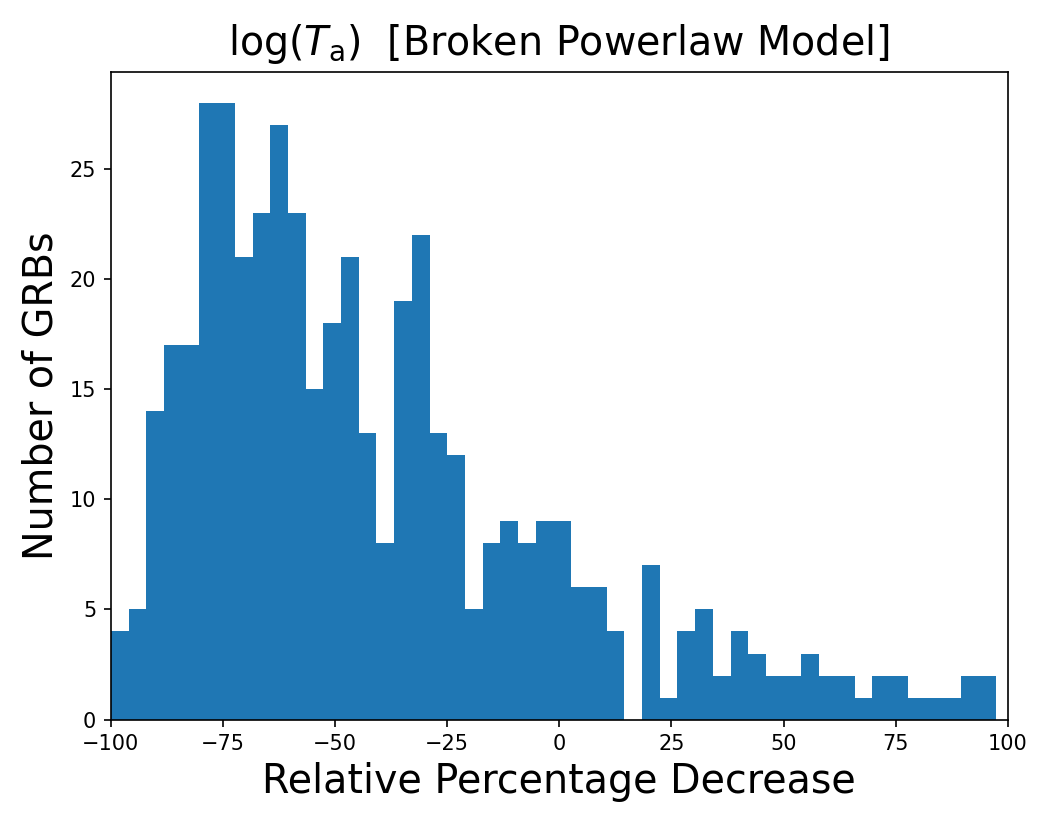}
\includegraphics[width=.25\textwidth, height=.17\textwidth]{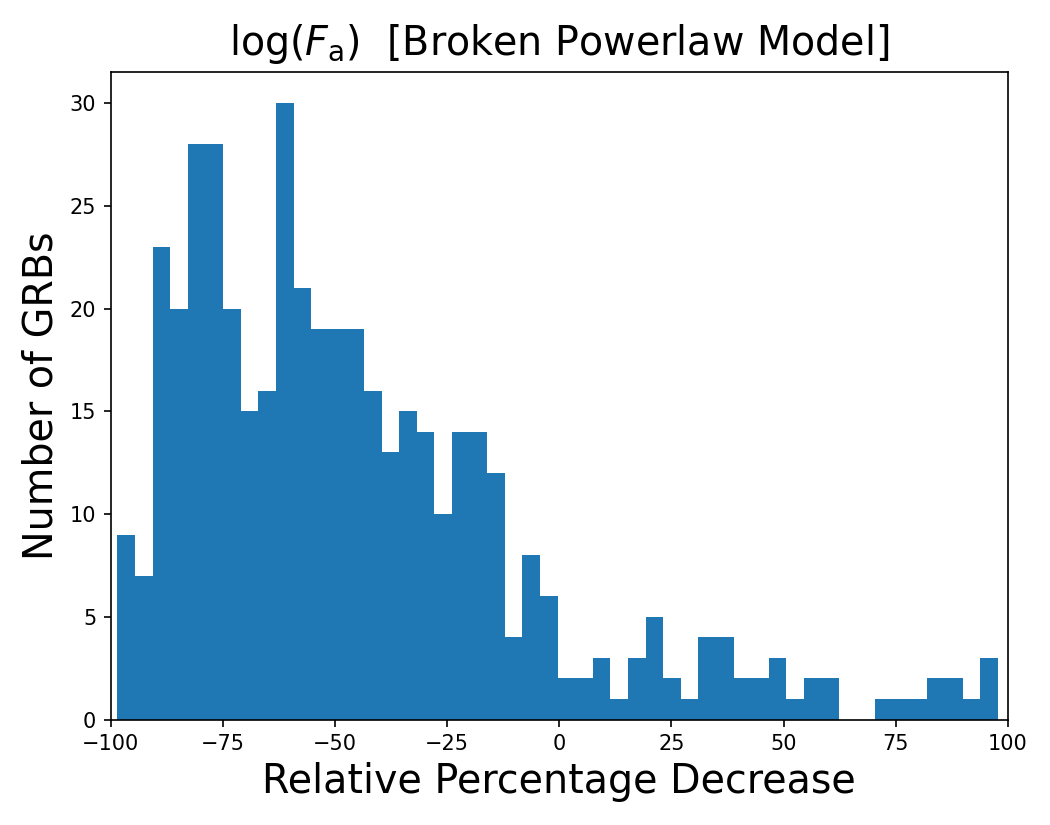}
\includegraphics[width=.25\textwidth, height=.17\textwidth]{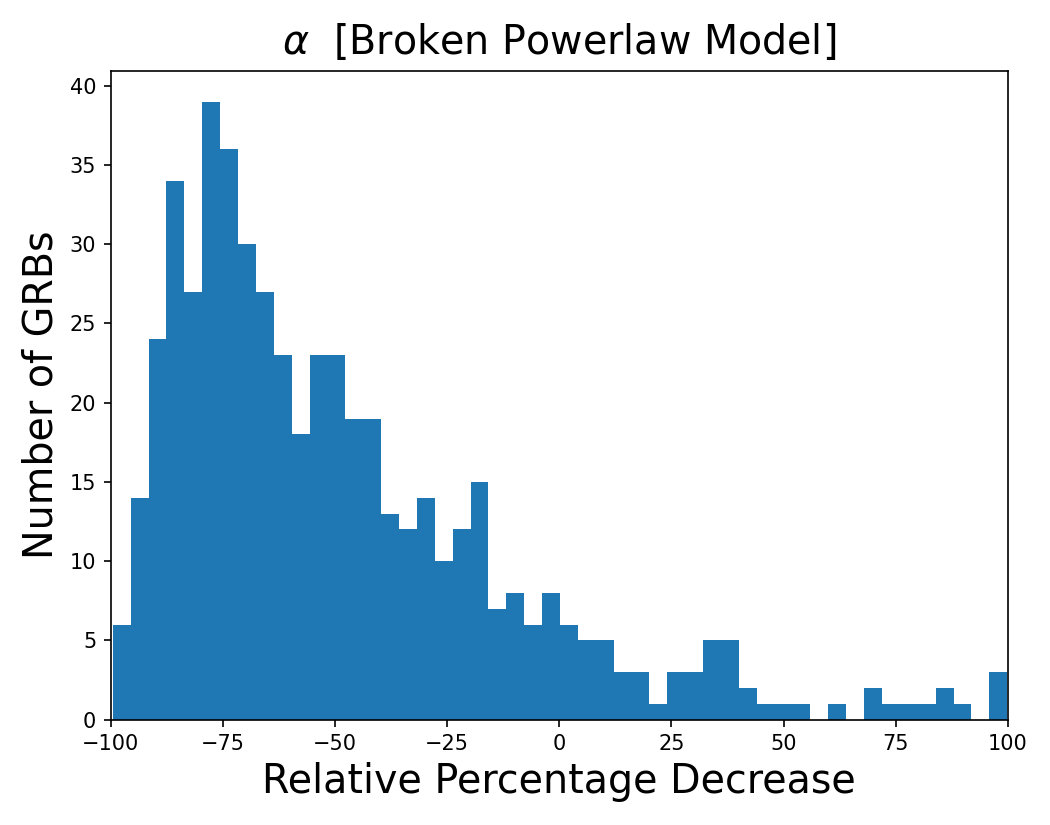} \\

\includegraphics[width=.25\textwidth, height=.17\textwidth]{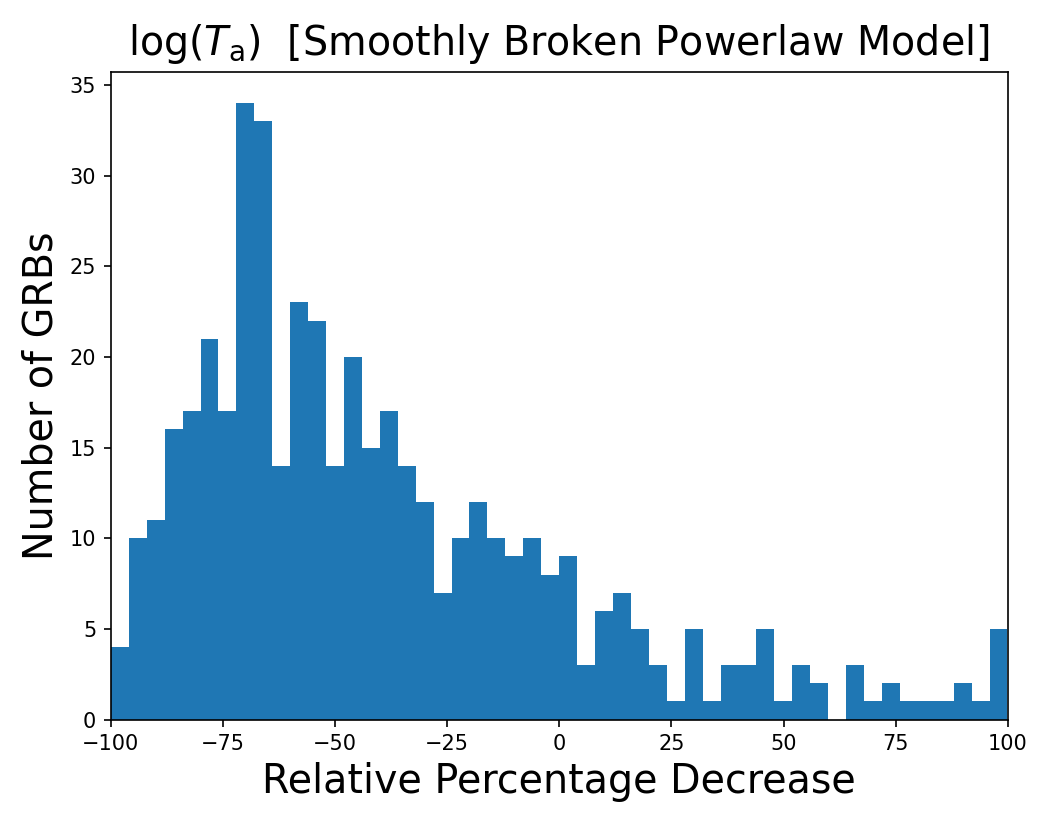}
\includegraphics[width=.25\textwidth, height=.17\textwidth]{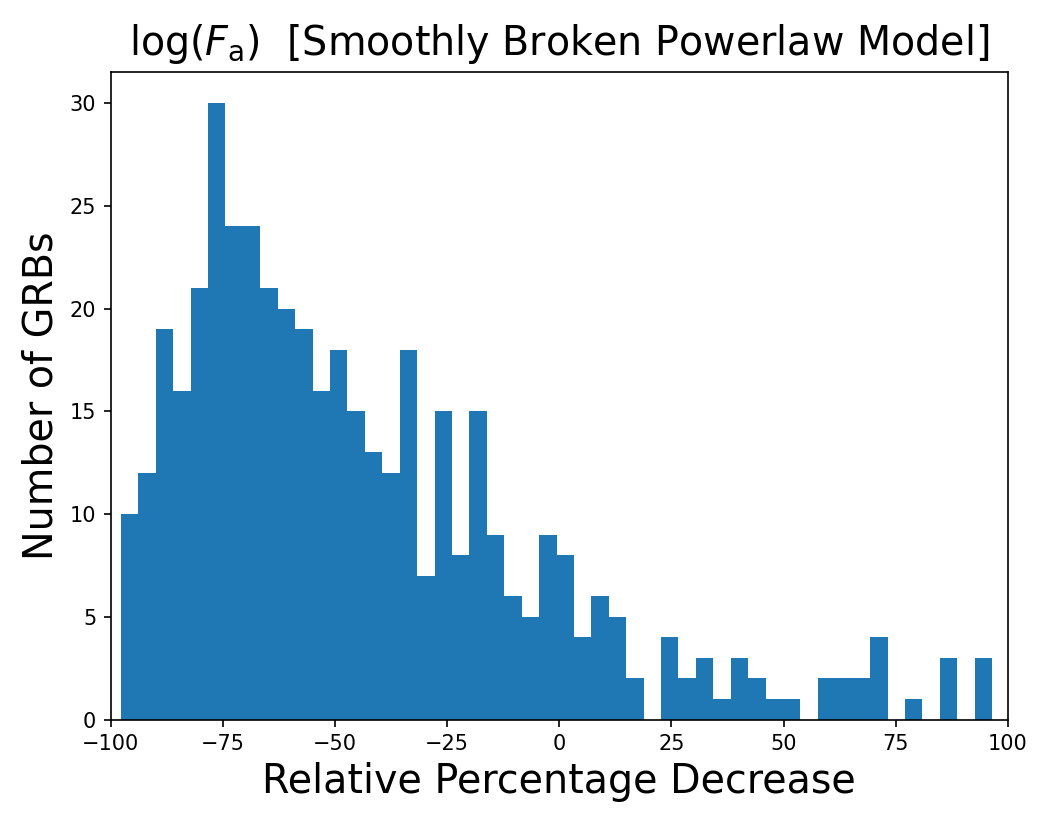}
\includegraphics[width=.25\textwidth, height=.17\textwidth]{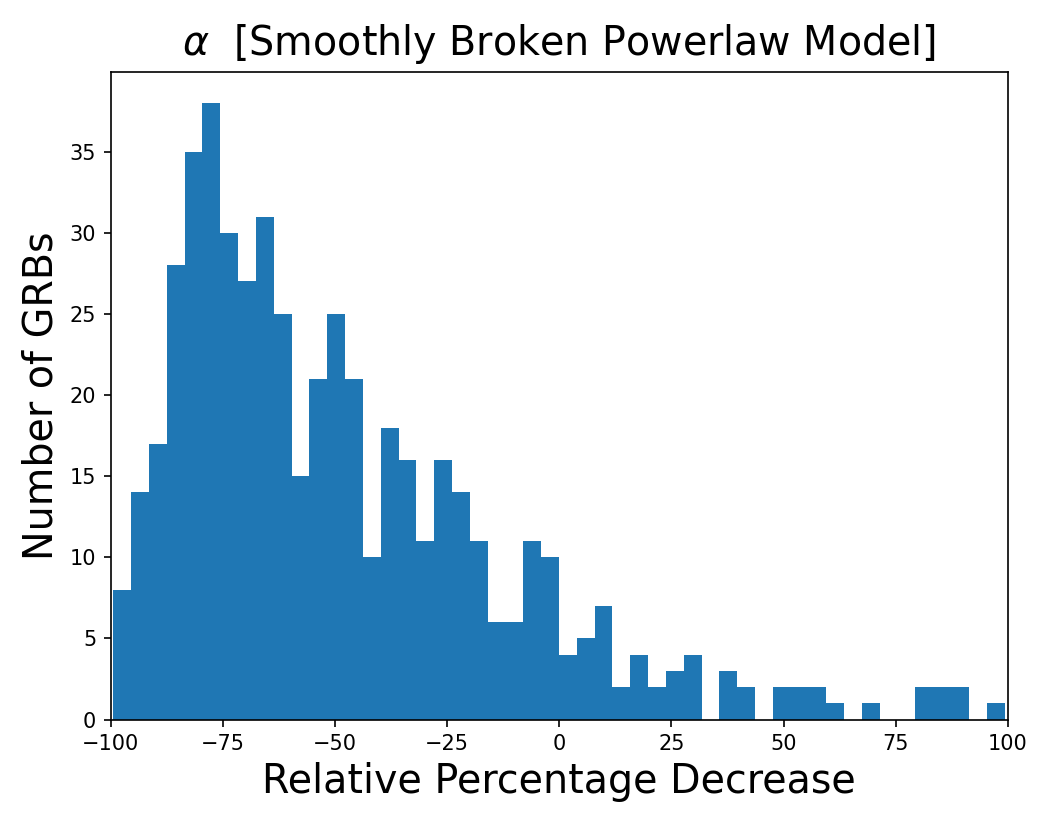} \\

\includegraphics[width=.25\textwidth, height=.17\textwidth]{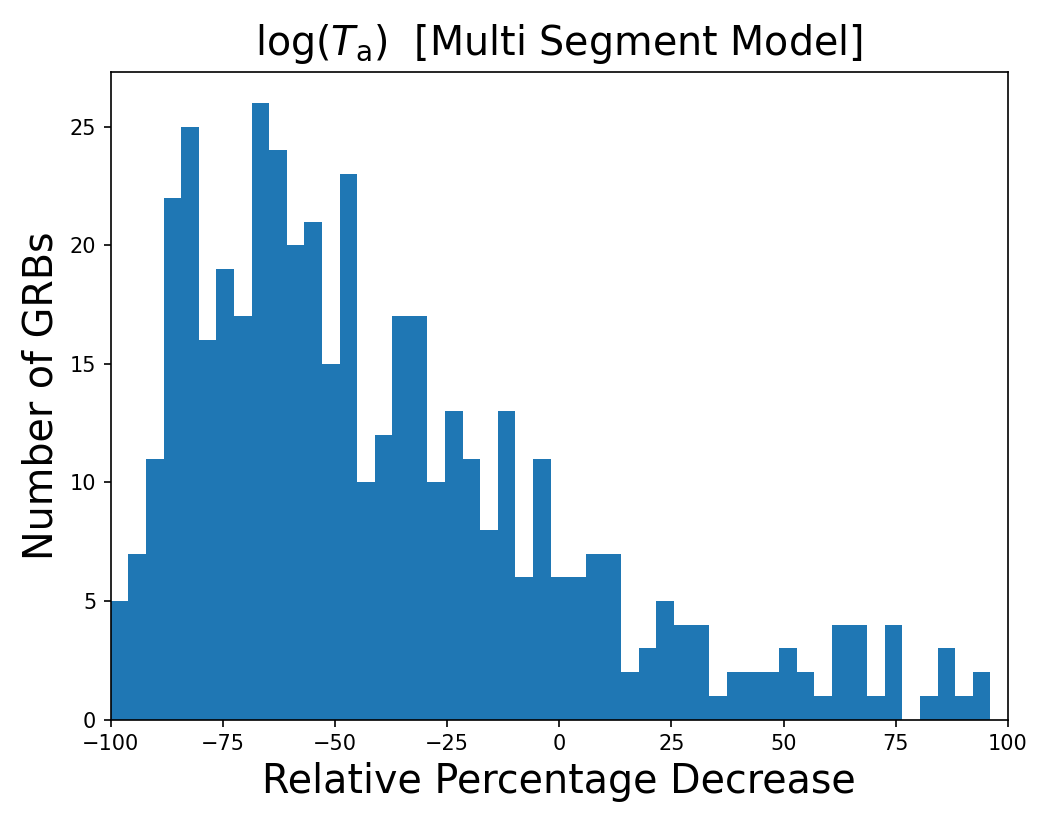}
\includegraphics[width=.25\textwidth, height=.17\textwidth]{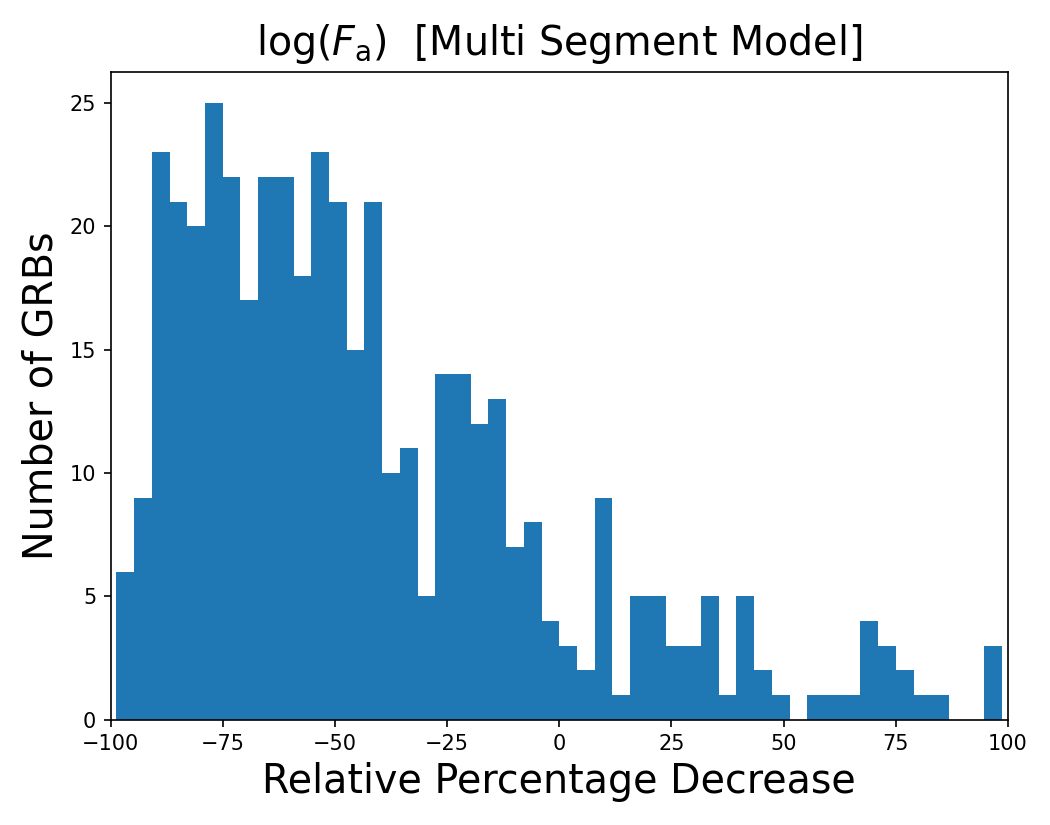}
\includegraphics[width=.25\textwidth, height=.17\textwidth]{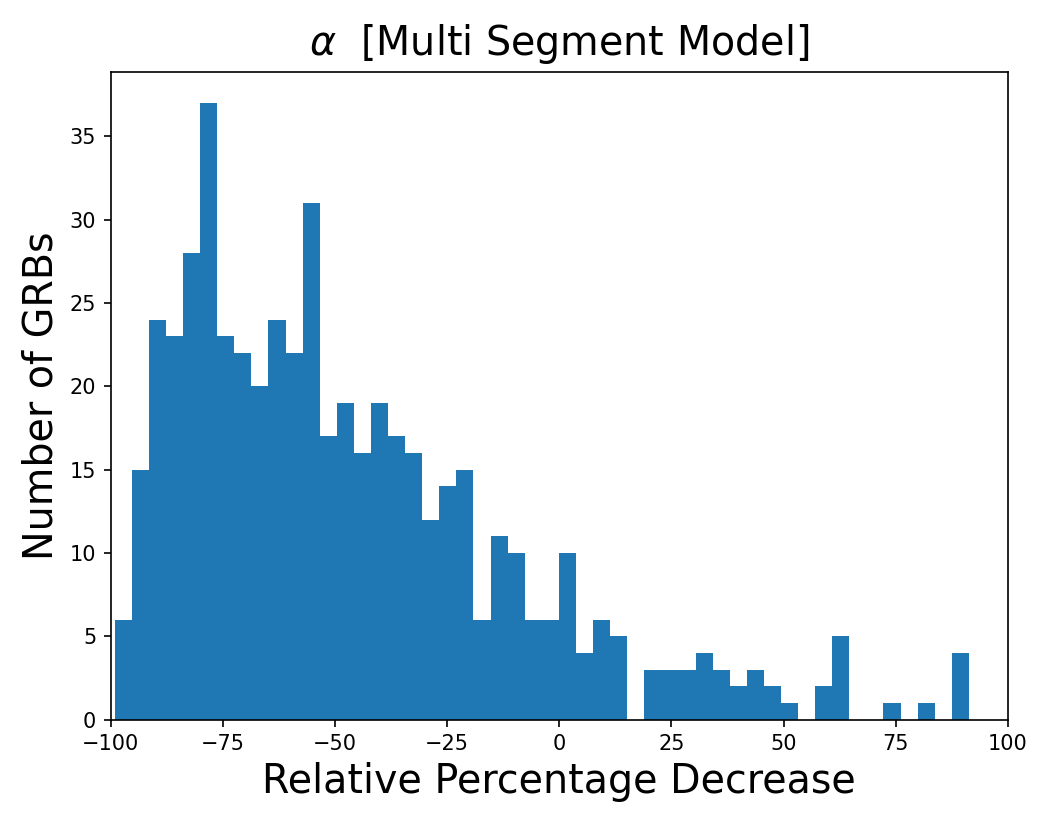} \\

\includegraphics[width=.25\textwidth, height=.17\textwidth]{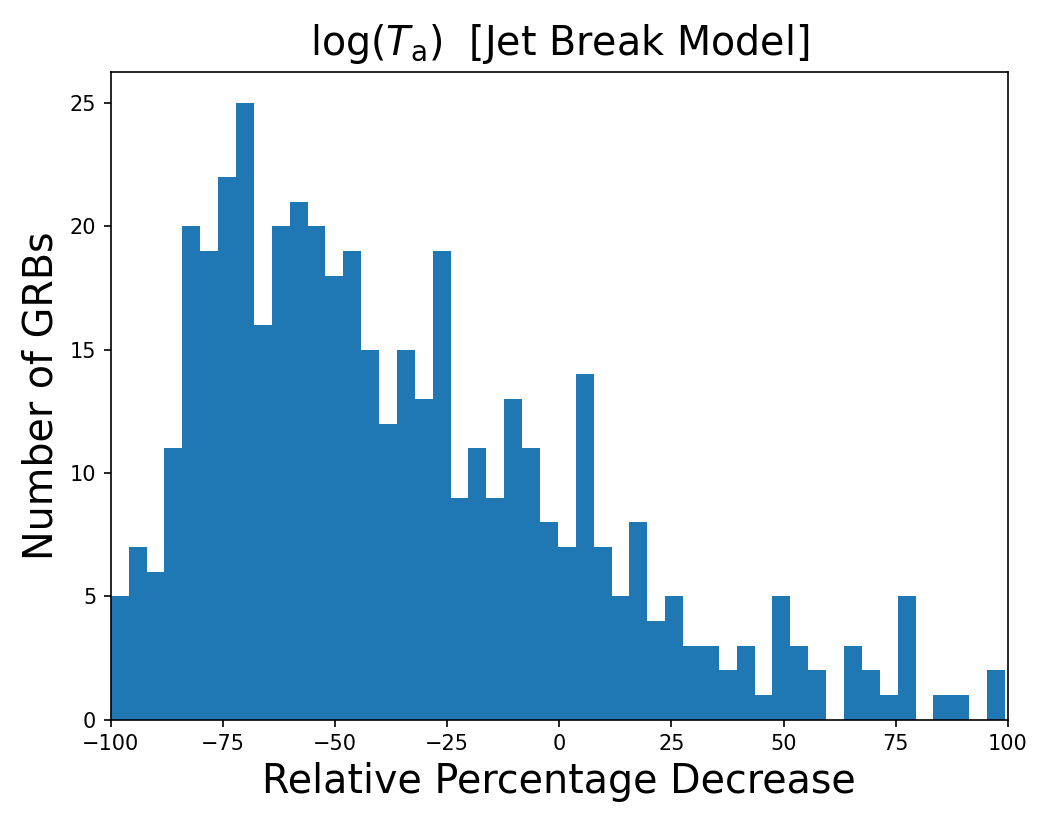}
\includegraphics[width=.25\textwidth, height=.17\textwidth]{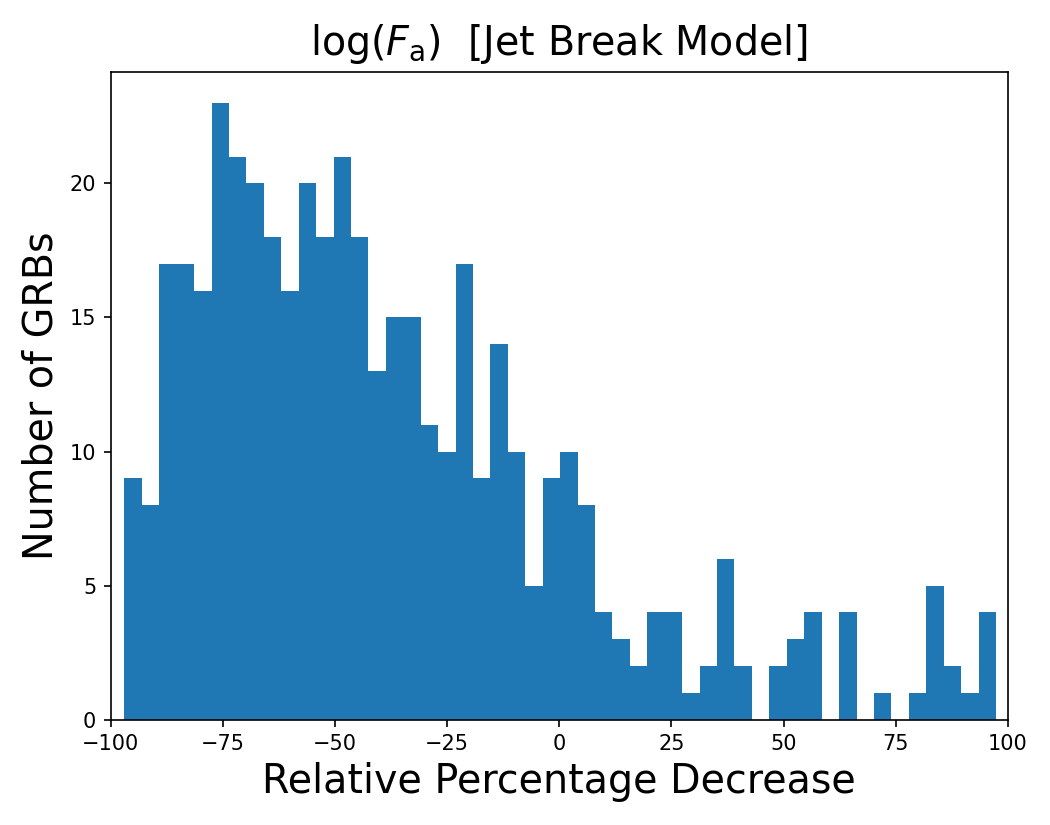}
\includegraphics[width=.25\textwidth, height=.17\textwidth]{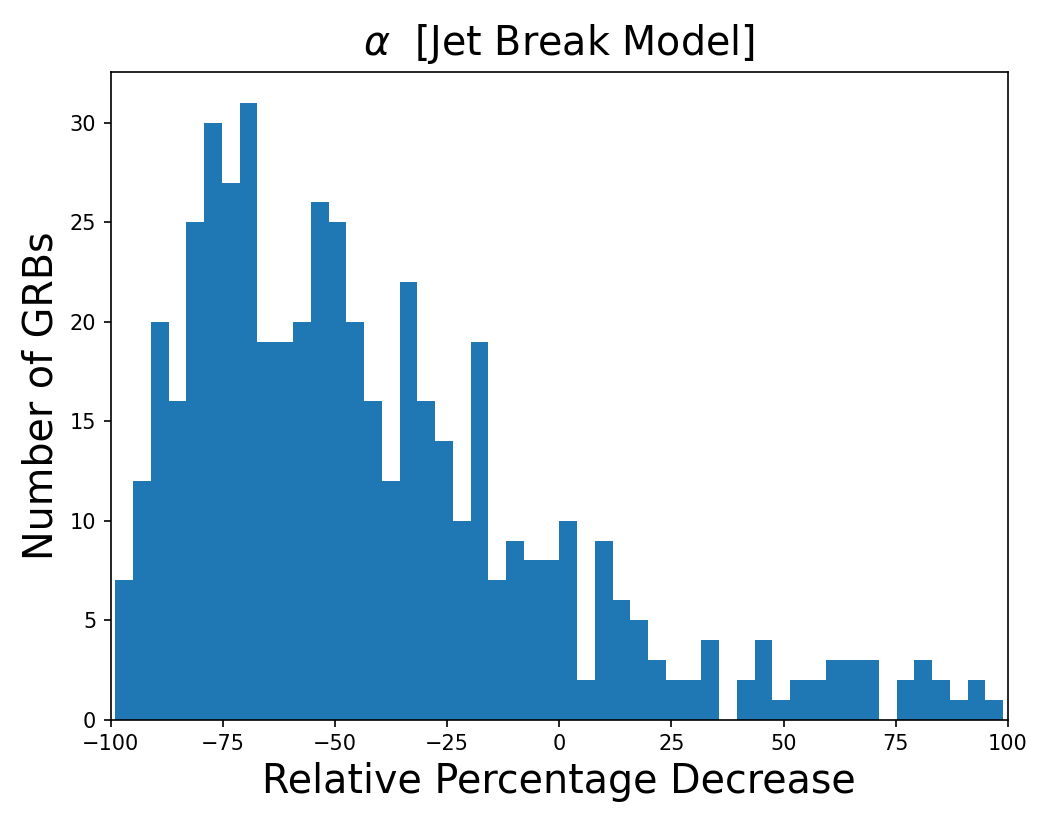} \\

\end{tabular}
\end{adjustbox}
\end{figure*}


\begin{table*}[htp] 
\centering
\renewcommand{\arraystretch}{1.1}
\begin{tabular}{|l|c|c|}
\hline 
\textbf{Prior Selected} & \textbf{N GRBs} & \textbf{\% of sample}  \\ 
\textbf{(lowest reduced $\chi^2$)} &  &   \\ 
\hline

Simple PL & 390 & 64.7 \%  \\
Energy Injection & 85 & 14.1 \%  \\
Jet Break Model & 46 & 7.6 \%  \\
Broke PL & 40 & 6.6 \%  \\
Smoothly Broken PL & 16 & 2.7 \%  \\
Synchrotron & 15 & 2.5 \%  \\
Multi-Segment PL & 11 & 1.8 \%  \\ \hline
Total & 603 & 100 \%  \\
\hline

\end{tabular}
\caption{Sample distribution and relative percentage of GRB light-curves best described by each of the seven PINN analytical physics priors, ranked by minimum reduced chi-squared ($\chi^2$) across all evaluated GRB-prior pairs. \label{tab:reduced_chisq_PINN}}
\end{table*}


\begin{table*}[htbp]
\centering
\renewcommand{\arraystretch}{1.2}
\begin{tabular}{lcccc}
\toprule
Attribute & PINN & Siamese& ReFANN  & Quantile \\ 
\midrule
High uncertainty reduction               & Strong & Moderate     & Moderate     & Moderate \\ \hline
Good sparse-data performance             & Strong & Strong & Strong & Moderate \\ \hline
Almost zero outliers for Good GRBs       & Weak   & Strong & Strong & Strong \\ \hline
Low computational cost                   & Weak   & Moderate     & Moderate     & Strong \\ \hline
Resilience to noise / over-fitting       & Strong & Strong & Strong & Moderate \\ \hline
Captures complex patterns                & Strong & Strong & Strong & Moderate \\
(flares/breaks)                          &  &  &  &  \\ \hline
Models full predictive distribution      & Weak & Moderate     & Moderate     & Strong \\
\bottomrule
\end{tabular}
\caption{Comparison of the four newly introduced reconstruction models on qualitative performance attributes. 
Ratings are derived from the quantitative results in Table~\ref{tab:Framework} (uncertainty reduction and outlier fractions) and the uncertainty-quantification mechanisms in Table~\ref{tab:lambda_study}. 
Strong, moderate, and weak categories are relative to the other three models. \label{tab:model_attributes}}
\end{table*}

Among the seven, the broken PL prior achieves both the largest mean uncertainty reduction ($~-40.65\%$ for $\log_{10} T_a, -44.68\%$ for $\log_{10} F_a, ~-48.84\%$ for $\alpha$) and the lowest outlier incidence ($~14.79\%$, 16.82\%, and $3.33\%$, respectively), while the jet-break prior yields the smallest improvement ($-35.83\% / -37.40\% / -41.70\%$) together with one of the higher outlier rates. 
This ordering has a straight-forward physical explanation: the jet-break prior fixes the post-break slope to $\alpha_2 \in [2.0,\,3.0]$ (Section~\ref{Physics Prior: Seven Analytical Models}, Model~5), motivated by the fireball jet model \citep{PIRAN1999, Rhoads1999, Sari1999}, but only $\approx 12\%$ of Swift-XRT afterglows show compelling evidence for a genuine achromatic jet break \citep{Racusin2009}. 
Imposing this constraint on the remaining $\approx 88\%$ of the sample forces an incorrect late-time slope onto GRBs that never break in this way, which inflates the error-fraction tail.
The broken and smoothly broken PL priors instead leave both segment slopes free within the canonical decay hierarchy of \cite{Zhang2006} and \cite{Nousek2006}, and consequently adapt to a wider range of observed morphologies.

The 603 entries in Table~\ref{tab:reduced_chisq_PINN} correspond to GRB-prior pairs rather than distinct GRBs: NN of the 543 GRBs converged under more than one prior and are therefore counted once per converged prior in this $\chi^2$ ranking.
Ranking the seven reconstructions of each GRB by reduced $\chi^2$ (Section~\ref{Physics Prior: Seven Analytical Models}) shows that the simple PL prior is the best fit for 390 of 603 GRBs with a completed reconstruction ($64.7\%$), followed by the energy-injection ($14.1\%$), jet-break ($7.6\%$), broken PL ($6.6\%$), smoothly broken PL ($2.7\%$), synchrotron ($2.5\%$), and multi-segment ($1.8\%$) priors. 
This is consistent with the parent sample composition, in which $55.3\%$ of the GRBs are independently classified as ``Good GRBs" whose light-curves follow the W07 model closely, without flares, bumps, or additional breaks (Section~\ref{Dataset Selection}): a $\chi^2-$based ranking that never sees this classification recovers a similar morphological split. 

We note, however, that selecting the $\chi^2-$best prior per GRB does not automatically maximise the downstream parameter improvement; the resulting ensemble (evaluated on the same 603 GRB-prior pairs of Table~\ref{tab:reduced_chisq_PINN}, using each GRB's individually $\chi^2-$preferred prior) gives mean decreases of $~ -36.59\%~(\log_{10} T_a)$, $-39.43\%~(\log_{10} F_a)$, and $-44.81\% (\alpha)$, each several percentage points each below the broken PL prior evaluated on its own. 
Reduced $\chi^2$ measures how closely the reconstruction tracks the observed flux values, while the error-fraction reduction depends on how well the reconstruction constrains the local slope near $T_a$; a prior can fit the data well while still over- or under-estimating that slope. 
We therefore report the full per-prior breakdown in Table~\ref{tab:reduced_chisq_PINN} rather than a single aggregated PINN statistic.
Figure~\ref{fig:percentage-dec} (columns 1–3) shows the resulting distribution of the relative percentage decrease in $\log_{10} T_a$, $\log_{10} F_a$, and $\alpha$ for all seven priors, respectively. 
All seven distributions are concentrated below zero, confirming that the reconstruction reduces the W07 parameter uncertainty for the majority of the sample; the tails extending beyond $100\%$ correspond to the outlier fractions quoted above and are dominated by GRBs classified as flares/bumps, or flares/bumps + multiple breaks ($24.06\%$ and $7.5\%$ of the sample, respectively; Section~\ref{Dataset Selection}), whose morphology is not fully captured by any single seven-parameter analytical prior.

ReFANN and the Siamese architecture both produce substantial reductions in the W07 parameter error fractions. 
ReFANN, which reconstructs the light-curves without assuming a predefined functional form \citep{Wang2020}, reduces the error fractions by $-19.83\%$ for $\log_{10}T_a$, $-21.43\%$ for $\log_{10}F_a$, and $-26.51\%$ for $\alpha$, with improvements observed for 522 of the 545 GRBs. 
The corresponding outlier fractions remain below $4.1\%$ for all three parameters. 
The Siamese architecture, which combines a shared-weight dual-branch encoder with a physics-informed monotonicity regularisation and MC Dropout for uncertainty estimation, achieves slightly larger reductions of $-22.88\%$, $-22.83\%$, and $-26.93\%$ for $\log_{10}T_a$, $\log_{10}F_a$, and $\alpha$, respectively. 
These improvements are obtained for 524 GRBs, while the corresponding outlier fractions remain below $3.7\%$ for all parameters.

The PQR model, which imposes no physics or smoothness prior and instead bounds the reconstruction via three independently fitted quantile levels (Section~\ref{sec:quantile}), achieves the smallest but still substantial reductions of the four new models: $-18.92\%$ for $\log_{10} T_a$, $-20.04\%$ for $\log_{10} F_a$, and $-24.42\%$ for $\alpha$, with improvements recorded for 522 of the 545 GRBs. 
Its outlier fractions (2.94\%, 4.04\%, and 1.28\%) are the lowest of all four models, reflecting the conservative, distribution-free nature of the pinball-loss objective, which trades a smaller average gain for resistance to the long-tailed residuals that drive PINN and, to a lesser extent, ReFANN and Siamese into the outlier regime.

Among the four new models introduced in this paper, the PINN is particularly suited to “Good GRBs” whose light-curves closely follow a canonical four-segment template (\cite{Zhang2006}), because the PL and broken PL priors correctly embed the expected slope hierarchy (steep decay $\to$ plateau $\to$ normal decay $\to$ post-jet). 
For GRBs with flares, the PINN’s data loss is sufficiently flexible to fit individual flare points locally, while the physics prior prevents the reconstruction from extrapolating the flare structure into unobserved gaps. 
The outlier rate (defined as GRBs where the relative percentage increase in any parameter exceeds $100\%$) is reported in Table~\ref{tab:reconstruction_comparison}.

\section{Discussion} \label{section:Discussion}

The four reconstruction methods introduced in this paper represent different modelling strategies, ranging from the physics-informed PINN to the purely data-driven PQR model, with ReFANN and the Siamese NN lying between these two by incorporating physics-inspired regularisation. Their comparative performance helps us understand how different levels of physical guidance affect reconstruction accuracy, uncertainty estimation, and robustness across GRB light-curves with different morphologies.

The PINN framework achieves the largest absolute parameter uncertainty reductions, up to $-44.68\%$ for $\log_{10} F_a$ and $-48.84\%$ for $\alpha$ under the broken PL prior, yet carries an outlier rate of $3$--$20\%$ depending on the parameter and prior, roughly an order of magnitude above the $\lesssim 4\%$ seen for the three data-driven models. 
This is the expected signature of a physics-constrained model applied indiscriminately to a morphologically heterogeneous sample: when a GRB's true decay does not follow the assumed analytical form, the prior actively penalises the correct local slope and inflates the error-fraction tail.
Gradient flow pathologies of precisely this kind have been studied in the PINN literature by \citet{Wang2021}, who showed that the physics gradient can overwhelm the data gradient early in training when the two loss components are improperly normalised. 
Our $\lambda$ warm-up and EMA normalisation scheme (Section~\ref{Training Procedure}) directly addresses these pathologies.
 
The data-driven models carry no such morphological penalty and degrade more gracefully on complex afterglows, at the cost of a smaller average improvement, consistent with the bias-variance trade-off. 
The Siamese model partially bridges this gap: its weight-sharing constraint and learnable reference anchor \citep{Wang2017} act as a soft structural prior without committing to any specific decay law, which is why it achieves the best data-driven performance of the three ($-22.88\%/ -22.83\%/ -26.93\%$ for $\log_{10} T_a/\,\log_{10} F_a/\,\alpha$) while maintaining the second-lowest outlier rate ($4.2\%$).

 
The classical PINN formulation of \citet{Raissi2019} penalises deviations of the network derivative from the ODE right-hand side, $d\hat{y}/d\tau \approx dP/d\tau$. 
As detailed in §~\ref{Training Procedure}, the classical ODE derivative form 
produced degenerate flat reconstructions in our setting; \cite{Wang2021} provide the theoretical basis for why such gradient pathologies arise.
This failure mode is consistent with the gradient flow analysis of \citet{Wang2021}, who found that derivative-form physics losses introduce large, poorly scaled gradients that can dominate the data loss and drive the network toward unphysical solutions.
 
Our value-space physics loss, $\mathcal{L}_\mathrm{phys} = N^{-1}\sum_i [\hat{y}_i - P(\tau_i)]^2$, avoids this by penalising deviations in absolute flux rather than in slope. 
This allows the PINN to deviate from the physics model wherever the data demand it while still being guided toward the analytical curve in sparsely observed gaps. 
The scientific consequence is that the reconstructed light-curve is not forced to adopt the prior's exact slope between observed points, a crucial property for GRBs with flares or bumps that locally violate the assumed PL form. 
\citet{Cuomo2022} and \citet{Karniadakis2021} both note that value-space constraints are appropriate when the physics model is approximate, and the data should retain priority.

Comparing our models against the results of the full series, the broken PL PINN prior (-40.65\% for $\log T_a$, -44.68\% for $\log F_a$, -48.83\% for $\alpha$) is comparable to, and for $\log F_a$ marginally exceeds, the CSS model of \cite{Gupta2026} (-44.2\%, -44.5\%, -50.4\%), which represents the strongest purely data-driven result in the series. 
The PINN broken PL prior also clearly surpasses QSS (-43.5\% -43.2\%/ -48.3\%), which was the leading result prior to \cite{Gupta2026}. 
A critical distinction, however, is that the CSS model achieves these reductions without any physical prior; it optimises a smoothness penalty on the fourth derivative of a cubic spline, which can capture the overall decaying shape of a light-curve but has no knowledge of the underlying synchrotron-emission physics. 
The PINN, by contrast, embeds the governing afterglow decay law directly into the training objective, meaning that its reconstructions in data-sparse gaps are physically interpretable: the filled-in flux values follow the solution of the selected analytical model rather than an unconstrained smooth interpolant. 
This is important for further physical analysis, including closure-relation tests, magnetar spin-down constraints, and the identification of cooling breaks. In these applications, the reconstructed light-curve shape within the gaps provides useful information, rather than only the uncertainties in the endpoint parameters.

The Attention U-Net and MLP ($\approx -37\%$ to $-41\%$) remain the closest data-driven competitors from the earlier series. 
ReFANN and the Siamese model ($\approx -20\%$ to $-27\%$) achieve reconstruction performance comparable to Polynomial Curve Fitting and CNN--BiLSTM \citep{Kaushal2026}. They also clearly outperform Isotonic Regression, BNN, DGP, and TCN, while maintaining substantially lower outlier rates and without requiring a fixed functional template.

Crucially, the earlier data-driven methods (DGP, BNN, and TCN), which showed weaker reductions ($-5\%$ to $-18\%$ in at least one parameter), rely on explicit or implicit smoothness assumptions that are not always well suited to GRB afterglows with abrupt temporal transitions. 
In contrast, ReFANN's iterative residual feedback \citep{Wang2020} and the Siamese model's shared-weight representation learning 
makes no such assumption. 
This flexibility likely contributes to their superior performance over DGP and BNN, despite their use of relatively simple architectures.

Comparing against contemporary work in the literature beyond this series, \citet{Srinivasaragavan2020} fitted broken PL models to 
545 Swift-XRT afterglows and derived closure-relation constraints with median $\alpha$ uncertainties of $0.05-0.15$; our reconstruction reduces these fractional uncertainties by $24-49\%$ depending on the method, substantially expanding the usable sample for closure-relation tests. 
Similarly, the Gaussian Process reconstructions of \citet{Aigrain2023}, which represent the state of the art for smooth, continuous time-series functions, are not directly applicable to GRB light-curves due to their multi-segment morphology and non-stationary covariance structure; our methods address this limitation by design.
 
 
The ordering of PINN priors by reconstruction performance (BPL $>$ SBPL $>$ MSM $>$ PL $>$ EI $>$ Synchrotron $>$ JBM, for the mean percentage decrease across parameters) reflects the degree to which each prior's assumed morphology matches the actual distribution of afterglow shapes in the 545-GRB sample. 
The broken PL prior's dominance is attributable to its two free slope parameters and a single free break time, which together can represent the plateau to normal-decay transition, the single most common morphological feature in Swift-XRT afterglows, without imposing any specific physical mechanism. 
The fireball model of \citet{PIRAN1999}, as refined by \citet{Zhang2006} and \citet{Nousek2006}, predicts precisely this two-segment structure for afterglows dominated by a single emission component; the BPL prior encodes this prediction in its parametric form.

The energy-injection prior \citep{Dai1998, Zhang2001} similarly allows a free effective slope but adds a free injection parameter $q$, which in practice provides less parametric flexibility than two independent slopes for afterglows whose temporal evolution is not dominated by continuous energy injection. 
The jet-break prior \citep{Rhoads1999, Sari1999} underperforms because only $\approx 12\%$ of Swift-XRT afterglows display a compelling achromatic jet break \citep{Racusin2009, Liang2008}; forcing $\alpha_2 \in [2.0,\,3.0]$ onto the remaining $88\%$ introduces a systematic bias. 
The synchrotron cooling-break prior \citep{Sari1998} fixes $\Delta\alpha = 0.25$, which is realised only in ISM environments and under specific energy-injection conditions \citep{Zhang2001, Dainotti2023b}.

The reduced$-\chi^2$ ranking, which favours the simple PL prior for $64.7\%$ of reconstructed GRBs, is consistent with the morphological composition of the sample: $55.3\%$ of GRBs are classified as ``Good GRBs'' whose light-curves follow the W07 four-segment template without additional features. 
This cross-check between the independent morphological classification (Section~\ref{Dataset Selection}) and the $\chi^2-$ranked prior selection is an important internal consistency test: a prior-selection algorithm that has no access to the morphological labels recovers a similar morphological split, suggesting both the classification and the PINN fitting are responding to the same underlying structure in the data.
 
 
The plateau parameters $\log_{10}T_a$, $\log_{10}F_a$ and the derived plateau luminosity $L_a$ are central to the two-dimensional Dainotti relation \citep{Dainotti2008}
and the three-dimensional fundamental plane \citep{Dainotti2016, Dainotti2017, Dainotti2020}.
Their measurement precision currently limits the cosmological constraining power of GRBs relative to Type Ia supernovae.
Since $T_a$ marks the end of the plateau, commonly associated with a long-lived central engine, a more precise $\log_{10}T_a$ also constrains the spin-down timescale of a possible newborn millisecond magnetar \citep{Zhang2001}.

\citet{Dainotti2022a} showed through simulations that a $47.5\%$ reduction in plateau-parameter uncertainties, 
combined with a doubling of the GRB sample via machine-learning redshift estimation, would allow GRBs to match the $\Omega_M$ precision of \citet{Betoule2014}: a goal projected to be achievable by $\sim$2030 with LCR applied, compared with $\sim$2038 at the unaided detection rate \citep{Kaushal2026}.
Preliminary estimates suggest that optimised sub-samples of plateau GRBs could reduce the uncertainty on $H_0$ by up to $35\%$ relative to current statistical constraints \citep{Dainotti2023c}.
The $19–49\%$ error-fraction reductions demonstrated here span the lower to the upper end of the approach to this threshold, with the broken-PL PINN prior ($\sim$49\% for $\alpha$) and the CSS model of \citet{Gupta2026} both exceeding 47.5\% for the post-plateau slope, while improvements for $\log T_a$ and $\log F_a$ remain just below this mark for most models.
Our reconstruction directly addresses the LCR component of this 
requirement.

These improvements bear on GRBs
as $\Lambda$CDM probes through the following propagation.
The plateau luminosity is related to the observed flux by
\begin{equation}
L_a = 4\pi\,D_L^2(z;\,\boldsymbol{\theta}_{\rm cos})\,F_a\,K,
\label{eq:La}
\end{equation}
where $D_L$ is the luminosity distance for cosmological parameters $\boldsymbol{\theta}_{\rm cos} = (\Omega_M,~H_0)$ and $K$ is the K-correction.
The rest-frame plateau duration is $T_{a,~\rm rest} = T_a/(1+z)$.
For a flat $\Lambda$CDM model,
\begin{equation}
D_L(z) = \frac{c(1+z)}{H_0}
\int_0^z 
\frac{dz'}{\sqrt{\Omega_M(1+z')^3 + (1-\Omega_M)}},
\label{eq:DL}
\end{equation}
and the Dainotti correlation, $\log L_a = a\log T_{a,~\rm rest} + b + \mathcal{N}(0,~\sigma_{\rm int})$ \citep{Dainotti2008, Dainotti2013}, which links these two observables.
Its free parameters $(a,\,b,\,\sigma_{\rm int})$ are fitted jointly with $\boldsymbol{\theta}_{\rm cos}$ via a circularity-free approach that avoids assuming a fiducial cosmology when computing $L_a$ \citep{Dainotti2022a}.

In this framework, reduced scatter in $\log F_a$ directly reduces the noise in $\log L_a$ via Eq.~(\ref{eq:La}), while reduced scatter in $\log T_a$ tightens the abscissa of the Dainotti plane.
The $19-49\%$ reductions in $\sigma(\log T_a)$ and $\sigma(\log F_a)$ demonstrated here therefore narrow each GRB's likelihood contribution, tightening the joint posterior on $(a,\,b)$.
The magnitude of this gain depends on how much of the total $L_a-T_a$ scatter is measurement error rather than intrinsic astrophysical dispersion: the intrinsic scatter is $\sigma_{\rm int} \approx 0.3-0.4$~dex \citep{Dainotti2011, Dainotti2013, Zhou2025}, while the measurement contribution, estimated from the $16^{\rm th}-84^{\rm th}$ percentile of the W07 parameter uncertainties across the 534 successfully fitted GRBs in our sample (Table~\ref{tab:EF/RC}), is $\sigma_{\rm meas} \approx 0.04-0.14$~dex for $\log T_a$ and $0.03-0.11$~dex for $\log F_a$.
Since $\sigma_{\rm int}$ dominates the error budget of the full sample, the per-source gain from reducing $\sigma_{\rm meas}$ alone is moderate; it is, however, more pronounced for the Platinum sub-sample of \citet{Dainotti2020}, where strict quality selection already suppresses both scatter components. 
It compounds across the full 545-source dataset, producing a measurable tightening of the ensemble posterior on the correlation parameters.

\citet{Mukherjee2026} showed that GRB distance moduli can be calibrated model-independently with neural networks; feeding reconstructed plateau parameters into $\Lambda$CDM inference is a natural next step.
GRBs alone cannot constrain $H_0$, which is degenerate with the correlation normalisation unless externally calibrated \citep{Khadka2021}; joint analyses of GRBs, quasars, BAO and SNe~Ia fill the redshift gap between SNe~Ia and the CMB, though the inferred parameters depend on the SN~Ia calibration adopted \citep{Dainotti2022, Bargiacchi2023}. 
The leverage of GRBs on the Hubble tension \citep{Verde2019, Planck2020, Dainotti2021, Riess2022}, therefore, lies in such joint fits at $z$ beyond the SN range \citep{Dainotti2026}.
Consistently, \citet{Zhou2025} found that adding 50 reconstructed points to 35 plateau light-curves left the intrinsic scatter and $\Omega_m=0.290\pm0.008$ unchanged within uncertainties, confirming that sample size matters more than light-curve sampling density. 
The gains reported here are therefore formal; quantifying their effect on $(\Omega_M,~H_0)$ requires a dedicated likelihood analysis with selection-bias correction via the Efron--Petrosian method \citep{Efron1992, Dainotti2013}, which we leave to future work.

\color{black}

 
The closure-relation analysis of \citep{Srinivasaragavan2020, Dainotti2021a, Dainotti2025} further demonstrated that improved constraints on $\alpha$ directly sharpen tests of fireball closure relations, which in turn constrain the electron spectral index $p$ and the circumburst medium density profile.
Our $24-49\%$ reduction in $\alpha$ uncertainties across the four models would enable a more decisive discrimination between ISM ($n = \mathrm{const}$) and stellar-wind ($n \propto r^{-2}$) environments for a larger fraction of the sample, with implications for GRB progenitor identification \citep{PIRAN1999}.

Beyond closure relations, plateau parameters also enter population studies of the long-GRB rate density, which serves as a tracer of the cosmic star-formation history. 
The observed rate departs from the star-formation rate density at low redshift, and proposed explanations include evolution of the luminosity function or of the jet opening angle, and a metallicity dependence of the progenitors \citep{LloydRonning2002, Petrosian2015, LloydRonning2019, Ghirlanda2022}. 
\citet{Khatiya2025} derived the rate density for GRBs with measured redshifts and X-ray plateaus.
They found that a simple evolution of the jet opening angle cannot reconcile it with the star-formation rate density, and \citet{Bal2025} extended the analysis to machine-learning-predicted redshifts \citep{Narendra2025}. 
Because these studies rely on plateau-derived quantities, the $\sim$19--45\% reductions in the uncertainties of $\log_{10}T_a$ and $\log_{10}F_a$ reported here, the largest for the broken-PL PINN prior, could reduce the uncertainty of the plateau luminosities entering such samples. 
We have not tested this because it would require repeating those analyses using reconstructed parameters.


 
This work introduces a physics-informed neural network (NN) approach for reconstructing GRB X-ray afterglow light-curves.
All previous ML applications in our earlier work
and in the broader GRB ML literature 
treat the physical model as a post hoc interpretive framework applied to the network output rather than as part of the training objective. 
The PINN instead embeds the governing differential equation directly into the optimisation process, so every weight update is guided by both data fidelity and physical consistency.

This is qualitatively different from the physics-inspired regularisation used in ReFANN, which penalises the squared gradient of the reconstructed flux to suppress oscillations, and in the Siamese model, which penalises positive flux gradients to encourage monotonic behaviour. 
These approaches impose generic shape constraints, whereas the PINN enforces the governing differential equation of the assumed afterglow model throughout training. 
Consequently, the PINN reconstruction in gap regions remains physically interpretable because it follows the solution of the governing differential equation rather than an unconstrained interpolation.

The principal current limitation is the single-prior assumption: each GRB is reconstructed under each of seven priors independently, and the per-prior statistics reported in Table~\ref{tab:reconstruction_comparison} reflect the full heterogeneity of the sample. 
A natural extension is a mixture-of-experts PINN \citep{Karniadakis2021, Cuomo2022} in which multiple physics branches contribute to the reconstruction with learned weights, enabling the network to interpolate between physical regimes, for example, transitioning smoothly from an energy-injection plateau to a post-jet decay within a single GRB. 
NN reconstruction of cosmological functions has been shown to benefit from analogous multi-component architectures by \citet{Dialektopoulos2023, Dialektopoulos2024}, providing a methodological blueprint for the GRB application.

Extending the PINN framework to multi-wavelength observations, incorporating optical and radio afterglow data alongside X-ray measurements, is a natural avenue for future work; the optical fundamental plane correlations of \citet{Dainotti2022b} provide a direct target for such an extension, and would enable closure-relation tests across multiple bands simultaneously \citep{Sari1998}, providing additional constraints on the cooling-break frequency, and reducing the degeneracy between the energy-injection and synchrotron priors.
 
 
A key differentiator among the four models is the mechanism used to construct the $95\%$ confidence interval. 
The PINN uses residual-based UQ: the epistemic uncertainty is estimated from the standard deviation of post-fit residuals at the observed timestamps, and the aleatoric component is obtained by interpolating the per-point observational uncertainties to the reconstruction grid and smoothing with a Gaussian kernel. 
This approach is conservative: the CI is wider where the PINN fit is poor and narrower where it closely tracks the data, avoiding the MC-dropout collapse issue that affects GELU networks in extrapolation regions \citep{Gal2016, Gal2017}.

ReFANN and Siamese use MC-Dropout with 200 stochastic forward passes \citep{Gal2016, Kendall2017}, combining epistemic uncertainty from the variance of predicted means with aleatoric uncertainty from a dedicated log-variance output head. This probabilistic framework provides uncertainty estimates for the reconstructed light-curves. However, because the epistemic component reflects uncertainty in the network weights rather than uncertainty arising from missing observations, the estimated confidence intervals may not fully capture the uncertainty associated with large observational gaps.

ReFANN's advantage lies in requiring no assumption about the underlying afterglow morphology at all: its squared-gradient regulariser only discourages oscillation, not any specific decay law, which is why its performance is comparable to the model-independent Polynomial Curve Fitting baseline of \cite{Kaushal2026} rather than to any physics-constrained method. 
This comes at the cost of over-smoothing sharp, narrow flares when the surrounding gap is wide, since the gradient penalty cannot distinguish a genuine flare from noise once observations are sparse.
The Siamese network's shared-weight, learnable-anchor design effectively lets the network measure each GRB's local behaviour relative to its own, self-learned reference point rather than to a fixed analytical curve, which is why it slightly outperforms ReFANN while keeping a comparably low outlier rate: the monotonicity penalty (Eq.~\ref{eq:regularisation}) is weaker than a full physics prior but still discourages the single most common unphysical failure mode (rising flux in a poorly sampled gap).

The PQR model is the most conservative of the four methods by design: because it bounds the reconstruction with three independently fit quantile levels rather than a single mean-and-variance model, it cannot be pulled toward an incorrect physical slope the way the PINN can, nor can it introduce spurious smoothness the way ReFANN's gradient penalty can. 
This robustness is precisely why it has both the lowest outlier fraction of the four models and the smallest mean uncertainty reduction: the pinball loss optimises for calibrated coverage of the observed scatter, not for the tightest possible bound on $T_a$ and $F_a$. 
It is therefore best suited to bulk, homogeneous-quality processing of large samples where a uniform, distribution-free uncertainty is preferable to per-GRB physical tailoring.

The PQR model CI is fundamentally distribution-free: the lower and upper bounds are the $\tau = 0.025$ and $\tau = 0.975$ quantile fits, with no assumption of Gaussianity. 
This is the most statistically conservative choice and is particularly appropriate for the long-tailed flux distributions typical of GRB afterglows. 
A fully rigorous coverage guarantee could be obtained via conformalised PQR model \citep{Romano2019}, which we note as a future extension.
 
Across all four methods, the CI widths reported in the representative light-curves of Figure~\ref{fig:Reconstrusted_grbs} confirm the expected pattern: narrower bands where observations are dense, wider bands in temporal gaps, with the PINN bands being morphologically the smoothest (because they follow the analytical prior) and the Quantile bands being the broadest (because the distribution-free construction is the most conservative).

The reconstruction and uncertainty-reduction method developed in this work can be used with real-time alert systems for future wide-field observatories, such as the Vera C. Rubin Observatory \cite{Ivezi2019}, SVOM \cite{Wei2016}, and Einstein Probe \cite{Yuan2022}.  
In such a deployment, the $\chi^2-$based prior-ranking procedure of Section~\ref{section:results} could be run automatically on early-time data to select a PINN physics prior on the fly, while joint XRT, optical, and radio coverage would allow the closure-relation and cooling-break tests discussed above to be performed across bands rather than in X-rays alone.


\section{Conclusion} \label{section:conclusion}

We present four new models for reconstructing GRB X-ray afterglow light-curves.
These models are tested on the same sample of 545 GRBs and with the W07 fitting procedure used in previous studies in this series.
All four models introduced here decrease the fractional uncertainties in the plateau parameters $\log_{10}T_a$, $\log_{10}F_a$, and $\alpha$ compared with the original observations. However, the level of improvement and the number of outliers vary between the methods, depending on how strongly each model is based on a particular physical or functional form.

We show that the proposed methods can effectively model GRB light-curves and reconstruct the missing temporal regions in the observed LCs. The main results are summarised below:

\begin{itemize}

\item The PINN framework, evaluated across seven physics-informed priors, achieves the largest uncertainty reductions among the four new models introduced in this work. 
Its best-performing prior (broken PL) reduces the error fraction by $-40.65\%$ for $\log_{10} T_a$,~ $-44.68\%$ for $\log_{10} F_a$, and $-48.84\%$ for $\alpha$, roughly twice the improvement obtained by ReFANN, Siamese, or Quantile (Table~\ref{tab:reconstruction_comparison}). 
This gain comes with a considerably higher incidence of outliers ($3-20\%$ depending on the parameter and prior) versus $\lesssim4\%$ for the other three models, reflecting a trade-off between the strength of the physics-informed regularisation and its robustness to GRBs that do not conform to any of the seven assumed decay morphologies.

\item A reduced-$\chi^2$ ranking of the seven priors favours the simple PL prior for $64.7\%$ of the reconstructed sample, consistent with the $55.3\%$ of the dataset independently classified as morphologically simple (``Good") GRBs (Section~\ref{Dataset Selection}), while the physically motivated jet-break prior, appropriate for only $\approx 12\%$ of Swift-XRT afterglows that show a genuine achromatic jet break \citep{Racusin2009}, is, correspondingly, both the least frequently selected and the weakest performing of the seven when applied indiscriminately.
While the broken PL PINN prior delivers the largest quantitative uncertainty reduction of the four models introduced here, its principal contribution is methodological: a physically constrained, differentiable reconstruction framework that complements the purely data-driven ReFANN, Siamese, and Quantile approaches.

\item The Siamese model achieves the best performance among the three data-driven models: 22.88\%, 22.83\%, and 26.93\% for $\log(T_a)$, $\log(F_a)$, and $\alpha$ respectively, with outlier fractions below 3.7\% for all three parameters. Its shared-weight, learnable-reference-anchor design acts as a soft structural prior without committing to a specific decay law, giving it the best balance of reduction and robustness among the models that make no explicit physical assumption.

\item The ReFANN model yields a significant reduction in uncertainty of $19.83\%$ for $\log T_a$, $21.43\%$ for $\log F_a$, and $26.51\%$ for $\alpha$. 
Based on the ranking matching Table~\ref{tab:reconstruction_comparison}, these reductions place ReFANN third among the four newly introduced models, ahead of PQR model but behind the PINN and Siamese architectures across all three parameters.
Its physics-inspired gradient regularisation makes no assumption about which afterglow morphology applies, which is also why it slightly over-smooths sharp, narrow features when the surrounding data gap is wide.

\item The PQR model provided a robust, distribution-free approach to light-curve reconstruction, achieving uncertainty reductions of $18.92\%$ for $\log T_a$, $20.04\%$ for $\log F_a$, and $24.42\%$ for $\alpha$ (Table~\ref{tab:reconstruction_comparison}), with the lowest outlier fractions of all four models (2.94\%, 4.04\%, and 1.28\%). By dynamically scaling its polynomial degree to the available data and minimising the Pinball Loss, the model remained resistant to over-fitting in sparsely sampled GRBs while still capturing the asymmetric dispersion characteristic of afterglow flux, at the cost of the smallest average uncertainty reduction, a deliberate trade of maximal gain for calibrated, assumption-free coverage. 
The combination of gap-aware point insertion with MC noise sampled from each GRB's own error distribution ensured that reconstructed points retained physically realistic statistical properties rather than introducing artificial smoothing bias.

\end{itemize}

Set against the broader five-paper reconstruction series, the broken-PL PINN prior achieves uncertainty reductions comparable to those of the CSS model introduced by \cite{Gupta2026}, the current strongest data-driven result in the series, while additionally providing physically interpretable gap-region reconstructions grounded in synchrotron afterglow theory. 
The Siamese network establishes a new best-in-class result among purely data-driven approaches that incorporate no smoothness or physical prior beyond a monotonicity penalty, and all four models introduced here clearly outperform the weaker neural baselines of the earlier series: BNN, DGP, and TCN.
The broken-PL PINN prior and the CSS model of \citet{Gupta2026} both exceed the $47.5\%$ uncertainty-reduction threshold identified by \citet{Dainotti2022a} as sufficient for GRBs to approach the $\Omega_M$ precision of the \citet{Betoule2014} SNe~Ia sample, with our $\alpha$ reductions crossing this mark under both models.
The pathway from these parameter improvements to quantitative $\Lambda$CDM constraints, through a circularity-free joint fit of the Dainotti correlation and the luminosity-distance relation, with Efron--Petrosian selection-bias correction, is laid out in Section~\ref{section:Discussion} and is planned as a dedicated follow-up analysis.

We show that the proposed methods can effectively model GRB light-curves and fill temporal gaps in the observations. By comparing the different approaches in a consistent and systematic way, this study identifies their main strengths and limitations and provides a useful reference for developing future machine-learning methods for GRB light-curve reconstruction.

\section*{CRediT authorship contribution statement}

\textbf{Ayush:} Conceptualisation, methodology, supervision, software, validation, investigation, formal analysis, writing (original draft \& review/editing).
\textbf{Ritik Kumar:} Conceptualisation, methodology, supervision, software, validation, investigation, formal analysis, writing (original draft \& review/editing).
\textbf{Maria G. Dainotti:} Conceptualisation, methodology, supervision, software, data curation, validation, investigation, writing review/editing, project administration, funding acquisition, and resources.
\textbf{Vipul Sharma:} Methodology, software, validation, investigation, writing the original draft, review \& editing.
\textbf{Amit Shukla:} Conceptualisation, methodology, supervision, validation, investigation, writing review/editing, project administration, funding acquisition, and resources.
\textbf{Dieter Hartmann:} Validation, investigation, writing review/editing.

\section*{Declaration of competing interest}

The authors declare that they have no known competing financial interests or personal relationships that could have appeared to influence the work reported in this paper.

\section*{Data availability}

Data will be made available upon reasonable request to the corresponding author. 

\section*{Acknowledgements}
The authors thank Khushi Gupta for her valuable insights and fruitful discussions in the construction of the framework. of this paper.
We are grateful to the Council of Scientific and Industrial Research (CSIR) for funding AG via a CSIR-JRF fellowship under grant 09/1022(13386)/2022-EMR-I. 
The URJA research group in the Department of Astronomy, Astrophysics, and Space Engineering at IIT Indore provided computational resources.
M.G.D. acknowledges the support of the JSPS Grant-in-Aid for Scientific Research (KAKENHI) (A), Grant Number JP25H00675.
For open access, the author has applied a CC-BY public copyright licence to any Author Accepted Manuscript (AAM) version arising from this submission.

\bibliographystyle{aastex701}
\bibliography{bibliography}

\end{document}